\documentclass{aa}  

\usepackage{graphicx}
\usepackage{booktabs}  

\usepackage{xcolor}

\usepackage{txfonts}
\begin{document}

   \title{Bar formation and secular evolution in live Milky Way-like galaxies}

   \author{Eva Durán-Camacho
          \inst{1,2}
          \and
          Sioree Ansar\inst{3,4}\fnmsep
          \and
          Paula D. López
          \inst{5}
          \and
          Johan H. Knapen
          \inst{1,2}
          \and
          Jesus Falcón-Barroso
          \inst{1,2}
          \and\\
          Adriana de Lorenzo-Cáceres
          \inst{2,1}
          \and
          Jairo Méndez-Abreu
          \inst{2,1}
          }

   \institute{Instituto de Astrof\'{i}sica de Canarias, V\'{i}a L\'{a}ctea S/N, E-38205 La Laguna, Spain\\
              \email{eva.duran@iac.es}
        \and
             Departamento de Astrof\'{i}sica, Universidad de La Laguna, E-38206 La Laguna, Spain
        \and
             Institute for Computational Cosmology, Department of Physics, Durham University, South Road, Durham DH1 3LE, UK
        \and
             Centre for Extragalactic Astronomy, Department of Physics, Durham University, South Road, Durham DH1 3LE, UK
        \and
             Departamento de Astronom\'{i}a, Universidad de La Serena, Av. Ra\'{u}l Bitr\'{a}n 1305, La Serena, Chile
             }

   \date{Received August 8, 2026; accepted November 18, 2026}

 
  \abstract
   {Bars and their boxy/peanut (B/P) shapes are key drivers of secular evolution in disc galaxies, redistributing angular momentum and reshaping both the radial and vertical structure of stellar discs.  Bars are present in disc galaxies spanning the full range of bulge content, from bulgeless to bulge-dominated hosts. Although bars are known to grow and slow down, the connection between these processes and the structural properties of the host galaxy remains poorly understood. }
   {We investigate how the formation and evolution of bars and B/P structures depend on the initial stellar mass and bulge fraction of Milky Way-like galaxies. We further explore the links between the kinematic settling of the disc, bar formation, resonance structure, and vertical thickening}
   {We analyse a suite of fifteen isolated Milky Way-like hydrodynamical simulations evolved for 3 Gyr. The models span a range of three stellar masses and five bulge fractions, and are generated with {\sc arepo}, including live stellar, bulge, and dark matter components that allow self-consistent evolution. We measure bar properties using Fourier decomposition techniques, determine resonance locations from circular and epicyclic frequencies, and quantify B/P evolution using diagnostics based on the vertical stellar distribution}
   {The timing of bar formation depends strongly on galaxy structure. Low-mass galaxies generally form bars only after the global disc kinematics have settled, whereas the most massive systems can develop bars while the stellar orbital distribution is still evolving. All bars slow down with time and grow both in strength and size, while remaining consistent with the slow-bar regime. B/P structures form shortly after the bar appears and continue growing throughout the simulations, with more massive and disc-dominated galaxies developing vertically thickened structures substantially earlier than lower-mass systems. Only one model exhibits clear evidence of a classical buckling event, while all others form B/P bulges through smooth secular evolution.}
   {The initial stellar mass and bulge fraction of these MW-like galaxies strongly regulate the timing of bar and B/P formation. Our results reveal a coherent evolutionary sequence linking bar growth, bar slowdown and the assembly of B/P structures, whose timing is set by the initial structure of the disc rather than by the settling of its global kinematics}


   \maketitle
\abstract

\abstract

%

\section{Introduction}

Stellar bars are among the most prominent elongated structures at the centres of disc galaxies, with observations indicating that more than half of nearby spiral galaxies host a central bar \citep{Sellwood1993,Eskridge2000,Knapen2000,MenendezDelmestre2007,Masters2011,Erwin2018}. These non-axisymmetric structures play a fundamental role in galaxy evolution by redistributing angular momentum between the disc, bulge, and dark matter halo, driving gas inflows, modifying stellar orbits, and reshaping both the radial and vertical structure of galaxies \citep{Athanassoula2002, kormendy2004,Athanassoula2013,Sellwood2014}. As a result, bars are now recognised as one of the primary drivers of secular galaxy evolution.

The Milky Way (MW) provides a particularly important laboratory for studying bar-driven evolution. Evidence for a Galactic bar was first inferred from gas kinematics \citep{de1964comparison} and later confirmed through infrared photometry, star counts, and stellar kinematics \citep{binney1991understanding,blitz1991direct,Dwek1995,stanek1994}. More recently, large spectroscopic and astrometric surveys, combined with \textit{Gaia} measurements, have constrained the bar length to approximately $4$–$5$\,kpc and measured pattern speeds in the range $\sim35$–$45~{\rm km\,s^{-1}\,kpc^{-1}}$ \citep{Bovy2019,Sanders2019,queiroz2021milky,Drimmel2023}. Observations further reveal that the inner MW exhibits a pronounced boxy/peanut (B/P) morphology, indicating that our Galaxy has experienced substantial bar-driven secular evolution \citep{McWilliam2010,Ness2012,wegg2015structure}. Notably, kinematic modelling of the Galaxy finds no evidence for a classical bulge component \citep{Shen2010}, limiting any contribution to less than $8\%$ of the disc mass, and indicating that the MW's bar-driven secular evolution has taken place in the absence of a dynamically hot spheroid.  

Numerical simulations have long shown that bars can arise spontaneously from global instabilities in dynamically cold stellar discs \citep{Hohl1971,Ostriker1973,Efstathiou1982}. Once formed, bars evolve through the redistribution of angular momentum within the galaxy. Angular momentum is emitted primarily by near-resonant material in the inner disc and absorbed by the outer disc, bulge, and dark matter halo, causing the bar to grow stronger and longer while simultaneously slowing down \citep{LyndenBell1972,Tremaine1984,Athanassoula2003,Bureau2005,Athanassoula2013,Sellwood2014,Fragkoudi2021}. This secular evolution has been observed in both idealised and cosmological simulations, where bars continue to evolve over several Gyr through the exchange of angular momentum between galactic components \citep{kataria2018study,rosas-guevara2022,Wheeler2023,Ghosh2024, Ansar2024, Ansar2025, Fragkoudi2025}. Simulations employing live dark matter haloes have demonstrated that the halo plays a central role in this process, acting as a reservoir that can absorb angular momentum at resonances and thereby substantially alter the growth rate, pattern speed, and long-term evolution of the bar \citep{Debattista2000,Athanassoula2002,MartinezValpuesta2006,Sellwood2014,Petersen2021}.

The pattern speed of a bar is a key dynamical quantity as it determines the locations of the principal resonances within the galaxy. In particular, the inner Lindblad resonance (ILR), corotation resonance (CR), and outer Lindblad resonance (OLR) regulate the transfer of angular momentum and influence the redistribution of stars and gas throughout the disc \citep{Contopoulus1989,binney2011galactic}. Observational measurements indicate that most nearby bars are consistent with so-called fast bars, for which the ratio $\mathcal{R}=R_{\rm CR}/R_{\rm bar}$ lies in the range $\sim1.0$–$1.4$ \citep{Corsini2011,Aguerri2015,Gadotti2015,Guo2019,cuomo2021bar,cuomo2024}. Understanding how bar strength, size, pattern speed, and resonance structure evolve together therefore provides important insight into the secular evolution of barred galaxies.

An important consequence of bar evolution is the formation of boxy/peanut-shaped (B/P) structures. These vertically thickened inner structures are observed in a large fraction of barred galaxies \citep{Lutticke2000I,Lutticke2000II,MendezAbreu2008II,Erwin2013,MendezAbreu2014,Erwin2016,Erwin2017,Ciambur2017,MendezAbreu2019,li2023bucklingII} and are now widely interpreted as the vertically extended inner regions of stellar bars \citep{Combes1990,Athanassoula2005,Fragkoudi2021,Ghosh2023}. Early numerical studies associated their formation with the buckling instability, in which the bar bends vertically and redistributes stars onto high-latitude orbits \citep{Raha1991,Merritt1994,MartinezValpuesta2006}. However, subsequent work has shown that B/P structures can also develop gradually through resonant orbit trapping and secular vertical heating without a strong buckling phase \citep{Quillen2002,Debattista2006,Fragkoudi2017,Sellwood.Gerhard.2020, Ghosh2024,PaulaLopez2025}. The relative importance of these pathways, and their dependence on galaxy structure, remains an active area of research.

Although the general picture of bar-driven secular evolution is well established, several important questions remain open. It is still unclear how the onset of the bar instability happens relative to the settling of the global disc kinematics, how the timing of bar and B/P formation depends on galaxy structure and to what extent bar strength, length, pattern speed, and resonance structure remain coupled throughout secular evolution. Furthermore, while stellar mass, bulge fraction, and halo properties are all expected to influence bar growth, their relative importance remains poorly constrained.

In this paper, we investigate the formation and secular evolution of bars and B/P structures in a suite of self-consistent MW-like hydrodynamical simulations. In Section~\ref{sec: numerical sims}, we summarise the properties of the models and the initial conditions analysed throughout this work. In Section~\ref{sec: disc formation}, we define the kinematic settling time of the stellar disc and examine how it varies across our suite of simulations. In Section~\ref{sec: bar properties}, we present the methodology used to characterise the bars, including measurements of bar strength, length, pattern speed, and resonance structure, together with a summary of the bar properties at the end of the simulations. In Section~\ref{sec: results}, we investigate the relationships between kinematic settling of the disc and bar formation, the evolution of bar dynamics, and the growth of B/P structures. Finally, we discuss the implications of our findings and summarise our main conclusions in Section~\ref{sec: conclusions}.


\section{Numerical simulations}
\label{sec: numerical sims}

To investigate the formation and secular evolution of stellar bars in MW-like galaxies, we analyse a suite of fifteen self-consistent $N$-body simulations spanning a range of stellar masses, bulge fractions, and dark matter halo concentrations. The simulations were originally introduced by \citet{DuranCamacho2024} (see also \citealt{DuranCamacho2026}). These were constructed to reproduce key structural and kinematic properties of the MW. In this section, we briefly summarise the numerical methodology and the construction of the initial galaxy models, referring the reader to the original work for a complete description of the simulation setup and validation tests.

\subsection{Numerical method}

All simulations considered in this study were generated using the moving-mesh hydrodynamical code {\sc arepo} \citep{springel2010pur,pakmor2016improving}\footnote{{\sc arepo}'s public repository can be found at \url{https:// arepo-code.org/}}. {\sc arepo} combines the advantages of Eulerian grid-based methods and Lagrangian approaches by solving the hydrodynamic equations on a moving Voronoi mesh, while gravitational forces are computed using a TreePM scheme inherited from the {\sc gadget} family of codes. This allows the stellar, gaseous, and dark matter components to evolve self-consistently within a fully live gravitational potential.

The simulations follow the evolution of isolated MW-like galaxies represented by stellar, gaseous, and dark matter components. The dark matter is modelled with an Hernquist profile, while the stellar component consists of a spherical bulge, also following a Hernquist profile, and a disc adopting an exponential surface density profile. The gaseous component is distributed in an exponential disc that initially follows the stellar density distribution. In all models, the gas mass is fixed to $20\%$ of the total stellar mass. The stellar and dark matter components are represented by live potentials that evolve self-consistently throughout the simulation. No external or analytical gravitational potential is imposed, allowing non-axisymmetric structures such as spiral arms and stellar bars to emerge naturally from dynamical instabilities within the disc \citep[see also][]{pettitt2015morphology,baba2017eventful,iles2022differences}. 

As the gas contributes only a small ($\sim 1\%$) fraction of the total galaxy mass (i.e.  $\sim 20\%$ of the stellar disc mass), it is treated as non-self-gravitating cells and its self-gravity is neglected. This approximation has a negligible impact on the large-scale dynamics governing bar formation while substantially reducing the computational cost. The gas is evolved using isothermal equations of state with a sound speed of $c_{\rm s}=10{\rm kms^{-1}}$, corresponding to a temperature of approximately $10^4\,{\rm K}$. This suppresses small-scale fragmentation and allows the simulations to be evolved for several Gyr, enabling the study of the formation and long-term secular evolution of stellar bars. 

The gas component is refined according to mass and volume based criteria, with a target cell mass of $10^{3}\,{\rm M_\odot}$. Gas cells are constrained to volumes between $27\,{\rm pc}^3$ and $1.25\times10^{8}\,{\rm pc}^3$, corresponding to cubes with side lengths of 3 and 500 pc, respectively. The stellar particles have a mass resolution of $8\times10^{3}\,{\rm M_\odot}$ and a gravitational softening length of 50 pc, while the dark matter particles have a mass resolution of $9\times10^{5}\,{\rm M_\odot}$ and a softening length of 300 pc. Gas cells follow {\sc arepo}’s adaptive gravitational softening scheme, such that their softening length varies in response to changes in cell mass and volume. The simulations are evolved within a cubic domain of side length 100 kpc using periodic boundary conditions. Since the gaseous disc remains confined to the central $\sim30$ kpc of the computational domain, these boundaries do not influence the evolution of the galaxy. The most evolved models contain approximately $2.3$–$2.6\times10^{7}$ gas cells.

\subsection{Initial conditions}
\label{sec: intitial conditions}
The initial conditions were generated using the {\sc makenewdisk} code \citep{springel2005modelling} and are identical to those presented in \cite{DuranCamacho2024} and \cite{DuranCamacho2026}. The simulation suite consists of fifteen MW-like galaxy models designed to explore the influence of stellar mass and central mass concentration on the formation and secular evolution of stellar bars. The models span three values of the total stellar mass, $\rm{M}_{\star}=4.25$, $5.25$, and $6.25\times10^{10}\, {\rm M_{\odot}}$, motivated by observational estimates of the MW stellar mass \citep[e.g.][]{mcmillan2016mass,cautun2020milky}. For each stellar mass, we consider five bulge mass fractions, $B_{F}=M_b^\star/M_\star$, ranging from 25\% to 5\% in decrements of 5\%. This results in a total sample of fifteen models, and their initial rotation curves can be seen in Fig.~\ref{fig:rotation_curve}.

The model grid was designed to span a representative range of MW analogues while maintaining a common global mass budget. This allows us to examine how stellar mass and bulge fraction affect the subsequent formation, growth, and secular evolution of stellar bars. All models adopt a disc scale length of $h=3$ kpc. The disc scale height and bulge scale length are defined relative to $h$, resulting in fixed values of $0.29$ kpc and $0.35$ kpc, respectively. The total galaxy mass is also held constant throughout the suite, with $M_{\rm total}=9.52\times10^{11}\,{\rm M_\odot}$. Similarly, the disc gas mass is fixed at $M_{\rm d}^{\rm gas}=8.6\times10^{9}\,{\rm M_\odot}$, while the gaseous halo mass is assumed to be 40\% of the disc gas mass, corresponding to $M_{\rm h}^{\rm gas}=3.44\times10^{9}\,{\rm M_\odot}$. Variations in stellar mass are hence compensated by corresponding changes in the dark matter halo mass, while the halo concentration parameter is adjusted to reproduce realistic MW-like rotation curves. The resulting halo concentrations range from $c=8-10$ in the least massive models to $c=3-4$ in those with the highest stellar mass, so the most strongly self-gravitating stellar discs also sit in the least centrally concentrated haloes. Since the central density varies with both bulge fraction and halo concentration, the local dynamical time, which scales as $\rho(R)^{-1/2}$, is not identical across the suite, and absolute formation times should be compared with this in mind.

The initial conditions are constructed to be in approximate dynamical equilibrium, following the moment-based approach of \citet{springel2005modelling} \citep[see also][for similar results solving Jeans equations]{pfenniger1991structure,pfenniger1993}. We note that equilibrium in this case does not imply stability, as a disc stable against all classes of perturbation would not form bars or spiral structure. The remaining parameters defining each galaxy model are summarised in Appendix~\ref{appendix initial conditions}, including a further discussion of their initial stability. For a complete description of the initial condition generation and equilibrium construction procedure, we refer the reader to \citet{DuranCamacho2024} and \citet{springel2005modelling}.

\section{Kinematic settling of the stellar disc}
\label{sec: disc formation}

The formation of a stellar bar requires the presence of a sufficiently massive and rotationally supported disc. Consequently, before investigating the onset of bar formation, it is useful to determine when each galaxy develops a well-defined stellar disc. Comparing the disc formation time with the subsequent bar formation time allows us to assess whether bars emerge shortly after the disc becomes dynamically established or whether a significant delay exists between these two evolutionary stages.

In order to measure the disc formation time, we quantify the degree of rotational support. And for that, we employ the corotation parameter, $\kappa_{\rm co}$, introduced by \citet{Correa2017}. This parameter measures the fraction of the stellar kinetic energy associated with ordered rotation about the disc symmetry axis and is defined as

\begin{equation}
    \kappa_{\rm CO} = \frac{K_{\rm rot}}{K} = \frac{1}{K} \sum^{L_{z,i}>0}_{i}\frac{1}{2} m_i \left(\frac{L_{z,i}} {m_iR_i} \right)^2,
\end{equation}

where $m_i$ is the mass of stellar particle $i$, $L_{z,i}$ is its angular momentum about the $z$-axis, and $R_i$ is its projected galactocentric radius in the disc plane. The sum is performed over all stellar particles rotating in the same sense as the net disc rotation, while $K$ denotes the total stellar kinetic energy. Physically, $\kappa_{\rm co}$ measures the relative importance of ordered rotation compared to random motions. Galaxies dominated by rotationally-supported discs exhibit high values of $\kappa_{\rm co}$, whereas spheroid-dominated systems have substantially lower values. 

In cosmological simulations, $\kappa_{\rm co}$ is often used to identify the epoch at which a disc first forms \citep[e.g.][]{Dillamore2024}. However, our simulations differ in that they are initialised with an equilibrium stellar disc already in place (see Sect.~\ref{sec: intitial conditions}). Consequently, the concept of a disc formation time is not directly applicable.

Although the stellar disc is present from the beginning of the simulations, the system initially undergoe a period of rapid evolution. As shown in Appendix~\ref{appendix initial conditions}, every model begins with a minimum Toomre $Q$ parameter below unity, so the stellar discs are locally unstable to axisymmetric perturbations and heat-up until $Q$ is restored to order unity. The gaseous disc settles vertically over a comparable period, and the initial velocity distribution adjusts to the exact potential of the discretised model. The models therefore possess an approximate equilibrium, but not a stable one. During this period, the degree of rotational support, quantified by $\kappa_{\rm co}$, evolves rapidly before approaching a nearly constant value. We therefore use $\kappa_{\rm co}$ not to identify the formation of the disc itself, but to determine when its global dynamical state reaches a quasi-steady configuration. Physically, $\kappa_{\rm co}$ measures the fraction of kinetic energy invested in ordered co-rotating motion, such that decreases in $\kappa_{\rm co}$ indicate the conversion of rotational support into more dispersion-dominated or non-circular stellar orbits. This evolution is expected not only during the initial relaxation of the galaxy, but also during the subsequent growth of a stellar bar, which redistributes stars onto elongated bar-supporting orbits. We therefore define the kinematic settling time, as the first snapshot after which $\kappa_{\rm co}$ enters a sustained plateau, identified by a Savitzky–Golay smoothing of the time series followed by the detection of a prolonged interval ($\geq500$ Myr) over which the local slope remains both small and approximately constant. This provides a reproducible reference time against which the onset of bar formation can be compared.

\begin{figure}
    \centering
    \includegraphics[width=\columnwidth]{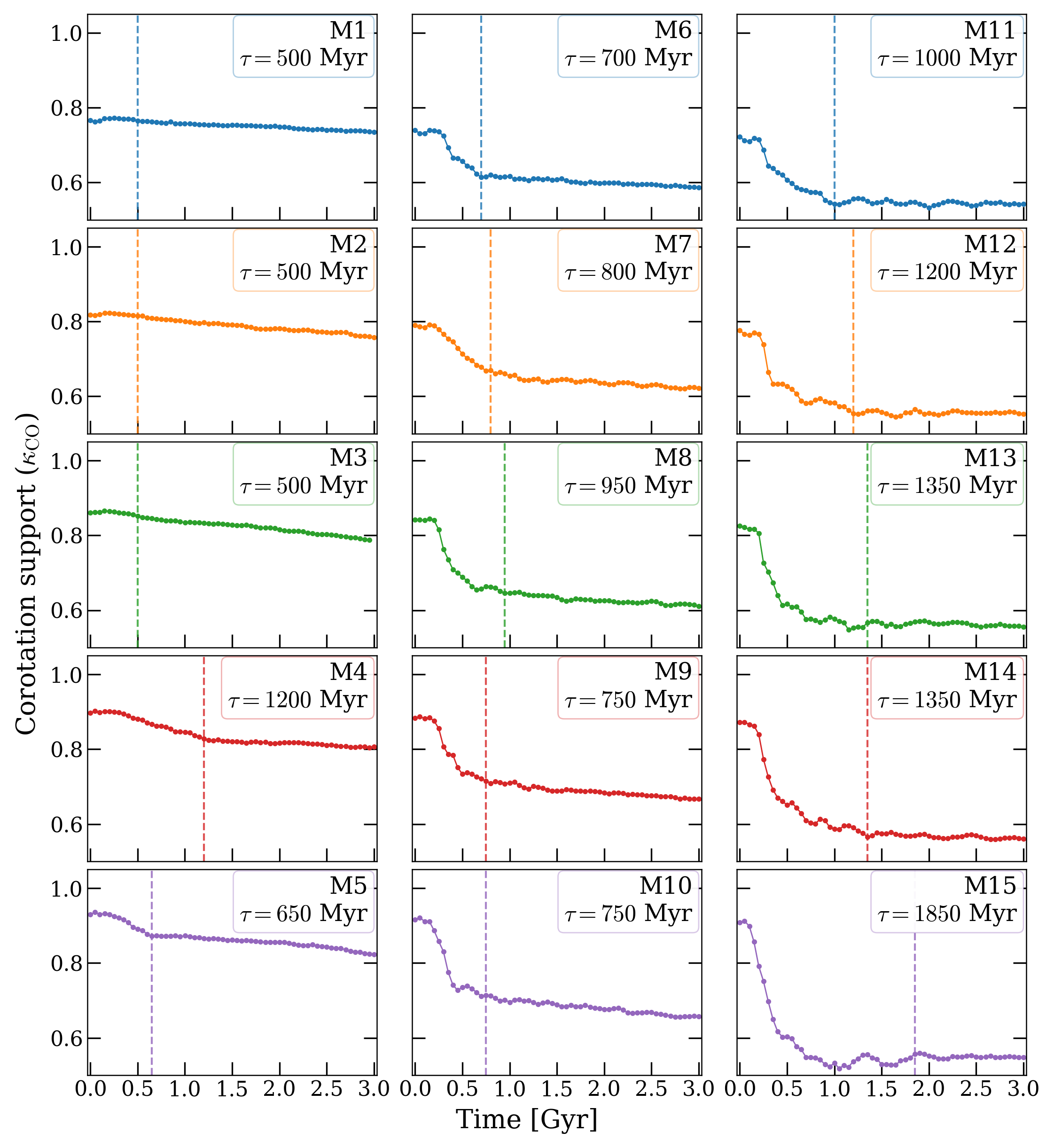} 
    \caption{
        Time evolution of the corotation parameter, $\kappa_{\rm co}$, for the fifteen MW-like galaxy models. The panels are arranged as in Fig.~\ref{fig:rotation_curve}: columns correspond to increasing initial stellar mass, while rows correspond to decreasing stellar bulge mass fraction. The vertical dashed line in each panel marks the kinematic settling time, $\tau$, defined as the first snapshot after which $\kappa_{\rm co}$ reaches a quasi-steady plateau. The value of $\tau$ is indicated in the upper-right corner of each panel.
    }
    \label{fig:kappa_CO}
\end{figure}

Figure~\ref{fig:kappa_CO} shows the evolution of the corotation parameter for all fifteen galaxy models. The layout follows that of Fig.~\ref{fig:rotation_curve}, with columns corresponding to increasing stellar mass and rows corresponding to decreasing bulge fraction. The vertical dashed line in each panel marks the kinematic settling time, $\tau$.

All models exhibit an initial relaxation phase characterised by a decline in $\kappa_{\rm co}$, indicating a redistribution of kinetic energy from ordered rotation into random motions. Following this phase, $\kappa_{\rm co}$ approaches a nearly constant value and fluctuates only weakly around a plateau, marking the onset of a quasi-steady disc configuration. The duration of this relaxation phase varies substantially across the model suite.

A clear trend is observed with stellar mass. The least massive galaxies (left column) reach a quasi-steady configuration within $\sim0.5$–$1.2$ Gyr and retain high levels of rotational support, with plateau values of $\kappa_{\rm co}\approx0.75$–$0.85$. In contrast, the most massive galaxies (right column) undergo a more prolonged relaxation phase, requiring up to $\sim1.9$ Gyr to settle, and settle at significantly lower values of $\kappa_{\rm co}\approx0.55$–$0.60$. This lower level of rotational support is closely associated with the earlier formation of strong stellar bars in these systems. As the bars grow, they redistribute stars from nearly circular disc orbits onto elongated bar-supporting and other non-circular orbits through the exchange of angular momentum, naturally reducing the fraction of kinetic energy contained in ordered rotation. The lower plateau values of $\kappa_{\rm co}$ in the most massive galaxies therefore likely reflect both their more prolonged initial relaxation and the stronger dynamical influence of the bars that subsequently develop.

The bulge fraction also influences the settling timescale. At fixed stellar mass, models with larger bulges begin the simulations with systematically lower values of $\kappa_{\rm co}$, reflecting the larger contribution of the dispersion-supported bulge to the total stellar kinetic energy. These systems also generally reach a quasi-steady state earlier than their low-bulge counterparts. This behaviour is particularly evident in the highest-mass galaxies, where $\tau$ increases from $\sim1.0$ Gyr for $B_F=25\%$ to nearly $1.9$ Gyr for $B_F=5\%$. Conversely, galaxies with smaller bulges undergo a more prolonged decline in $\kappa_{\rm co}$ as stronger bars develop and progressively redistribute stars from nearly circular disc orbits onto elongated bar-supporting and other non-circular orbits. The measured settling times therefore reflect both the initial dynamical contribution of the bulge and the subsequent secular evolution driven by bar growth.

Overall, these results show that the evolution of the global rotational support depends strongly on both stellar mass and bulge fraction. The measured settling time, $\tau$, reflects the combined effects of the initial relaxation of the galaxy and the subsequent redistribution of stellar orbits associated with bar growth. Comparing this kinematic settling time with the onset of bar formation therefore provides a useful way of assessing whether bars emerge while the stellar component is still undergoing significant dynamical evolution or only after the global orbital distribution has reached a quasi-steady state.

\section{Bar properties}
\label{sec: bar properties}

The formation of a stellar bar is only the first stage of its evolution. Once formed, bars continue to grow, exchange angular momentum with the surrounding galaxy, and undergo structural and dynamical transformations over several Gyr. In order to characterise these evolutionary processes, measurements of morphology and dynamics of the bar are required. In this section, we describe the methods used to quantify the main properties of the bars present in our simulations.

\subsection{Bar strength and formation time}
\label{sec: bar strength}

The strength of a stellar bar is one of its most fundamental properties, as it quantifies the amplitude of the non-axisymmetric perturbation imposed on the galactic potential. Stronger bars are generally more efficient at redistributing angular momentum, driving radial migration, funnelling gas towards the central regions, and inducing secular evolution within the host galaxy \citep[e.g.][]{Athanassoula2003,kormendy2004,Athanassoula2013}. Consequently, measuring the bar strength and determining the epoch at which a bar first forms are essential for understanding the subsequent dynamical evolution of the galaxy.

Bar identification is based on both the amplitude ($A_2$), used as a proxy for bar strength, and phase ($\phi_2$) of the $m=2$ Fourier mode of the stellar surface density distribution. A stellar bar corresponds to a coherent elongated structure and therefore exhibits an approximately constant phase angle over a significant radial extent, whereas spiral arms are characterised by a rapidly varying phase angle. We therefore identify the bar formation time, $\tau_{\rm bar}$, only when both a sufficiently large Fourier amplitude ($A_2>0.25$) and a coherent phase profile are present. The precise implementation of this criterion is described in Appendix~\ref{appendix bar detection}. 

\begin{figure}
    \centering
    \includegraphics[width=\columnwidth]{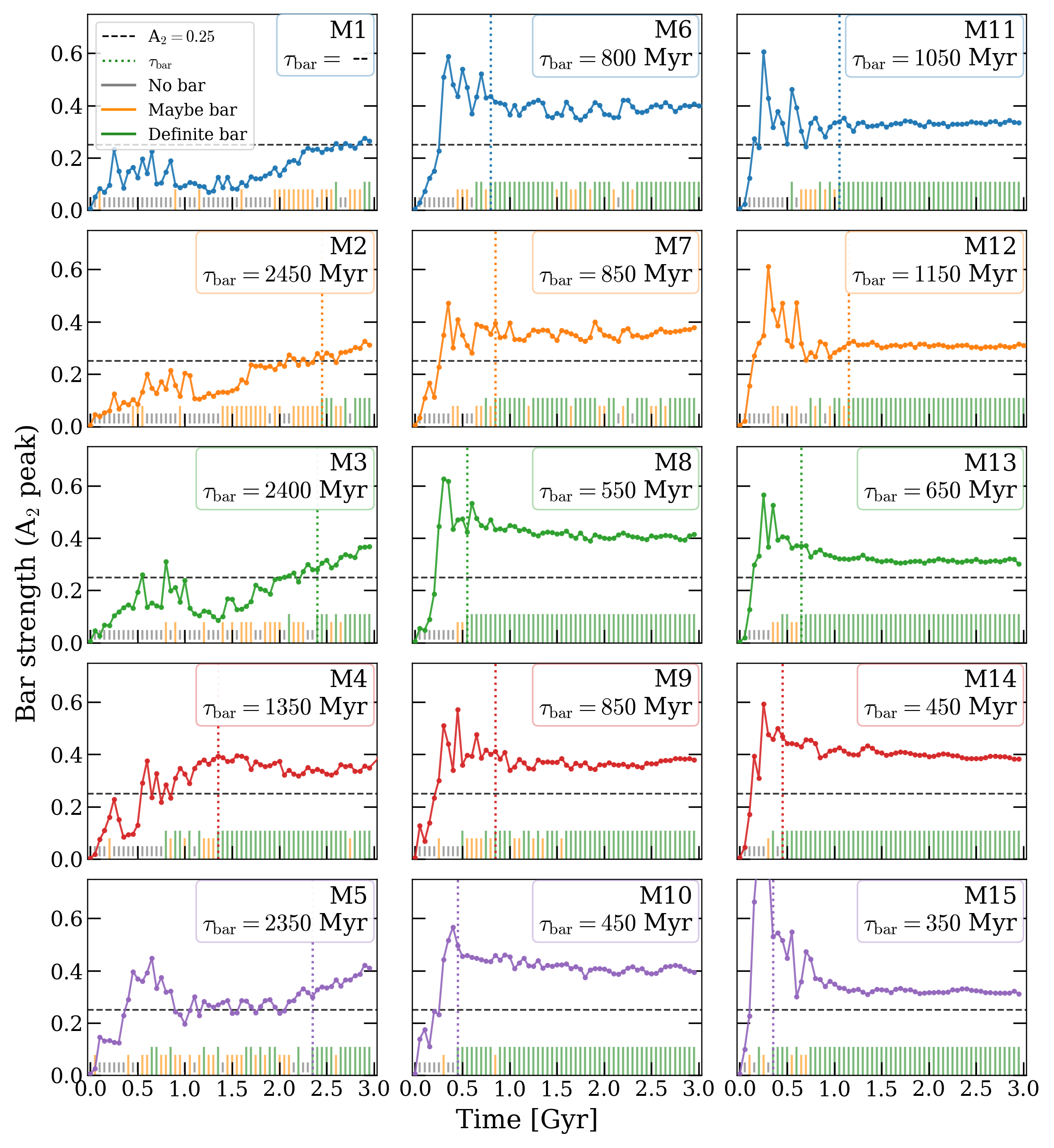} 
    \caption{
        Evolution of the bar strength, $A_2$, for the fifteen simulations in our model suite. Panels are arranged as in Fig.~\ref{fig:rotation_curve} and Fig.~\ref{fig:kappa_CO}: columns correspond to increasing initial stellar mass, while rows correspond to decreasing stellar bulge mass fraction. The coloured curve shows the temporal evolution of the peak $m=2$ Fourier amplitude measured within the inner 8 kpc. The horizontal dashed line marks $A_2=0.25$, the threshold adopted to identify a strong bar. The coloured bars at the bottom of each panel indicate the bar classification based on the phase-angle coherence criterion: grey denotes snapshots with no bar, orange denotes uncertain bar detections, and green denotes definite bars. The vertical dotted line marks the bar formation time, $\tau_{\rm bar}$. The value of $\tau_{\rm bar}$ is reported in the upper-right corner of each panel.
    }
    \label{fig:bar strength}
\end{figure}

Figure~\ref{fig:bar strength} shows the evolution of the bar strength for all $15$ models. The bar formation time, $\tau_{\rm bar}$, determined using the criteria described above, is indicated by the vertical dashed line in each panel and reported in the upper-right corner. The coloured markers at the bottom of each subplot show the bar classification for individual snapshots, with green corresponding to a clear bar, orange to a possible bar, and grey to an unbarred system.

Most models develop a well-defined bar within the 3 Gyr duration of the simulations, with the exception of M1 which forms a comparatively weak, late-forming bar. Its bar strength continues to rise throughout the run and only approaches the adopted threshold near the end of the simulation, indicating a delayed and still-ongoing bar growth phase. We include M1 in the following analysis, treating it as a weak/late-forming case.

The bar formation times reveal a strong dependence on both stellar mass and bulge fraction. At fixed stellar mass, galaxies with smaller bulges generally form bars earlier than those with larger bulges, meaning that massive bulges delay bar formation. This behaviour is particularly evident for the most massive galaxies, where $\tau_{\rm bar}$ decreases from $\sim1.1$ Gyr for $B_F=25\%$ to only $\sim0.35$ Gyr for $B_F=5\%$. This trend is consistent with the stabilising influence of centrally concentrated bulges, which suppress the growth of global non-axisymmetric instabilities \citep{binney2011galactic}.

A similarly strong dependence is observed with stellar mass. The least massive galaxies (M1–M5) either form bars very late or fail to develop a clear bar within the simulated time interval. In contrast, intermediate- and high-mass galaxies rapidly become bar unstable, with most bars forming within the first Gyr of evolution. These systems also exhibit a characteristic rapid growth phase followed by a decline to a nearly constant bar strength, indicative of the transition from dynamical bar growth to secular evolution. This behaviour suggests that increasing the stellar mass enhances the self-gravity of the disc, making it more susceptible to global bar-forming instabilities and accelerating the growth of the $m=2$ mode. The dependence on stellar mass appears stronger than that on bulge fraction, suggesting that the self-gravity of the stellar disc plays a primary role in regulating the onset of the bar instability.

Interestingly, despite the large variation in formation times, the final bar strengths are comparatively similar across the barred population, typically settling at $A_2 \sim 0.3$–$0.4$. The primary effect of varying the stellar mass and bulge fraction therefore appears to be on the onset of the bar instability rather than on the long-term amplitude of the bar itself.

A comparison between the kinematic settling time and the bar formation time reveals that, in several models, bars begin to develop while the disc is still undergoing dynamical relaxation. We investigate this connection further in Sect.~\ref{sec: disc vs bar formation time}.

\subsection{Bar length}
\label{sec: bar length}

\begin{figure}
    \centering
    \includegraphics[width=\columnwidth]{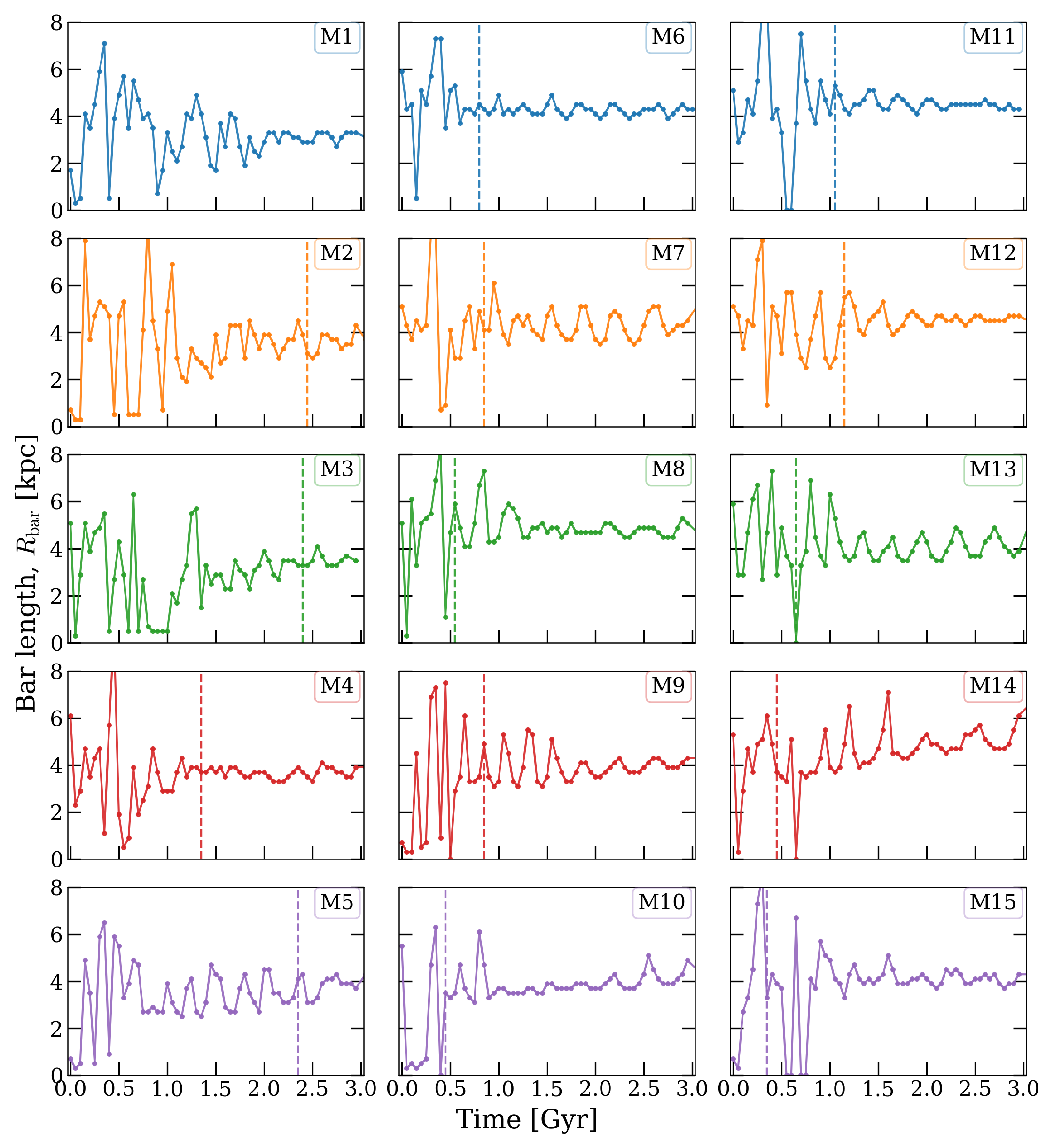} 
    \caption{
        Evolution of the bar length, $R_{\rm bar}$, for the fifteen simulations. Panels are arranged as in Fig.~\ref{fig:rotation_curve}, with columns corresponding to increasing initial stellar mass and rows to decreasing bulge mass fraction. We include the corresponding bar formation times $\tau_{\rm bar}$ for each model as a vertical dashed line.
    }
    \label{fig:bar length}
\end{figure}

To calculate the bar length at each time snapshot we first search for the radius of the peak of $A_2(r)$ within a maximum radius of $r<10$ kpc. We define the bar length to be the radius at which $A_2$ decreases by 65\% of the peak value, for a radius $r>r_{\rm peak}$. We also check if the bar phase $\phi_2$ remains constant within $\pm 3^{\circ}$ of the phase at peak $A_2$. The radius of a sudden change in the bar phase $\phi_2$ will be marked as the end of the bar. For a detailed discussion on different methods to calculate bar length see \citet{Athanassoula.Misiriotis.2002} and \citet{Ansar2025}.

We compute this quantity for all models and times, regardless of whether a given snapshot has already been classified as hosting a definite bar according to our formation criteria (Sect.~\ref{appendix bar detection}). This allows us to track the growth of non-axisymmetric structure and identify cases of gradual or delayed bar growth that might otherwise be excluded from the analysis. Figure~\ref{fig:bar length} shows the evolution of the bar length for all fifteen simulations, where we include the corresponding bar formation times as vertical dashed lines. While M1 does not strictly meet our bar formation time criteria, it develops a subtle non-axisymmetric structure toward the end of the simulation. We include its radial evolution for completeness, marking its region as pre-bar phase. 

The bar length is initially highly variable, reflecting both the transient relaxation of the disc and the rapid growth phase associated with bar formation. Once a well-defined bar is established, the measured lengths settle towards nearly constant values and evolve more gradually over the remainder of the simulation.

Although the bar formation times span almost 2 Gyr across the model suite, the final bar lengths exhibit a much narrower distribution. Most models converge towards bar lengths of approximately $4$--$5$ kpc by the end of the simulations, suggesting that the initial stellar mass and bulge fraction primarily regulate the onset of the bar instability rather than the eventual size of the bar. A weak trend is present whereby the most massive galaxies tend to host slightly longer bars than the least massive systems, although the effect is considerably smaller than that observed for the bar formation times. Similarly, no strong systematic dependence on bulge fraction is apparent once the bar has fully developed.

Several models exhibit low-amplitude oscillations in bar length during the secular phase, particularly after the bar has reached its maximum extent. These fluctuations may partly arise from uncertainties in identifying the precise bar end, where the transition between the bar and surrounding disc is gradual. However, they may also reflect genuine dynamical evolution, as the outer bar continuously exchanges angular momentum with neighbouring disc material and resonant stellar populations. The amplitude of these variations remains small compared to the overall bar size, indicating that the bars have largely reached a quasi-steady state after their initial growth phase.

\subsection{Pattern speed}
\label{sec: pattern speed}

The bar pattern speed, $\Omega_{\rm bar}$, quantifies the angular rotation rate of the bar and is a key dynamical quantity for characterising barred galaxies. Together with the circular and epicyclic frequencies, it determines the location of the main resonances and therefore controls where the bar exchanges angular momentum with the disc, bulge, and dark matter halo.

We measure $\Omega_{\rm bar}$ using the single-snapshot Fourier method developed by \citet{Dehnen2023}. This technique estimates the instantaneous pattern speed directly from the stellar positions and velocities in an individual simulation snapshot, allowing independent measurements of $\Omega_{\rm bar}$ and its uncertainty at each output time. A full description of the implementation and bar-identification procedure is provided in Appendix~\ref{appendix pattern speed}.

\begin{figure*}
    \centering
    \includegraphics[width=\textwidth]{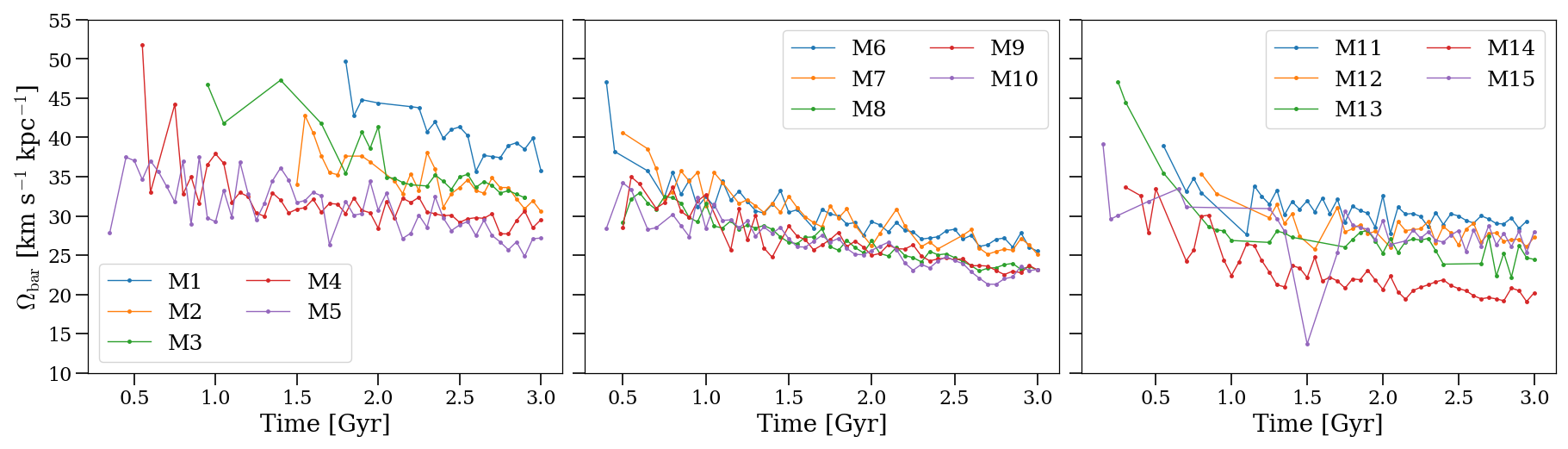} 
    \caption{
        Evolution of the bar pattern speed, $\Omega_{\rm bar}$, for the fifteen simulations. The three panels correspond to the three initial stellar masses, increasing from left to right. Different coloured curves show models with different $B_F$. In all cases, the bars slow down with time, with most models converging towards $\Omega_{\rm bar}\sim20$–$30~{\rm km\,s^{-1}\,kpc^{-1}}$ by the end of the simulations.
    }
    \label{fig:pattern speed}
\end{figure*}

Figure~\ref{fig:pattern speed} shows the time evolution of the bar pattern speed for the fifteen models, grouped according to their initial stellar mass. In all cases, $\Omega_{\rm bar}$ decreases with time, indicating that the bars systematically slow down during their secular evolution. This behaviour is expected for live galaxy models, where bars lose angular momentum to the outer disc and dark matter halo through resonant interactions.

The decline is generally strongest shortly after bar formation, when the bar is still growing rapidly and angular momentum exchange is most efficient. At later times, the evolution becomes more gradual and most models approach a slowly varying pattern speed. By the end of the simulations, the majority of the bars have pattern speeds in the range $\Omega_{\rm bar}\sim20$–$35~{\rm km\,s^{-1}\,kpc^{-1}}$, despite their different initial conditions and formation times.

The intermediate-mass models exhibit the most homogeneous evolution, with all five bars converging towards remarkably similar late-time pattern speeds of $\Omega_{\rm bar}\sim23$–$26~{\rm km\,s^{-1}\,kpc^{-1}}$. In contrast, the low- and high-mass groups show substantially larger scatter. The low-mass models retain the highest pattern speeds on average (median $\sim30.9~{\rm km\,s^{-1}\,kpc^{-1}}$ at $t\approx 2.9$ Gyr), largely due to the persistently fast bar in M1, and may also be more sensitive to transient non-axisymmetric structures because their bars form later and are generally weaker. The high-mass models span the widest range of final values, from $\sim20~{\rm km\,s^{-1}\,kpc^{-1}}$ in M14 to nearly $30~{\rm km\,s^{-1}\,kpc^{-1}}$ in M11, suggesting a stronger influence of the central mass concentration and bar growth history. This indicates that the final bar pattern speed is not determined solely by the initial stellar mass, but also by the subsequent secular evolution of the bar and the efficiency with which it redistributes angular momentum throughout the galaxy.

Overall, the results show that bar slowdown is a robust outcome across the full model suite. The initial stellar mass and bulge fraction influence the early evolution and the amount of scatter between models, but most bars evolve towards comparable pattern speeds by $t\sim3$ Gyr.

\subsection{Bar resonances}

The secular evolution of barred galaxies is strongly influenced by resonances, where the orbital frequencies of stars and gas become proportional to the bar pattern speed \citep{Binney2008}. In particular, the ILR, CR, and OLR regulate the exchange of angular momentum between the bar and the surrounding galaxy and are thought to play an important role in shaping central structures \citep[e.g.][]{Athanassoula2003, Sellwood2014}.

We determine the locations of these resonances by combining the bar pattern speed measured using the method described in Section~\ref{sec: pattern speed} with the circular and epicyclic frequencies derived from the gravitational potential of each simulation snapshot. In the epicyclic approximation, CR occurs where $\Omega=\Omega_{\rm bar}$, while the ILRs and OLR correspond to the locations where $\Omega-\kappa/2=\Omega_{\rm bar}$ and $\Omega+\kappa/2=\Omega_{\rm bar}$, respectively. Further details of the resonance calculation are provided in Appendix~\ref{appendix resonances}.

\begin{figure}
    \centering
    \includegraphics[width=\columnwidth]{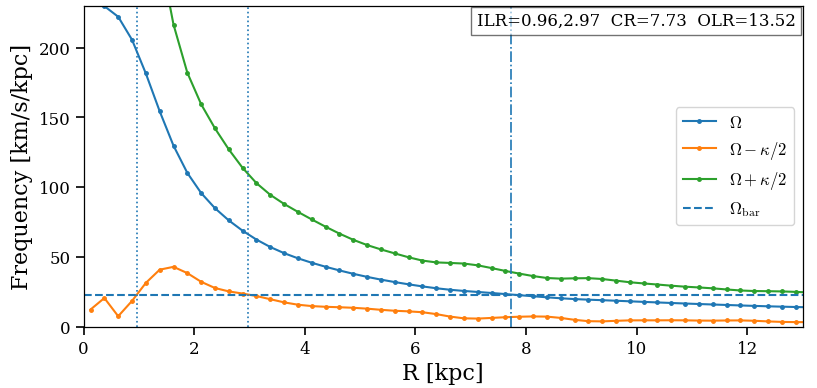} 
    \caption{
        Example resonance calculation for model M8 at $t=3$ Gyr. The blue curve shows the circular frequency $\Omega$, while the orange and green curves show $\Omega-\kappa/2$ and $\Omega+\kappa/2$, respectively. The horizontal dashed line marks the measured bar pattern speed, $\Omega_{\rm bar}$. The vertical dotted lines indicate the radii where the resonance conditions are satisfied, corresponding to the ILR, CR and OLR. 
    }
    \label{fig: resonances}
\end{figure}

An example is shown in Figure~\ref{fig: resonances} for M8 at $t=3$ Gyr. The blue, orange, and green curves show the circular frequency, $\Omega$, and the combinations $\Omega-\kappa/2$ and $\Omega+\kappa/2$, respectively, while the horizontal dashed line indicates the measured $\Omega_{\rm bar}$. For this particular model, the resonance radii are identified at ${\rm ILR}=0.96$ and $2.97$ kpc, indicating the presence of two inner Lindblad resonances. Corotation occurs at ${\rm CR}=7.7$ kpc, while the outer Lindblad resonance is located at ${\rm OLR}=13.5$ kpc.

The resonance radii measured at the final simulation time for all models are summarised in Table~\ref{tab: summary}. A common feature of all models is the presence of multiple ILRs. In several cases, the frequency curve $\Omega-\kappa/2$ exhibits a non-monotonic behaviour in the central regions, producing two or more intersections with the bar pattern speed. Multiple ILRs are typically associated with a strong central mass concentration \citep{athan92a,binney2011galactic}. Notably, we find this behaviour even in our lowest $B_{F}$ models, suggesting that the bar itself, rather than the initial bulge alone, contributes substantially to the central mass concentration responsible for this effect. Such multiple ILRs can significantly affect the orbital structure and gas inflow within the inner few kpc.

The corotation radius typically lies between $\sim5$ and $9$ kpc, substantially beyond the end of the bar. Comparing the measured bar lengths with the corotation radii yields values of $R_{\rm CR}/R_{\rm bar}\sim1.5$–$2.2$, indicating that all bars in our model suite are dynamically consistent with so-called \emph{slow bars}, for which corotation is expected to lie just beyond the bar end \citep{Debattista2015,Fragkoudi2021}. 

A clear trend is also visible between pattern speed and resonance location. Models with slower bars generally exhibit larger corotation and OLR. For example, M14 has one of the slowest pattern speeds ($\Omega_{\rm bar}\approx20.5~{\rm km\,s^{-1}\,kpc^{-1}}$) and correspondingly the largest corotation radius ($R_{\rm CR}=8.54$ kpc), whereas the faster bars of M1 and M11 have CR below $\sim6.5$ kpc. This behaviour follows directly from the resonance condition $\Omega(R)=\Omega_{\rm bar}$ and reflects the secular slowdown of bars discussed in the previous section.

Overall, despite significant differences in stellar mass, bulge fraction, bar formation time, and pattern speed, the resonance structure of the models is remarkably similar. All galaxies develop well-defined ILR, CR, and OLR resonances, providing the dynamical framework through which angular momentum redistribution and secular evolution proceed.

\subsection{Summary of bar properties}
\label{sec: summary}

\begin{table*}
    \centering
    \caption{Summary of bar properties and resonance radii at a time of $3$Gyr. The columns list the kinematic settling time $\tau$, bar formation time $\tau_{\rm bar}$, bar strength $A_2$, bar radius $R_{\rm bar}$, bar pattern speed $\Omega_{\rm bar}$, and the ILR, CR, and OLR radii at this time. Times are given in Myr, radii in kpc, and $\Omega_{\rm bar}$ in km s$^{-1}$ kpc$^{-1}$.}
    \label{tab: summary}
    \begin{tabular}{ccccccccc}
    \hline
    Model & $\tau$ & $\tau_{\rm bar}$ & $A_2$ & $R_{\rm bar}$ & $\Omega_{\rm bar}$ & ILR & CR & OLR \\
    \hline
    M1 & 500 & $> 3000$ & 0.27 & 3.30 & 38.5 & 1.05, 1.69 & 4.87 & 8.67 \\
    M2 & 500 & 2450 & 0.31 & 4.30 & 30.9 & 0.98, 1.80 & 6.20 & 11.70 \\
    M3 & 500 & 2400 & 0.37 & 3.50 & 32.3 & 1.02, 1.63 & 5.86 & 11.19 \\
    M4 & 1200 & 1350 & 0.35 & 3.90 & 30.6 & 1.03, 1.67 & 6.08 & 11.86 \\
    M5 & 650 & 2350 & 0.41 & 3.70 & 24.8 & 0.35, 0.41, 0.88, 1.72 & 8.21 & 13.66 \\
    M6 & 700 & 800 & 0.41 & 4.30 & 27.8 & 0.32, 0.44, 0.82, 2.27 & 6.49 & 11.66 \\
    M7 & 800 & 850 & 0.38 & 4.50 & 27.1 & 0.33, 0.43, 0.82, 2.32 & 6.53 & 11.65 \\
    M8 & 950 & 550 & 0.41 & 5.10 & 23.2 & 0.96, 2.97 & 7.73 & 13.52 \\
    M9 & 750 & 850 & 0.38 & 4.30 & 22.8 & 0.25, 0.53, 0.73, 2.34 & 7.78 & 13.56 \\
    M10 & 750 & 450 & 0.39 & 4.90 & 23.6 & 0.28, 0.48, 0.81, 2.63 & 7.47 & 13.46 \\
    M11 & 1000 & 1050 & 0.33 & 4.30 & 28.4 & 0.31, 0.44, 0.84, 2.58 & 6.37 & 10.30 \\
    M12 & 1200 & 1150 & 0.32 & 4.70 & 27.0 & 1.00, 2.64 & 6.57 & 10.75 \\
    M13 & 1350 & 650 & 0.30 & 3.90 & 26.2 & 0.29, 0.47, 0.82, 3.09 & 6.60 & 10.60 \\
    M14 & 1350 & 450 & 0.38 & 6.10 & 20.5 & 0.28, 0.45, 0.86, 3.46 & 8.54 & 13.52 \\
    M15 & 1850 & 350 & 0.31 & 4.30 & 28.0 & 1.06, 2.44 & 6.45 & 9.81, 10.28, 10.70 \\
    \hline
    \end{tabular}
\end{table*}

Table~\ref{tab: summary} summarises the principal bar properties measured for all models at the final simulation time of $3$ Gyr. For each galaxy, we report he kinematic settling time, $\tau$, the bar formation time, $\tau_{\rm bar}$, the bar strength, $A_2$, the bar radius, $R_{\rm bar}$, the bar pattern speed, $\Omega_{\rm bar}$, and the locations of the principal resonances derived from the measured pattern speed and frequency curves.

The table provides a compact overview of the diversity of bar evolution across the model suite. Although the galaxies span a broad range of bulge fractions, several common trends emerge. With the exception of M1, all models form a clear stellar bar during the simulated period. M1 instead develops a weak, late-forming bar whose strength only approaches our formation threshold near the end of the simulation. We nonetheless include its properties in the following comparison, as it still presents a non-axisymmetric structure qualitatively similar to that of the other models and provides a useful low-mass-concentration endpoint for the trends discussed below. The resulting bars exhibit comparable strengths, typically reaching $A_2 \sim 0.3$–$0.4$, despite forming at markedly different times. Similarly, the final bar lengths lie within a relatively narrow range of $R_{\rm bar}\sim3$–$6$ kpc, although the most extended bar develops in M14.

The bar pattern speeds show greater diversity, spanning a range of $\Omega_{\rm bar}\sim20$–$40$ km s$^{-1}$ kpc$^{-1}$ at late times. Nevertheless, all bars have slowed substantially from their initial values. The resonance radii vary accordingly, with corotation typically located at $R_{\rm CR}\sim5$–$9$ kpc and the OLR occurring near $R_{\rm OLR}\sim9$–$14$ kpc. Many models also exhibit multiple ILRs, reflecting the complex central frequency structure that develops during secular evolution.

\section{Results \& Discussion}
\label{sec: results}

The diversity of the model suite provides an opportunity to explore the physical processes governing bar formation and secular evolution across a broad range of galaxy properties. In the previous section, we quantified the principal properties of the stellar bars that develop in our simulations. We now investigate how these quantities relate to one another and how they depend on the initial stellar mass and bulge fraction of the host galaxy, aiming to identify the key mechanisms that regulate the growth and long-term evolution of stellar bars.

\subsection{Does the bar form before or after the global disc kinematics settle?}
\label{sec: disc vs bar formation time}

The relationship between kinematic settling of the disc and bar formation provides insight into the physical conditions required for the onset of the bar instability. A common classical view of bar instability is that stellar bars develop from global non-axisymmetric instabilities in discs that have already reached a near-equilibrium state \citep{Ostriker1973,Binney2008}. However, recent studies suggest that bar formation may also occur while galaxies are still undergoing significant structural evolution \citep{Ghosh2024,Fragkoudi2021}, raising the question of how the onset of the bar instability is ordered relative to the settling of the disc's global kinematics.

The simulations presented here allow us to investigate this question directly by comparing the kinematic settling time, $\tau$, defined in Section~\ref{sec: disc formation}, with the bar formation time, $\tau_{\rm bar}$, defined in Section~\ref{sec: bar strength}. We can determine whether bars typically emerge after the disc has settled or whether they can develop while the galaxy is still dynamically evolving through the relative timing of these two events across our model suite. 

We note that the two timescales are not entirely independent. As mentioned in Sect.~\ref{sec: disc formation}, the growth of the bar itself lowers $\kappa_{\rm CO}$. Hence, a model that forms a bar early-on continues to lose rotational support for longer and reaches its $\kappa_{\rm CO}$ plateau later. As such, the comparison presented here should be interpreted as relating the onset of bar formation to the epoch at which the global stellar kinematics reach a quasi-steady state, rather than to an entirely independent measure of disc relaxation.

\begin{figure}
    \centering
    \includegraphics[width=\columnwidth]{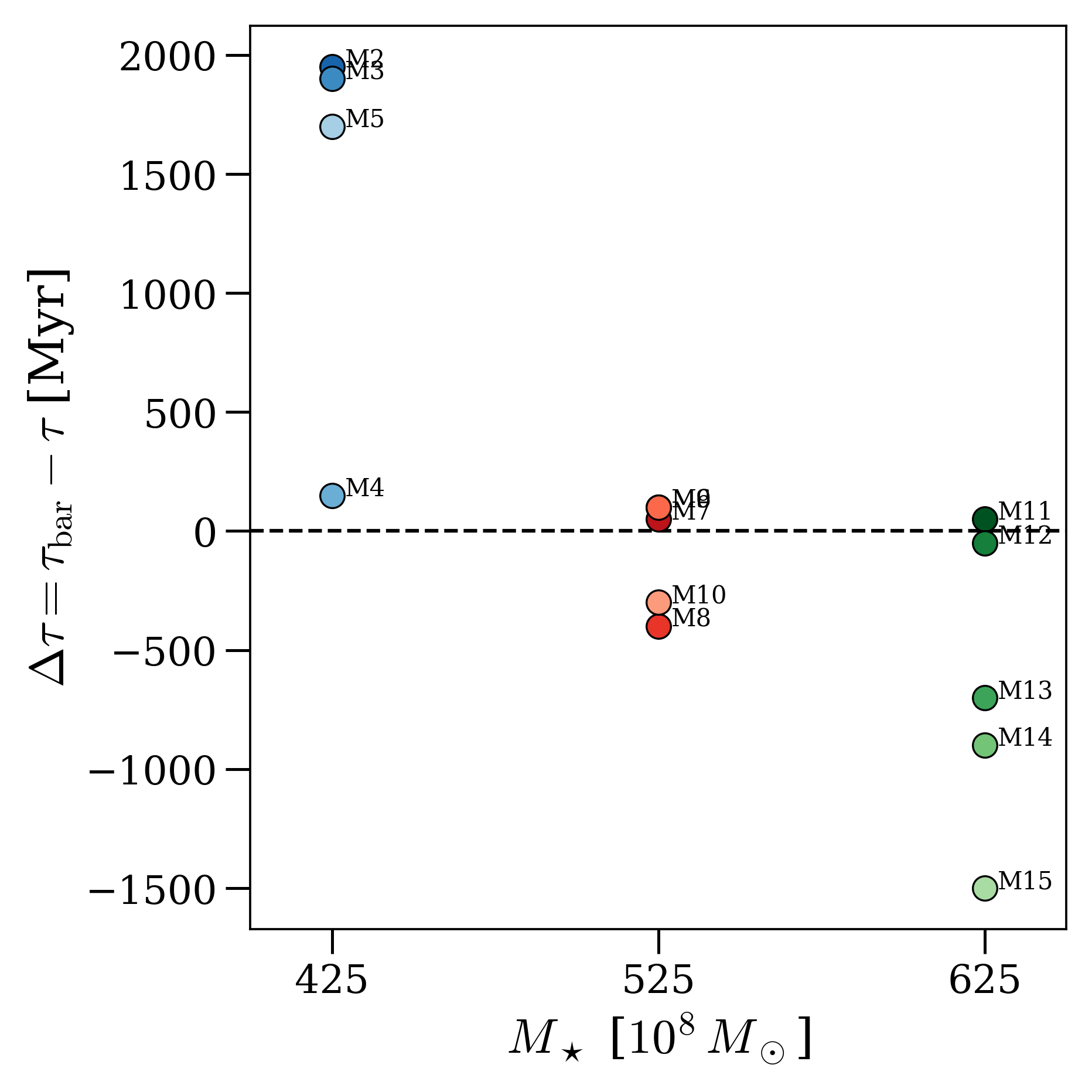} 
    \caption{
        Time difference $\Delta\tau=\tau_{\rm bar}-\tau$ as a function of initial stellar mass. Colours denote the initial stellar mass groups. Within each mass bin, darker shades correspond to larger initial bulge fractions. Model M1 is excluded as it does not satisfy the bar-identification criterion within the simulated time.
    }
    \label{fig: tau vs tau bar}
\end{figure}

We show in Fig.~\ref{fig: tau vs tau bar} the corresponding time difference, $\Delta\tau=\tau_{\rm bar}-\tau$, as a function of the initial stellar mass, with the dashed line indicating equality between the two timescales. Points above this line correspond to models in which the global disc kinematics settle before the bar forms, whereas points below the line indicate systems where the bar forms while the disc is still undergoing dynamical relaxation. 

A clear mass-dependent trend emerges. In the lowest-mass models, all barred galaxies lie above the 1:1 line, indicating that the global stellar kinematics reach a quasi-steady configuration before the onset of bar formation. M1 is not included as it does not satisfy our bar-identification criterion within the simulated time. The intermediate-mass models lie close to the 1:1 relation, suggesting that the establishment of the bar and the settling of the global stellar kinematics occur on comparable timescales. Within this mass bin, the models with lower bulge fractions tend to lie below the line, indicating that bar growth begins while the stellar orbital distribution is still evolving.

The highest-mass models show the opposite behaviour to the low-mass systems. Most of them lie well below the 1:1 line, with negative values of $\Delta\tau$, meaning that the bar forms before the global kinematics have fully settled. This effect is strongest for the lowest bulge fractions, particularly M13–M15. This indicates that high stellar disc mass and low central mass concentration promote a rapid bar instability, even while the galaxy is still dynamically settling. Conversely, larger bulges delay bar growth by stabilising the inner disc against global non-axisymmetric modes.

These results suggest that the bar instability does not have to wait for the global stellar kinematics to settle. Instead, the relative timing between disc relaxation and bar growth depends strongly on the balance between disc self-gravity and initial stellar mass. In the lowest-mass systems, the bar instability develops only after the disc has settled into a quasi-steady configuration, whereas in the most massive models the bar often emerges while the disc is still dynamically evolving. The transition from positive to negative values of $\Delta\tau=\tau_{\rm bar}-\tau$ with increasing stellar mass therefore indicates a fundamental change in the way bar instabilities are triggered across the model suite.

This behaviour is broadly consistent with classical theoretical expectations that massive, self-gravitating discs are intrinsically more susceptible to global non-axisymmetric instabilities \citep{binney2011galactic}. We report the initial values of the parameters $\varepsilon$ of \citet{Efstathiou1982} and $T_{\rm rot}/|W|$ of \citet{Ostriker1973} in Table~\ref{table B}. The resulting numbers order the suite in the same sense as the measured bar formation times (Spearman $\rho = +0.84$ and $-0.73$ respectively, the opposite signs reflecting that instability corresponds to low $\varepsilon$ but high $T_{\rm rot}/|W|$). Therefore, the models that form bars earliest are precisely those whose initial conditions produce the most strongly self-gravitating discs. In such systems, the strong self-gravity of the stellar disc promotes the rapid growth of the $m=2$ mode, allowing a bar to form on timescales comparable to, or even shorter than, those associated with the overall dynamical relaxation of the galaxy. Conversely, dynamically hot components such as massive bulges and centrally concentrated dark matter halos increase the stability of the inner disc, delaying the onset of bar formation by reducing the effectiveness of swing amplification and suppressing the growth of global modes \citep{Sellwood1984,Athanassoula1986,Athanassoula2003,kataria2018study}.

The systematic dependence on bulge fraction observed within each stellar-mass bin reinforces this interpretation. At fixed stellar mass, models with larger bulges consistently form bars later than their low-bulge counterparts, while the most disc-dominated systems exhibit the largest negative values of $\Delta\tau$. This finding supports the idea that bulges regulate the timing of bar formation primarily through their stabilising influence on the inner potential rather than by determining the eventual bar strength \citep{Sellwood2014,Athanassoula2013}.

Our results also have implications for the interpretation of bar formation in more realistic galaxy environments. Cosmological simulations have shown that bars can form over a wide range of epochs and under a variety of evolutionary conditions, including periods of ongoing mass accretion and secular restructuring \citep[e.g.][]{Fragkoudi2021,rosas-guevara2022}. The fact that several of our most massive models develop bars early-on suggests that the bar instability can grow before the global orbital distribution has settled, provided the disc is sufficiently self-gravitating. In contrast, lower-mass systems appear to require a longer period of relaxation before the bar instability can grow efficiently.

Overall, this comparison indicates that the onset of bar formation is governed by a competition between disc self-gravity, which promotes the instability, and central mass concentration, which suppresses it. The timing of bar formation therefore emerges as a sensitive diagnostic of the internal dynamical state of the galaxy rather than a simple consequence of the settling of the disc alone.

\subsection{Bar slowdown and corotation evolution}
\label{sec: bar slowdown}

The evolution of the corotation radius relative to the bar length provides a useful way of quantifying the secular evolution of barred galaxies. As a bar loses angular momentum to the outer disc, bulge, and dark matter halo, its pattern speed decreases and the corotation radius moves outwards. At the same time, bars are expected to grow through the trapping of stellar orbits, increasing their radial extent \citep{Athanassoula2003,MartinezValpuesta2006,Sellwood2014}. The ratio $\mathcal{R}=R_{\rm CR}/R_{\rm bar}$ thus offers a direct measure of how bar growth compares to bar slowdown, and is commonly employed as a diagnostic tool for bar dynamics \citep{Merrow2026}. 

For self-consistent stellar bars, corotation is expected to lie outside the bar end, implying $\mathcal{R}>1$ \citep{Contopoulus1980howfarbars}. Bars with $1.0\lesssim\mathcal{R}\lesssim1.4$ are commonly classified as fast bars, while systems with $\mathcal{R}>1.4$ are conventionally referred to as slow bars \citep{Debattista2000,Corsini2011,Aguerri2015,Fragkoudi2021}. 

\begin{figure}
    \centering
    \includegraphics[width=\columnwidth]{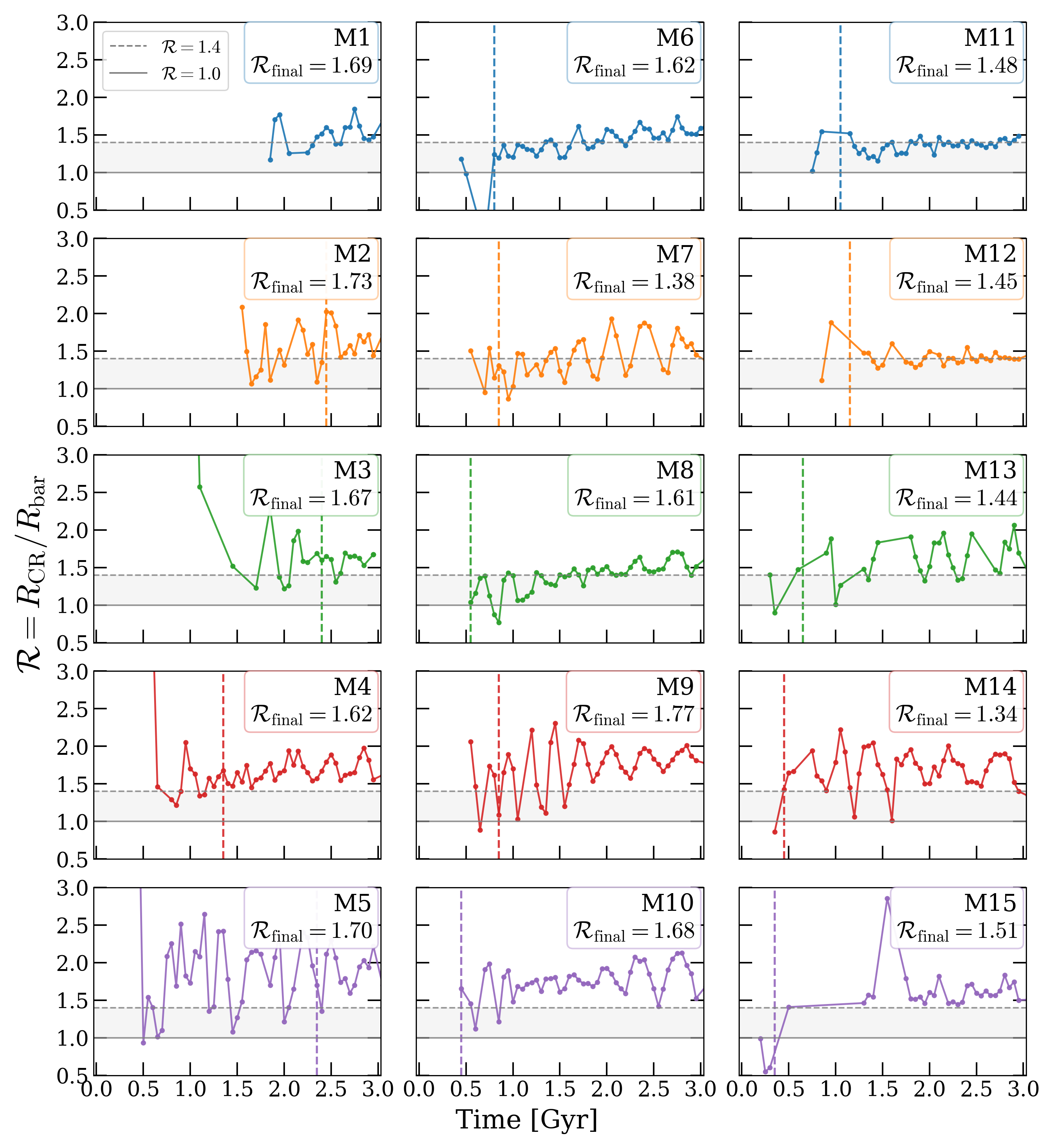} 
    \caption{
        Evolution of the ratio $\mathcal{R}=R_{\rm CR}/R_{\rm bar}$ for the fifteen simulations. The three columns correspond to increasing initial stellar mass, while different colours indicate different bulge fractions within each mass bin. The solid and dashed horizontal lines mark $\mathcal{R}=1.0$ and $\mathcal{R}=1.4$, respectively. Values of $\mathcal{R}$ at latest time are shown on right top for each model. Dashed vertical lines mark the corresponding $\tau_{\rm bar}$ value for each model.
    }
    \label{fig: R_ratio_evolution}
\end{figure}

In our case, Figure~\ref{fig: R_ratio_evolution} shows the evolution of $\mathcal{R}$ for all models. The panels are arranged as in previous figures, with columns corresponding to increasing initial stellar mass and rows to decreasing bulge fraction. The solid horizontal line marks $\mathcal{R}=1$, while the dashed line marks the conventional division between fast and slow bars at $\mathcal{R}=1.4$. The grey shaded region highlights the fast-bar regime, $1<\mathcal{R}<1.4$. For reference, the final value of $\mathcal{R}$ measured at $3$Gyr is reported in the upper-right corner of each panel. We include vertical dashed lines indicating the corresponding values of $\tau_{\rm bar}$ for each model. 

We note that $\mathcal{R}$ is computed at every snapshot for which both a bar radius and a corotation radius can be robustly measured. Consequently, the first $\mathcal{R}$ measurements do not necessarily coincide with the bar formation times defined in Section~\ref{sec: bar strength}, which are based on the bar-strength criterion adopted throughout this work. The earliest values should therefore be interpreted as describing the evolution of an emerging non-axisymmetric structure rather than a fully developed bar.

The evolution of $\mathcal{R}$ shows that most bars settle into values between $\sim [1.4,1.8]$ by the end of the simulations. This indicates that many of our bars lie close to, or slightly above, the conventional fast-bar boundary. The scatter is largest during the early stages of bar formation, when both $R_{\rm bar}$ and $R_{\rm CR}$ are rapidly evolving and the bar length is more difficult to define robustly. At later times, $\mathcal{R}$ fluctuates around a quasi-steady value, suggesting that the outward migration of corotation caused by bar slowdown is partly compensated by the secular growth of the bar.

We find tentative evidence that low-mass models exhibit larger values of $\mathcal{R}$, which may also be linked to their more concentrated dark matter haloes. The median values at the end of the simulations are $\mathcal{R}\approx1.69, 1.62, 1.45$ for the low-, intermediate-, and high-mass models, respectively. These first models have $c=8$--10, compared to $c=3$--4 for the most massive galaxies, and therefore contain a larger relative central halo contribution. A denser inner halo can absorb angular momentum from the bar more efficiently through resonant interactions, increasing the effective dynamical friction, slowing the bar, and moving CR outwards \citep{Debattista2000,Athanassoula2003,Sellwood2014barslowdown}. However, as halo concentration varies together with stellar mass and $B_F$ in our simulation suite, we cannot isolate the halo effect independently. Therefore, although it remains qualitatively consistent with theoretical expectations for bar-halo angular momentum exchange, the  observed trend should be interpreted as suggestive evidence rather than as a direct measurement of halo-driven slowdown.

Our choice of method for measuring bar length (Sect.~\ref{sec: bar length}) directly affects the derived values of $\mathcal{R}$. \cite{Michel-Dansac2006} compared several bar-length estimators against the resonance positions in both observed galaxies and $N$-body models. They found that methods based on Fourier decomposition of the surface density tend to trace the ultraharmonic resonance rather than corotation, and therefore return shorter bars than ellipse fitting does. As our $R_{\rm bar}$ is defined from the decline of the $m = 2$ Fourier amplitude, it falls into this first category, and our $\mathcal{R}$ values are likely biased high.

Overall, $\mathcal{R}$ values show that generally low-mass models have larger ratios, whereas most massive models lie closer to the fast-bar regime. This suggests that bars in more massive discs grow efficiently enough to keep pace with the outward movement of corotation, while weaker or later-forming bars remain shorter relative to their corotation radius. This behaviour is consistent with the standard picture in which bars lose angular momentum to the outer disc and dark matter halo, slow down, and grow in length during secular evolution \citep{Athanassoula2003,Debattista2000,Sellwood2014barslowdown}.

\subsection{Secular growth of stellar bars}
\label{sec: secular growth}

The previous sections have shown that the bars in our simulations exhibit a wide range of formation times, pattern speeds, strengths, and sizes. An important question is whether these quantities trace different aspects of the same secular growth process. If bar evolution is primarily driven by angular-momentum redistribution, then stronger bars would be expected to be both larger and slower than weaker systems. In this section we aim to asses the extent to which the structural and dynamical evolution of these bars remain coupled across the different galaxy models.

\begin{figure*}
    \centering
    \includegraphics[width=\textwidth]{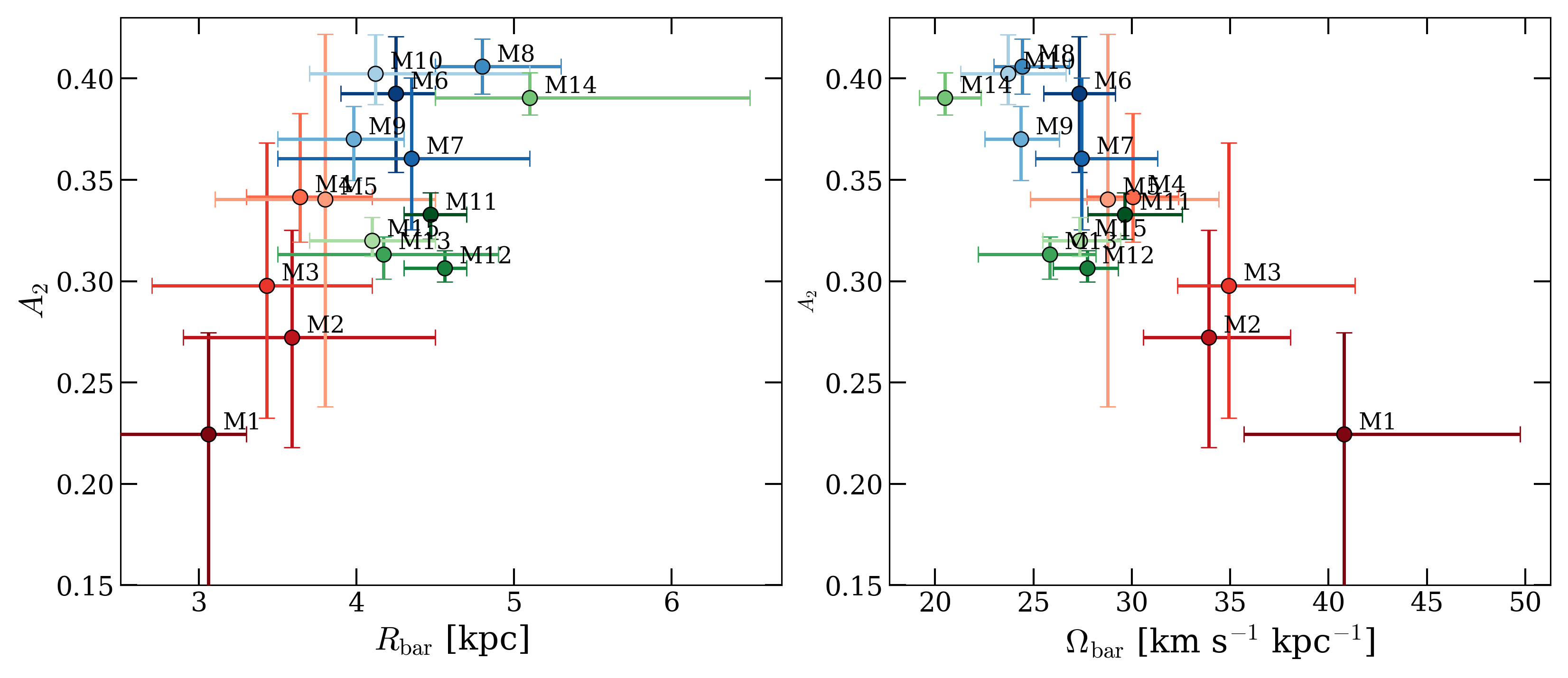} 
    \caption{
        Mean bar properties measured during the final $1$ Gyr of evolution for all models. 
        Left panel shows the relation between bar strength, $A_2$, and bar radius, $R_{\rm bar}$, while right panel compares $A_2$ with the bar pattern speed, $\Omega_{\rm bar}$. Symbols correspond to the mean values measured over the last $1$ Gyr of each simulation. Error bars indicate the maximum deviations from these mean values during the same period.Colours identify the three initial stellar-mass bins used throughout this work: low-mass models are shown in orange-red tones, intermediate-mass models in blue tones, and high-mass models in green tones. Within each mass bin, darker shades correspond to larger initial bulge fractions. 
    }
    \label{fig: A2 vs Rbar}
\end{figure*}

Figure~\ref{fig: A2 vs Rbar} compares the mean bar strength, bar radius, and pattern speed measured during the last $1$Gyr of evolution for all models. The symbols indicate the time-averaged values, while the error bars show the maximum deviations from the mean over this period. In contrast to previous figures, colours are used here to emphasise the mass grouping of the models: orange, blue, and green symbols correspond to the low-, intermediate-, and high-mass galaxies, respectively, while darker shades indicate larger bulge fractions within each mass bin.

The left panel of Fig.~\ref{fig: A2 vs Rbar} shows a clear positive correlation between bar strength and bar radius, with longer bars generally corresponding to stronger bars. The weakest bars in the sample, such as M1–M3, have both the smallest radii and lowest $A_2$ values, whereas the strongest bars are among the largest systems. This trend is consistent with a secular growth scenario in which bars capture progressively larger numbers of stellar orbits and extend further out into the disc as they evolve.

An even clearer relation is observed between bar strength and pattern speed on the right panel of Fig.~\ref{fig: A2 vs Rbar}. Models with the strongest bars systematically exhibit the lowest pattern speeds, while weaker bars tend to rotate more rapidly. The strongest bars ($A_2\gtrsim0.38$) are found almost exclusively at $\Omega_{\rm bar}\lesssim30~{\rm km\,s^{-1}\,kpc^{-1}}$, whereas the weakest bars correspond to the fastest rotators. The anti-correlation between $A_2$ and $\Omega_{\rm bar}$ appears stronger than the correlation between $A_2$ and $R_{\rm bar}$, suggesting that bar slowdown is closely linked to bar growth. This behaviour naturally arises from the secular redistribution of angular momentum. As the bar transfers angular momentum to the outer disc and dark matter halo through resonant interactions, it loses angular momentum, slows down, and simultaneously grows both in amplitude and radial extent. The correlations seen in Figure~\ref{fig: A2 vs Rbar} are therefore consistent with the expected evolutionary sequence in which bars become longer, stronger, and slower over time. 

Our results are also in good agreement with previous numerical studies that have linked bar growth to angular-momentum exchange and bar slowdown \citep{Athanassoula2003,MartinezValpuesta2006,Fragkoudi2021}. They are also broadly consistent with observational studies that find stronger bars to be systematically longer and more evolved systems \citep{Elmegreen2007,MenendezDelmestre2007,Erwin2019}. In particular, our results agree qualitatively with \citet{Cuomo2020}, who similarly found that stronger bars rotate at lower pattern speeds in a sample of 77 nearby barred galaxies, though with substantially more dispersion than in our simulations. While the scatter in our simulations indicates that stellar mass and bulge fraction influence the detailed evolutionary pathways of individual galaxies, the overall trends suggest that the coupling between bar growth and bar slowdown is a robust outcome across the full model suite.

The distribution of models in Fig.~\ref{fig: A2 vs Rbar} also reveals secondary trends with both stellar mass and bulge fraction. The most massive galaxies generally host the longest bars, while the lowest-mass systems preferentially occupy the short-bar regime. In addition, models with larger bulge fractions tend to develop somewhat weaker bars, consistent with the stabilising effect of a stronger central mass concentration. Nevertheless, the dominant trend remains the strong correlation between bar strength and bar size, indicating that the largest bars are also the most dynamically evolved. Taken together, these results support a picture in which the present-day structural properties of a bar are closely connected to its dynamical evolution. The strongest bars are not simply larger systems, but are also those that have undergone the greatest degree of secular evolution and angular-momentum redistribution.

\subsection{Boxy/Peanut bulges and bar buckling}

Our results so far have demonstrated that the bars in our simulations undergo significant secular evolution after their formation. As they grow in strength and length, and progressively lose angular momentum, their evolution is no longer confined to the plane of the galaxy. A key consequence of this long-term evolution is the development of vertical instabilities, which can thicken the inner bar and produce the characteristic boxy/peanut shaped (B/P) bulges observed in many edge-on disc galaxies \citep{Combes1990,Raha1991,Merritt1994,Merrifield1995,Athanassoula2005,MartinezValpuesta2006,MendezAbreu2008II,MendezAbreu2014,Erwin2016,Erwin2017,MendezAbreu2019,Kruk2019}.

The formation of B/P bulges is generally associated with the buckling instability of stellar bars, during which the bar temporarily bends out of the disc plane and redistributes stellar orbits onto vertically extended families \citep{Raha1991,Merritt1994,MartinezValpuesta2006}. Subsequent secular evolution can further strengthen the peanut structure through resonant trapping of stars even in the absence of a strong buckling event \citep{Quillen2002,Athanassoula2005,Fragkoudi2017, Sellwood.Gerhard.2020,Lokas2026}. Having established how the bar properties vary across our model suite, we now investigate the emergence of B/P structures and their connection to bar growth, bar slowdown, and buckling events.

We quantify the strength of the B/P structure following the approach introduced by \citet{Ghosh2024} for idealised $N$-body simulations \citep[see][for a similar approach with the Auriga cosmological simulations]{PaulaLopez2025}. For each snapshot, we compute a radial profile of the normalised vertical thickness, $S(R,t)$, within a narrow slice around the bar major axis. The resulting profile allows both the radial extent and maximum strength of the B/P structure to be tracked throughout the evolution of each simulation. Full details of the calculation and an illustrative example are provided in Appendix~\ref{app:BP_method}.

Using the radial profiles $S(R,t)$, we characterise the B/P structure through two quantities. The B/P strength is defined as the maximum value of the profile, $S_{\rm b/p}(t)=\max_R[S(R,t)]$, which quantifies the degree of vertical thickening relative to the initial disc. Second, we define the characteristic B/P radius, $R_{\rm b/p}$, as the radius at which this maximum occurs. Together, these quantities trace the growth and radial extent of the vertically thickened structure throughout the simulation.

\begin{figure}
    \centering
    \includegraphics[width=\columnwidth]{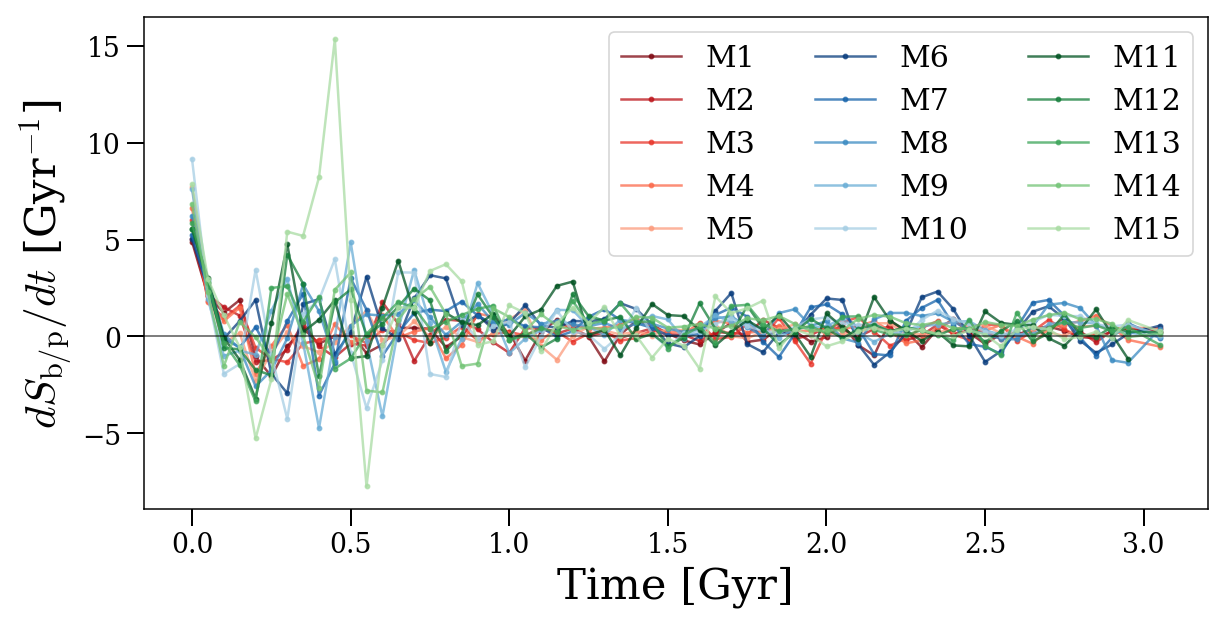} 
    \caption{
        Evolution of the B/P growth rate, ${\rm d}S_{\rm b/p}/{\rm d}t$, for all models. Colours indicate stellar-mass bins (red/orange: low mass; blue: intermediate mass; green: high mass), with darker shades corresponding to larger $B_F$. The horizontal dashed line marks ${\rm d}S_{\rm b/p}/{\rm d}t=0$. 
    }
    \label{fig: dSbp/dt}
\end{figure}

In addition, the temporal evolution of $S_{\rm b/p}$ provides information on the formation mechanism of the B/P bulge. Rapid changes in the B/P strength, quantified through its time derivative ${\rm d}S_{\rm b/p}/{\rm d}t$, can indicate episodes of enhanced vertical heating. Figure~\ref{fig: dSbp/dt} shows this temporal evolution for all models. Most simulations exhibit relatively modest fluctuations around zero after an initial phase of rapid vertical evolution, indicating a gradual and secular growth of the B/P structure. However, model M15 stands out by displaying a pronounced positive peak at $t\sim 450$--$700$ Myr, significantly larger than those seen in the rest of the sample.

This rapid increase in $S_{\rm b/p}$ suggests a short-lived episode of enhanced vertical heating and thickening. Visual inspection of the corresponding edge-on stellar density maps for all 15 models reveal a clear vertical asymmetry only on M15 during this period, with the bar bending away from the disc mid-plane. Such bending modes are a characteristic signature of the bar buckling instability, and therefore we interpret the strong peak in ${\rm d}S_{\rm b/p}/{\rm d}t$ as evidence for a buckling event in M15. In addition, from the evolution of the bar strength shown in Figure~\ref{fig:bar strength}, we can see that around the same time interval, M15 exhibits a sudden decrease in its $A_2$ evolution, departing from the otherwise smooth secular growth observed in most models. Such transient weakening or irregular evolution of the bar amplitude has frequently been associated with buckling episodes in numerical simulations \citep{MartinezValpuesta2006,Smirnov2018,Lokas2019,Lokas2020}. The simultaneous occurrence of a rapid increase in B/P strength, a vertically asymmetric morphology, and a temporary perturbation in the bar strength evolution strongly suggests that M15 is the only model to undergo a classical buckling phase.

The fact that M15 is the only model for which we identify a clear buckling signature suggests that, in our simulations, buckling requires a combination of high stellar disc mass and low central mass concentration. M15 belongs to the highest-mass set of models and has the lowest $B_F$ in that mass bin, making it one of the most strongly disc-dominated systems in the sample. This is consistent with the idea that buckling is more likely to occur in systems that form strong bars rapidly, since the vertical instability is generally associated with bars that have become sufficiently strong and vertically anisotropic \citep{MartinezValpuesta2006,Lokas2019}. However, this trend should be interpreted with caution, as only one model in our suite shows an unambiguous buckling event, and we cannot establish a statistically robust threshold in either stellar mass or $B_F$. Instead, our results suggest that high disc self-gravity and weak central mass concentrations may create favourable conditions for buckling. This interpretation is broadly consistent with previous numerical work showing that stronger bars are more susceptible to buckling, while additional dynamically hot or centrally concentrated components can reduce the amplitude of the vertical distortion or delay the instability \citep{Lokas2020,Wheeler2023,McClure2025}. This is also qualitatively consistent with observational studies finding that the incidence of B/P structures increases with stellar mass in barred galaxies \citep{Erwin2017}.

\begin{figure}
    \centering
    \includegraphics[width=\columnwidth]{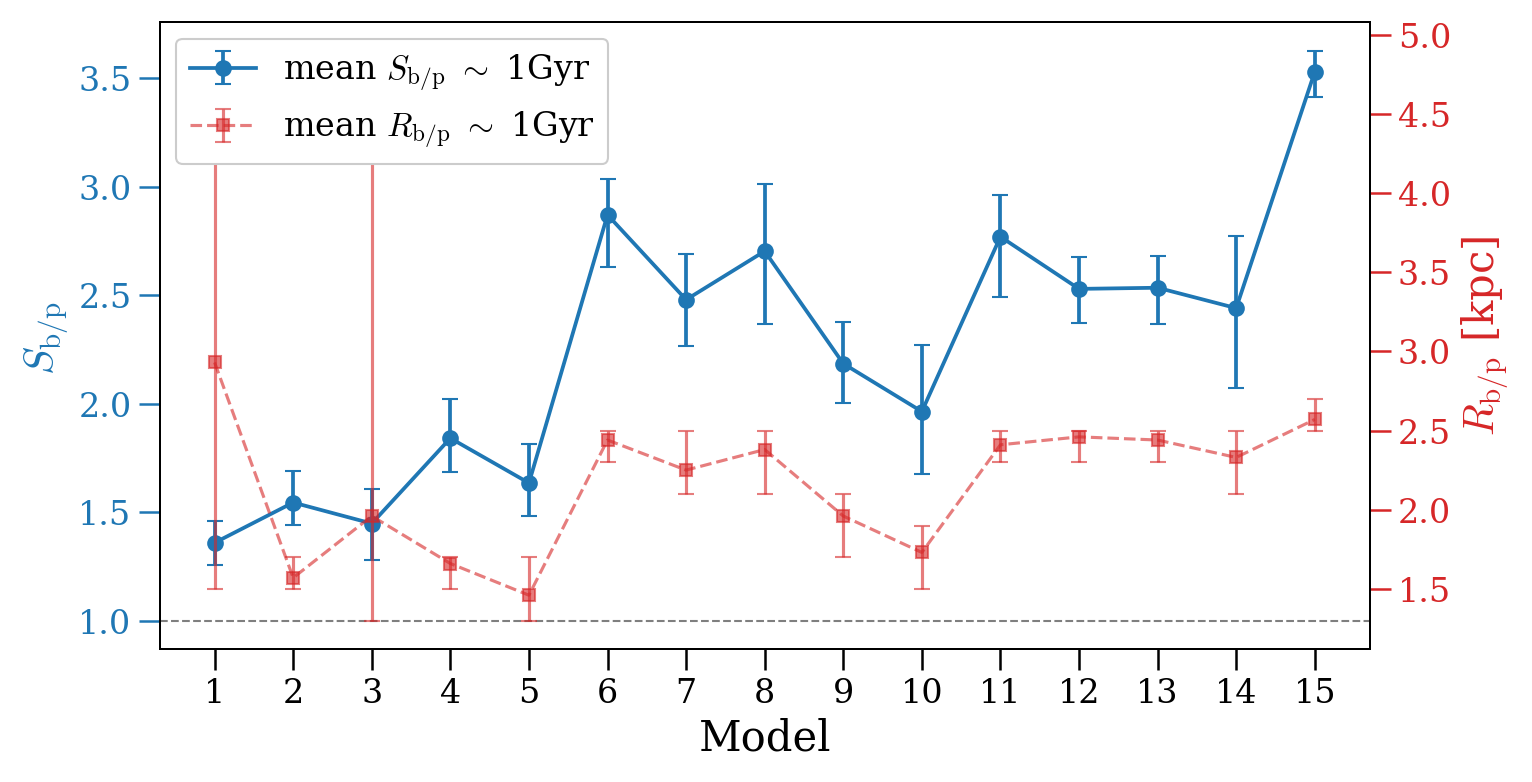} 
    \caption{
        Mean B/P strength, $S_{\rm b/p}$ (blue circles; left axis), and mean B/P radius, $R_{\rm b/p}$ (red squares; right axis), measured over the final $1$ Gyr of evolution for all models. Error bars indicate the maximum deviations from the mean during this period. The horizontal dashed line marks $S_{\rm b/p}=1$, corresponding to the absence of vertical enhancement relative to the initial disc. 
    }
    \label{fig: Sbp Rbp}
\end{figure}

Figure~\ref{fig: Sbp Rbp} shows the mean B/P strength, $S_{\rm b/p}$, and B/P radius, $R_{\rm b/p}$, measured during the last $1$ Gyr of evolution. A clear trend is that both quantities generally increase with stellar mass. The low-mass models (M1--M5) host the weakest B/P structures, with $S_{\rm b/p}\sim1.3$--1.9 and $R_{\rm b/p}\lesssim2$ kpc, whereas the intermediate- and high-mass systems develop substantially stronger and more extended vertically thickened structures. The most extreme case is M15, which reaches $S_{\rm b/p}\approx3.5$ and $R_{\rm b/p}\approx2.6$ kpc. This behaviour is consistent with the expectation that more massive discs form stronger bars, which are able to drive more efficient vertical heating and trap larger numbers of stars onto resonant orbit families that support the peanut morphology \citep{Athanassoula2005,MartinezValpuesta2006,Fragkoudi2015}.

A second notable trend is the close correspondence between $S_{\rm b/p}$ and $R_{\rm b/p}$. Models hosting stronger B/P structures also tend to exhibit larger B/P radii, suggesting that the growth of the vertically thickened component is accompanied by an outward expansion of the region affected by bar-driven vertical resonances. This mirrors the positive correlation between bar strength and bar length discussed in Sect.~\ref{sec: secular growth} and supports a picture in which the radial and vertical growth of bars are closely coupled during secular evolution \citep{Athanassoula2013,Ghosh2021,Ghosh2024}. Bars that undergo more substantial angular-momentum redistribution therefore become not only longer and stronger in the disc plane, but also develop larger and more prominent peanut-shaped bulges.

Although the overall trend is primarily driven by stellar mass, there are also indications that the $B_F$ plays a secondary role. Within a given mass bin, models with lower $B_F$ frequently lie towards larger values of both $S_{\rm b/p}$ and $R_{\rm b/p}$, whereas systems with more prominent bulges tend to host weaker vertical structures. This behaviour is particularly apparent in the highest-mass group, where M15 develops the strongest and most extended B/P bulge, while M11 and M12 remain comparatively weaker. Such behaviour is consistent with the idea that centrally concentrated components can reduce the efficiency of bar growth and vertical instabilities by increasing the central dynamical support of the galaxy \citep{Athanassoula2003,Debattista2006,Smirnov2021}. Nevertheless, the scatter within each mass bin indicates that the final B/P properties depend not only on the initial structural parameters but also on the detailed evolutionary history of each bar.

This interpretation also links naturally to the previous discussion of buckling. While M15 is the only model showing clear evidence for a classical buckling event, all barred galaxies in the sample develop B/P structures to some degree. This supports the growing view that buckling is not a necessary condition for B/P formation, and that vertically thickened bar structures can also arise through the gradual secular trapping of stellar orbits at vertical resonances \citep{Quillen2002,Debattista2006,Erwin2016,Fragkoudi2017,Ghosh2024}. The occurrence of buckling in M15 therefore appears to represent an extreme manifestation of a more general process of bar-driven vertical evolution that operates throughout the model suite.

\begin{figure}
    \centering
    \includegraphics[width=\columnwidth]{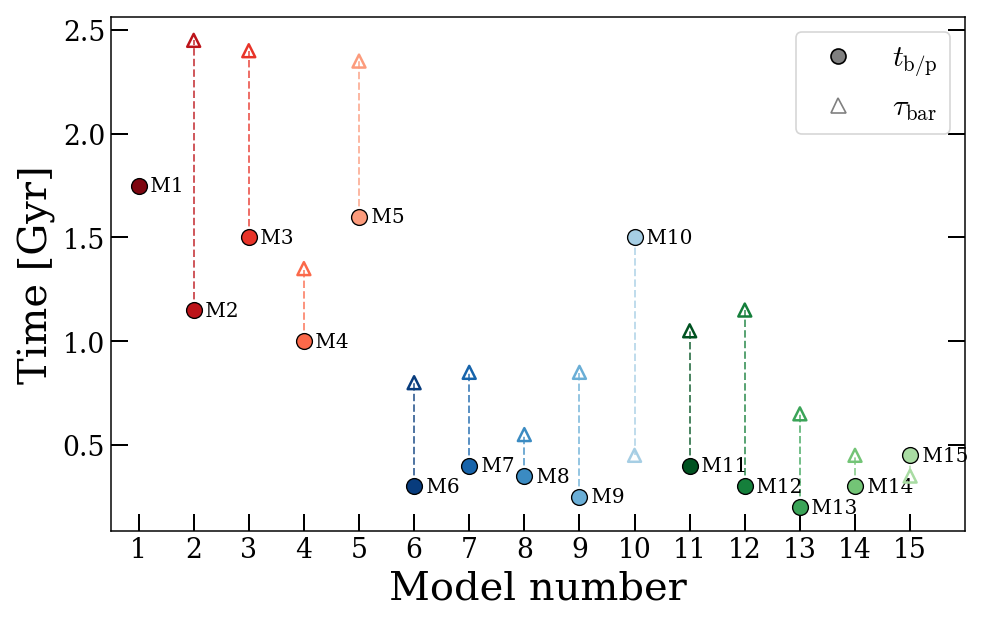}
    \caption{
        Boxy/peanut formation times, $t_{\rm b/p}$ (circles), determined using the criterion described in Appendix~\ref{appendix:BPformation}, compared to the bar formation times, $\tau_{\rm bar}$ (open triangles), for each model. Dashed lines connect the two times for each model. Colours follow the same scheme adopted throughout this work: red, blue, and green correspond to the low-, intermediate-, and high-stellar-mass bins, respectively, while lighter shades indicate lower $B_F$ within each mass bin.
    }
    \label{fig:tbp_models}
\end{figure}

We can also determine when the B/P structure becomes established at a time $t_{\rm b/p}$ using the radial shape of the profile $S(R,t)$, following the procedure described in Appendix~\ref{appendix:BPformation}. Visual inspection of the profile evolution shown in Figure~\ref{fig:BP visual vs code} demonstrates that the formation times identified by this automated method are in good agreement with the first appearance of a visually recognisable peanut-shaped morphology. 

Figure~\ref{fig:tbp_models} shows the resulting B/P formation times for all models, compared with their corresponding bar formation times, $\tau_{\rm bar}$. Note that $t_{\rm b/p}$ is computed across all snapshots, independently of the bar formation time. In cases where $t_{\rm b/p} < \tau_{\rm bar}$ (or for models like M1), indicates that early vertical density signatures are picked up before the central non-axisymmetry satisfies our bar formation time criterion, pointing to a weak or early-stage bar. A clear dependence on the initial stellar mass is evident. The intermediate- and high-mass models generally develop B/P structures within the first $\sim0.2$--$0.45$ Gyr of evolution, whereas the low-mass models require substantially longer times, with B/P formation occurring between $\sim1$ and $1.8$ Gyr. This mass dependence closely follows that of bar formation itself, and in most models $t_{\rm b/p}$ precedes $\tau_{\rm bar}$, indicating that vertical buckling begins while the bar is still in its early, sub-threshold growth phase. This supports the view that B/P structures are not simply a delayed consequence of a fully formed bar, but rather trace the earliest stages of bar-driven secular evolution, developing in tandem with, and in some cases ahead of, the bar's own growth.

Within the intermediate- and high-mass samples, the dependence on $B_F$ is comparatively weak, with most models forming their B/P structures at broadly similar times despite spanning a range of bulge fractions. In contrast, the low-mass models display significantly larger scatter, suggesting that the slower growth of the bar delays the onset of the vertical resonances and orbital trapping responsible for building the peanut structure. The delayed formation of the B/P component in these models is therefore likely a consequence of their slower overall secular evolution.

Finally, the B/P structures generally appear around the time of bar formation, and in most models even before the bar formally satisfies our detection criteria, but well before the bars reach their final strengths. This does not imply the absence of a bar prior to $t_{\rm b/p}$; rather, it suggests that a bar-like non-axisymmetry is already present and dynamically capable of driving vertical buckling before it grows strong enough to meet our formal threshold. The onset of vertical thickening therefore occurs at a very early stage in the bar evolutionary sequence, while the subsequent increase in $S_{\rm b/p}$ and $R_{\rm b/p}$ reflects continued secular growth of the peanut structure over several Gyr. Similar behaviour has been reported in both idealised and cosmological simulations, where B/P bulges emerge shortly after the formation of the bar and continue to strengthen as the bar evolves \citep{MartinezValpuesta2006,Debattista2006,Fragkoudi2017,Ghosh2024}.

Overall, the results indicate that the formation and subsequent evolution of B/P structures are closely linked to the secular growth of the bar. More massive galaxies develop B/P bulges earlier, and by the end of the simulations they generally host both stronger and more extended peanut structures. The growth of $S_{\rm b/p}$ and $R_{\rm b/p}$ over time further suggests that B/P bulges continue to evolve long after their initial appearance, reflecting the ongoing redistribution of stars onto vertically extended bar-supporting orbits. Although all models develop a measurable B/P component, only M15 exhibits clear evidence for a classical buckling event, implying that buckling is not a necessary condition for B/P formation in our simulations. Instead, the majority of the B/P structures appear to grow gradually, without breaking the mid-plane symmetry, through secular evolution, consistent with the picture in which vertical resonances and orbit trapping play a dominant role in shaping the inner three-dimensional structure of barred galaxies \citep{Combes1990,Athanassoula2005,Fragkoudi2017,Ghosh2024}. Taken together, these results suggest that the stellar mass and central mass concentration primarily regulate the timescale and efficiency of B/P growth, while the detailed evolutionary pathway, whether through gradual secular thickening or a buckling episode, depends on the dynamical state of the bar.

\section{Conclusions}
\label{sec: conclusions}

We  analyse the formation and secular evolution of bars and B/P bulges in a suite of fifteen live isolated MW-like simulations spanning a range of initial stellar masses, bulge fractions, and dark matter halo concentrations. Using a combination of bar strength measurements, pattern speed determinations, resonance analysis, and vertical structure diagnostics, we investigate how the structural and dynamical properties of bars evolve over a period of $3$ Gyr. Our main findings can be summarised as follows:

\begin{enumerate}

    \item The relative timing of bar formation and settling of the global disc kinematics depends strongly on galaxy structure. In the low-mass, bulge-dominated models, the stellar kinematics reach a quasi-steady state within $\sim 0.5-1.2$ Gyr, and the bar appears up to $\sim 2$ Gyr later. In the most massive, disc-dominated models, the bar grows within $\sim 0.35-0.65$ Gyr, while $\kappa_{\rm CO}$ is still declining. In those models the bar itself sustains the decline, by transferring stars onto non-circular orbits, so the disc's late settling reflects bar growth rather than a slower relaxation of the initial conditions.  \\

    \item Bars systematically slow down during their secular evolution. All models exhibit a decline in pattern speed with time, consistent with angular-momentum transfer from the bar to the outer disc and dark matter halo. 
    
    \item Bar strength, size, and dynamical state are closely connected. Stronger bars are generally longer and rotate more slowly, consistent with the classical picture of secular bar growth. As bars transfer angular momentum and evolve, they increase in length while their pattern speeds decrease. The resulting corotation-to-bar-radius ratios remain close to the fast-bar regime for most models.
    
    \item B/P structures form early and continue growing throughout the simulations. The B/P formation times closely track the formation of the bar itself, with more massive galaxies developing vertically thickened structures substantially earlier than low-mass systems. After formation, both the strength and radial extent of the B/P structure continue to increase over several Gyr, indicating that vertical evolution remains active long after the initial appearance of the peanut morphology.
    
    \item Among the fifteen simulations, only M15 shows unambiguous evidence for a buckling event, characterised by a rapid increase in B/P strength, a temporary weakening of the bar, and a clear vertical asymmetry of the stellar distribution. This event produces the strongest B/P structure measured out of all the models in our sample. M15 corresponds to the most massive and lowest-$B_F$ galaxy in the suite, suggesting that buckling is favoured in systems with strong disc self-gravity and weak central stabilisation. Nevertheless, all other models develop B/P structures through a more gradual increase in vertical thickness, supporting the view that buckling represents only one pathway to B/P formation.

    \item These trends reveal a coherent evolutionary sequence set by stellar mass and bulge fraction: more massive, disc-dominated systems (smaller $B_F$) form bars and B/P bulges earlier, sometimes before the global disc kinematics have fully settled, while lower-mass or more centrally concentrated galaxies (higher $B_F$) evolve more gradually. Once formed, bars across all models undergo qualitatively similar secular evolution, growing in size and strength while transferring angular momentum, slowing down, and driving continued vertical thickening. We note that, since the suite is constructed at fixed total mass with the halo concentration adjusted to preserve a realistic rotation curve, this sequence reflects the self-gravity of the disc relative to its halo rather than stellar mass alone.

\end{enumerate}

Our results thus link bar formation, slowdown and the growth of B/P structures into a single coherent sequence of secular evolution in disc galaxies. The settling of the disc's global kinematics however is not part of that sequence. It does not set the timing of bar formation, and in the most massive models it is the growth of the bar itself that prolongs the decline of $\kappa_{\rm CO}$. Future work will investigate how these trends connect to the formation of nuclear stellar discs and ansae structures, and will compare the evolutionary pathways identified here with those found in cosmological simulations and observations of nearby barred galaxies.


\begin{acknowledgements}
      This work was funded by the European Union (Widening Participation, ExGal-Twin, GA 101158446). The simulations presented here were performed using the supercomputing facilities at Cardiff University operated by Advanced Research Computing at Cardiff (ARCCA) on behalf of the Cardiff Supercomputing Facility and the HPC Wales and Supercomputing Wales (SCW) projects. This work used the DiRAC@Durham facility managed by the Institute for Computational Cosmology on behalf of the STFC DiRAC HPC Facility (www.dirac.ac.uk). The equipment was funded by BEIS capital funding via STFC capital grants ST/P002293/1, ST/R002371/1 and ST/S002502/1, Durham University and STFC operations grant ST/R000832/1. DiRAC is part of the National e-Infrastructure. 
      EDC would like to thank her cat Yang for all the support during the completion of this study.
      SA acknowledges support from a UKRI Future Leaders Fellowship (grant no. MR/X033740/1).
      PDL acknowledges financial support from ANID Fondo GEMINI 2024 32240021.
      JHK acknowledges grant PID2022-136505NB-I00 funded by MCIN/AEI/10.13039/501100011033 and EU, ERDF.
      J.F-B acknowledges support from the PID2022-140869NB-I00 grant from the Spanish Ministry of Science and Innovation.
      AdLC and JMA acknowledge financial support from the Spanish Ministry of Science, Innovation, and Universities (MICIU) through RYC2022-035838-I and PID2025-174546NB-I00 (TRAMWAYS project).
\end{acknowledgements}

%
%

\bibliographystyle{aa}
\bibliography{sorted_updated_references}

@ARTICLE{Aguerri2015,
       author = {{Aguerri}, J.~A.~L. and {M{\'e}ndez-Abreu}, J. and {Falc{\'o}n-Barroso}, J. and {Amorin}, A. and {Barrera-Ballesteros}, J. and {Cid Fernandes}, R. and {Garc{\'\i}a-Benito}, R. and {Garc{\'\i}a-Lorenzo}, B. and {Gonz{\'a}lez Delgado}, R.~M. and {Husemann}, B. and {Kalinova}, V. and {Lyubenova}, M. and {Marino}, R.~A. and {M{\'a}rquez}, I. and {Mast}, D. and {P{\'e}rez}, E. and {S{\'a}nchez}, S.~F. and {van de Ven}, G. and {Walcher}, C.~J. and {Backsmann}, N. and {Cortijo-Ferrero}, C. and {Bland-Hawthorn}, J. and {del Olmo}, A. and {Iglesias-P{\'a}ramo}, J. and {P{\'e}rez}, I. and {S{\'a}nchez-Bl{\'a}zquez}, P. and {Wisotzki}, L. and {Ziegler}, B.},
        title = "{Bar pattern speeds in CALIFA galaxies. I. Fast bars across the Hubble sequence}",
      journal = {\aap},
         year = 2015,
        month = apr,
       volume = {576},
          eid = {A102},
        pages = {A102},
          doi = {10.1051/0004-6361/201423383},
archivePrefix = {arXiv},
       eprint = {1501.05498},
 primaryClass = {astro-ph.GA},
       adsurl = {https://ui.adsabs.harvard.edu/abs/2015A&A...576A.102A}
}

@ARTICLE{Ansar2024,
       author = {{Ansar}, Sioree and {Das}, Mousumi},
        title = "{The Stellar Bar─Dark Matter Halo Connection in the TNG50 Simulations}",
      journal = {\apj},
         year = 2024,
        month = nov,
       volume = {975},
       number = {2},
          eid = {243},
        pages = {243},
          doi = {10.3847/1538-4357/ad7a6b},
archivePrefix = {arXiv},
       eprint = {2311.11998},
 primaryClass = {astro-ph.GA},
       adsurl = {https://ui.adsabs.harvard.edu/abs/2024ApJ...975..243A}
}

@ARTICLE{Ansar2025,
       author = {{Ansar}, Sioree and {Pearson}, Sarah and {Sanderson}, Robyn E. and {Arora}, Arpit and {Hopkins}, Philip F. and {Wetzel}, Andrew and {Cunningham}, Emily C. and {Quinn}, Jamie},
        title = "{Bar Formation and Destruction in the FIRE-2 Simulations}",
      journal = {\apj},
         year = 2025,
        month = jan,
       volume = {978},
       number = {1},
          eid = {37},
        pages = {37},
          doi = {10.3847/1538-4357/ad8b45},
archivePrefix = {arXiv},
       eprint = {2309.16811},
 primaryClass = {astro-ph.GA},
       adsurl = {https://ui.adsabs.harvard.edu/abs/2025ApJ...978...37A}
}

@ARTICLE{Athanassoula1986,
       author = {{Athanassoula}, E. and {Sellwood}, J.~A.},
        title = "{Bi-symmetric instabilities of the Kuz'min/Toomre disc.}",
      journal = {\mnras},
         year = 1986,
        month = jul,
       volume = {221},
        pages = {213-232},
          doi = {10.1093/mnras/221.2.213},
       adsurl = {https://ui.adsabs.harvard.edu/abs/1986MNRAS.221..213A}
}

@ARTICLE{athan92a,
author = {{Athanassoula}, E.},
title = "{Morphology of bar orbits.}",
journal = {\mnras},
year = 1992,
month = nov,
volume = {259},
pages = {328-344},
doi = {10.1093/mnras/259.2.328},
adsurl = {https://ui.adsabs.harvard.edu/abs/1992MNRAS.259..328A}
}

@ARTICLE{Athanassoula2002,
       author = {{Athanassoula}, E.},
        title = "{Bar-Halo Interaction and Bar Growth}",
      journal = {\apjl},
         year = 2002,
        month = apr,
       volume = {569},
       number = {2},
        pages = {L83-L86},
          doi = {10.1086/340784},
archivePrefix = {arXiv},
       eprint = {astro-ph/0203368},
 primaryClass = {astro-ph},
       adsurl = {https://ui.adsabs.harvard.edu/abs/2002ApJ...569L..83A}
}

@ARTICLE{Athanassoula.Misiriotis.2002,
       author = {{Athanassoula}, E. and {Misiriotis}, A.},
        title = "{Morphology, photometry and kinematics of N -body bars - I. Three models with different halo central concentrations}",
      journal = {\mnras},
         year = 2002,
        month = feb,
       volume = {330},
       number = {1},
        pages = {35-52},
          doi = {10.1046/j.1365-8711.2002.05028.x},
archivePrefix = {arXiv},
       eprint = {astro-ph/0111449},
 primaryClass = {astro-ph},
       adsurl = {https://ui.adsabs.harvard.edu/abs/2002MNRAS.330...35A}
}

@ARTICLE{Athanassoula2003,
       author = {{Athanassoula}, E.},
        title = "{What determines the strength and the slowdown rate of bars?}",
      journal = {\mnras},
         year = 2003,
        month = jun,
       volume = {341},
       number = {4},
        pages = {1179-1198},
          doi = {10.1046/j.1365-8711.2003.06473.x},
archivePrefix = {arXiv},
       eprint = {astro-ph/0302519},
 primaryClass = {astro-ph},
       adsurl = {https://ui.adsabs.harvard.edu/abs/2003MNRAS.341.1179A}
}

@ARTICLE{Athanassoula2005,
       author = {{Athanassoula}, E.},
        title = "{On the nature of bulges in general and of box/peanut bulges in particular: input from N-body simulations}",
      journal = {\mnras},
         year = 2005,
        month = apr,
       volume = {358},
       number = {4},
        pages = {1477-1488},
          doi = {10.1111/j.1365-2966.2005.08872.x},
archivePrefix = {arXiv},
       eprint = {astro-ph/0502316},
 primaryClass = {astro-ph},
       adsurl = {https://ui.adsabs.harvard.edu/abs/2005MNRAS.358.1477A}
}

@ARTICLE{Athanassoula2013,
       author = {{Athanassoula}, E. and {Machado}, Rubens E.~G. and {Rodionov}, S.~A.},
        title = "{Bar formation and evolution in disc galaxies with gas and a triaxial halo: morphology, bar strength and halo properties}",
      journal = {\mnras},
         year = 2013,
        month = mar,
       volume = {429},
       number = {3},
        pages = {1949-1969},
          doi = {10.1093/mnras/sts452},
archivePrefix = {arXiv},
       eprint = {1211.6754},
 primaryClass = {astro-ph.CO},
       adsurl = {https://ui.adsabs.harvard.edu/abs/2013MNRAS.429.1949A}
}

@ARTICLE{baba2017eventful,
author = {{Baba}, Junichi and {Morokuma-Matsui}, Kana and {Saitoh}, Takayuki R.},
title = "{Eventful evolution of giant molecular clouds in dynamically evolving spiral arms}",
journal = {\mnras},
year = 2017,
month = jan,
volume = {464},
number = {1},
pages = {246-263},
doi = {10.1093/mnras/stw2378},
archivePrefix = {arXiv},
eprint = {1609.00097},
primaryClass = {astro-ph.GA},
adsurl = {https://ui.adsabs.harvard.edu/abs/2017MNRAS.464..246B}
}

@ARTICLE{binney1991understanding,
author = {{Binney}, James and {Gerhard}, Ortwin E. and {Stark}, Antony A. and {Bally}, John and {Uchida}, Keven I.},
title = "{Understanding the kinematics of Galactic Centre gas.}",
journal = {\mnras},
year = 1991,
month = sep,
volume = {252},
pages = {210},
doi = {10.1093/mnras/252.2.210},
adsurl = {https://ui.adsabs.harvard.edu/abs/1991MNRAS.252..210B}
}

@BOOK{Binney2008,
       author = {{Binney}, James and {Tremaine}, Scott},
        title = "{Galactic Dynamics: Second Edition}",
         year = 2008,
       adsurl = {https://ui.adsabs.harvard.edu/abs/2008gady.book.....B}
}

@BOOK{binney2011galactic,
author = {{Binney}, James and {Tremaine}, Scott},
title = "{Galactic Dynamics: Second Edition}",
year = 2008,
adsurl = {https://ui.adsabs.harvard.edu/abs/2008gady.book.....B}
}

@ARTICLE{blitz1991direct,
author = {{Blitz}, Leo and {Spergel}, David N.},
title = "{Direct Evidence for a Bar at the Galactic Center}",
journal = {\apj},
year = 1991,
month = oct,
volume = {379},
pages = {631},
doi = {10.1086/170535},
adsurl = {https://ui.adsabs.harvard.edu/abs/1991ApJ...379..631B}
}

@ARTICLE{Bureau2005,
       author = {{Bureau}, M. and {Athanassoula}, E.},
        title = "{Bar Diagnostics in Edge-On Spiral Galaxies. III. N-Body Simulations of Disks}",
      journal = {\apj},
         year = 2005,
        month = jun,
       volume = {626},
       number = {1},
        pages = {159-173},
          doi = {10.1086/430056},
archivePrefix = {arXiv},
       eprint = {astro-ph/0403226},
 primaryClass = {astro-ph},
       adsurl = {https://ui.adsabs.harvard.edu/abs/2005ApJ...626..159B}
}

@ARTICLE{Bovy2019,
author = {{Bovy}, Jo and {Leung}, Henry W. and {Hunt}, Jason A.~S. and {Mackereth}, J. Ted and {Garc{\'\i}a-Hern{\'a}ndez}, Domingo A. and {Roman-Lopes}, Alexandre},
title = "{Life in the fast lane: a direct view of the dynamics, formation, and evolution of the Milky Way's bar}",
journal = {\mnras},
year = 2019,
month = dec,
volume = {490},
number = {4},
pages = {4740-4747},
doi = {10.1093/mnras/stz2891},
archivePrefix = {arXiv},
eprint = {1905.11404},
primaryClass = {astro-ph.GA},
adsurl = {https://ui.adsabs.harvard.edu/abs/2019MNRAS.490.4740B}
}

@ARTICLE{Ciambur2017,
       author = {{Ciambur}, Bogdan C. and {Graham}, Alister W. and {Bland-Hawthorn}, Joss},
        title = "{Quantifying the (X/peanut)-shaped structure of the Milky Way - new constraints on the bar geometry}",
      journal = {\mnras},
         year = 2017,
        month = nov,
       volume = {471},
       number = {4},
        pages = {3988-4004},
          doi = {10.1093/mnras/stx1823},
archivePrefix = {arXiv},
       eprint = {1706.09902},
 primaryClass = {astro-ph.GA},
       adsurl = {https://ui.adsabs.harvard.edu/abs/2017MNRAS.471.3988C}
}

@ARTICLE{Combes1990,
       author = {{Combes}, F. and {Debbasch}, F. and {Friedli}, D. and {Pfenniger}, D.},
        title = "{Box and peanut shapes generated by stellar bars.}",
      journal = {\aap},
         year = 1990,
        month = jul,
       volume = {233},
        pages = {82},
       adsurl = {https://ui.adsabs.harvard.edu/abs/1990A&A...233...82C}
}

@ARTICLE{Contopoulus1980howfarbars,
       author = {{Contopoulos}, G.},
        title = "{How far do bars extend}",
      journal = {\aap},
         year = 1980,
        month = jan,
       volume = {81},
       number = {1-2},
        pages = {198-209},
       adsurl = {https://ui.adsabs.harvard.edu/abs/1980A&A....81..198C}
}

@ARTICLE{Contopoulus1989,
author = {{Contopoulos}, G. and {Grosbol}, P.},
title = "{Orbits in barred galaxies}",
journal = {\aapr},
year = 1989,
month = nov,
volume = {1},
number = {3-4},
pages = {261-289},
doi = {10.1007/BF00873080},
adsurl = {https://ui.adsabs.harvard.edu/abs/1989A&ARv...1..261C}
}

@ARTICLE{cautun2020milky,
author = {{Cautun}, Marius and {Ben{\'\i}tez-Llambay}, Alejandro and {Deason}, Alis J. and {Frenk}, Carlos S. and {Fattahi}, Azadeh and {G{\'o}mez}, Facundo A. and {Grand}, Robert J.~J. and {Oman}, Kyle A. and {Navarro}, Julio F. and {Simpson}, Christine M.},
title = "{The milky way total mass profile as inferred from Gaia DR2}",
journal = {\mnras},
year = 2020,
month = may,
volume = {494},
number = {3},
pages = {4291-4313},
doi = {10.1093/mnras/staa1017},
archivePrefix = {arXiv},
eprint = {1911.04557},
primaryClass = {astro-ph.GA},
adsurl = {https://ui.adsabs.harvard.edu/abs/2020MNRAS.494.4291C}
}

@ARTICLE{Correa2017,
       author = {{Correa}, Camila A. and {Schaye}, Joop and {Clauwens}, Bart and {Bower}, Richard G. and {Crain}, Robert A. and {Schaller}, Matthieu and {Theuns}, Tom and {Thob}, Adrien C.~R.},
        title = "{The relation between galaxy morphology and colour in the EAGLE simulation}",
      journal = {\mnras},
         year = 2017,
        month = nov,
       volume = {472},
       number = {1},
        pages = {L45-L49},
          doi = {10.1093/mnrasl/slx133},
archivePrefix = {arXiv},
       eprint = {1704.06283},
 primaryClass = {astro-ph.GA},
       adsurl = {https://ui.adsabs.harvard.edu/abs/2017MNRAS.472L..45C}
}

@ARTICLE{corsini2011,
       author = {{Corsini}, E.~M.},
        title = "{Direct measurements of bar pattern speeds.}",
      journal = {Memorie della Societa Astronomica Italiana Supplementi},
         year = 2011,
        month = jan,
       volume = {18},
        pages = {23},
          doi = {10.48550/arXiv.1002.1245},
archivePrefix = {arXiv},
       eprint = {1002.1245},
 primaryClass = {astro-ph.GA},
       adsurl = {https://ui.adsabs.harvard.edu/abs/2011MSAIS..18...23C}
}

@ARTICLE{Cuomo2020,
       author = {{Cuomo}, V. and {Aguerri}, J.~A.~L. and {Corsini}, E.~M. and {Debattista}, V.~P.},
        title = "{Relations among structural parameters in barred galaxies with a direct measurement of bar pattern speed}",
      journal = {\aap},
         year = 2020,
        month = sep,
       volume = {641},
          eid = {A111},
        pages = {A111},
          doi = {10.1051/0004-6361/202037945},
archivePrefix = {arXiv},
       eprint = {2003.07455},
 primaryClass = {astro-ph.GA},
       adsurl = {https://ui.adsabs.harvard.edu/abs/2020A&A...641A.111C}
}

@ARTICLE{cuomo2021bar,
author = {{Cuomo}, Virginia and {Lee}, Yun Hee and {Buttitta}, Chiara and {Aguerri}, J. Alfonso L. and {Corsini}, Enrico Maria and {Morelli}, Lorenzo},
title = "{Bar pattern speeds in CALIFA galaxies. III. Solving the puzzle of ultrafast bars}",
journal = {\aap},
year = 2021,
month = may,
volume = {649},
eid = {A30},
pages = {A30},
doi = {10.1051/0004-6361/202040261},
archivePrefix = {arXiv},
eprint = {2103.00343},
primaryClass = {astro-ph.GA},
adsurl = {https://ui.adsabs.harvard.edu/abs/2021A&A...649A..30C}
}

@ARTICLE{cuomo2024,
       author = {{Cuomo}, Virginia and {Morelli}, Lorenzo and {Aguerri}, J. Alfonso L. and {Corsini}, Enrico Maria and {Debattista}, Victor P. and {Coccato}, Lodovico and {Pizzella}, Alessandro and {Boselli}, Alessandro and {Buttitta}, Chiara and {de Lorenzo-C{\'a}ceres}, Adriana and {Ferrarese}, Laura and {Gasparri}, Daniele and {Lee}, Yun Hee and {Mendez-Abreu}, Jairo and {Roediger}, Joel and {Zarattini}, Stefano},
        title = "{The dynamical state of bars in cluster dwarf galaxies: the cases of NGC 4483 and NGC 4516}",
      journal = {\mnras},
         year = 2024,
        month = feb,
       volume = {527},
       number = {4},
        pages = {11218-11232},
          doi = {10.1093/mnras/stad3945},
archivePrefix = {arXiv},
       eprint = {2312.14017},
 primaryClass = {astro-ph.GA},
       adsurl = {https://ui.adsabs.harvard.edu/abs/2024MNRAS.52711218C}
}

@ARTICLE{Dehnen2023,
       author = {{Dehnen}, Walter and {Semczuk}, Marcin and {Sch{\"o}nrich}, Ralph},
        title = "{Measuring bar pattern speeds from single simulation snapshots}",
      journal = {\mnras},
         year = 2023,
        month = jan,
       volume = {518},
       number = {2},
        pages = {2712-2718},
          doi = {10.1093/mnras/stac3184},
archivePrefix = {arXiv},
       eprint = {2211.00674},
 primaryClass = {astro-ph.GA},
       adsurl = {https://ui.adsabs.harvard.edu/abs/2023MNRAS.518.2712D}
}

@inproceedings{de1964comparison,
  title={Comparison of the Magellanic Clouds by other irregular barred spirals},
  author={De Vaucouleurs, G},
  booktitle={Symposium-International Astronomical Union},
  volume={20},
  pages={269--276},
  year={1964},
  adsurl = {https://ui.adsabs.harvard.edu/abs/1964IAUS...20..269D},
  organization={Cambridge University Press}
}

@ARTICLE{Debattista2000,
       author = {{Debattista}, Victor P. and {Sellwood}, J.~A.},
        title = "{Constraints from Dynamical Friction on the Dark Matter Content of Barred Galaxies}",
      journal = {\apj},
         year = 2000,
        month = nov,
       volume = {543},
       number = {2},
        pages = {704-721},
          doi = {10.1086/317148},
archivePrefix = {arXiv},
       eprint = {astro-ph/0006275},
 primaryClass = {astro-ph},
       adsurl = {https://ui.adsabs.harvard.edu/abs/2000ApJ...543..704D}
}

@ARTICLE{Debattista2006,
       author = {{Debattista}, Victor P. and {Mayer}, Lucio and {Carollo}, C. Marcella and {Moore}, Ben and {Wadsley}, James and {Quinn}, Thomas},
        title = "{The Secular Evolution of Disk Structural Parameters}",
      journal = {\apj},
         year = 2006,
        month = jul,
       volume = {645},
       number = {1},
        pages = {209-227},
          doi = {10.1086/504147},
archivePrefix = {arXiv},
       eprint = {astro-ph/0509310},
 primaryClass = {astro-ph},
       adsurl = {https://ui.adsabs.harvard.edu/abs/2006ApJ...645..209D}
}

@ARTICLE{Debattista2015,
       author = {{Debattista}, Victor P. and {Ness}, Melissa and {Earp}, Samuel W.~F. and {Cole}, David R.},
        title = "{A Kiloparsec-scale Nuclear Stellar Disk in the Milky Way as a Possible Explanation of the High Velocity Peaks in the Galactic Bulge}",
      journal = {\apjl},
         year = 2015,
        month = oct,
       volume = {812},
       number = {1},
          eid = {L16},
        pages = {L16},
          doi = {10.1088/2041-8205/812/1/L16},
archivePrefix = {arXiv},
       eprint = {1507.01433},
 primaryClass = {astro-ph.GA},
       adsurl = {https://ui.adsabs.harvard.edu/abs/2015ApJ...812L..16D}
}

@ARTICLE{Dillamore2024,
       author = {{Dillamore}, Adam M. and {Belokurov}, Vasily and {Kravtsov}, Andrey and {Font}, Andreea S.},
        title = "{Taking the Milky Way for a spin: disc formation in the ARTEMIS simulations}",
      journal = {\mnras},
         year = 2024,
        month = jan,
       volume = {527},
       number = {3},
        pages = {7070-7078},
          doi = {10.1093/mnras/stad3369},
archivePrefix = {arXiv},
       eprint = {2309.08658},
 primaryClass = {astro-ph.GA},
       adsurl = {https://ui.adsabs.harvard.edu/abs/2024MNRAS.527.7070D}
}

@ARTICLE{drimmel2023,
author = {{Gaia Collaboration} and {Drimmel}, R. and {Romero-G{\'o}mez}, M. and {Chemin}, L. and {Ramos}, P. and {Poggio}, E. and {Ripepi}, V. and {Andrae}, R. and {Blomme}, R. and {Cantat-Gaudin}, T. and {Castro-Ginard}, A. and {Clementini}, G. and {Figueras}, F. and {Fouesneau}, M. and {Fr{\'e}mat}, Y. and {Jardine}, K. and {Khanna}, S. and {Lobel}, A. and {Marshall}, D.~J. and {Muraveva}, T. and {Brown}, A.~G.~A. and {Vallenari}, A. and {Prusti}, T. and {de Bruijne}, J.~H.~J. and {Arenou}, F. and {Babusiaux}, C. and {Biermann}, M. and {Creevey}, O.~L. and {Ducourant}, C. and {Evans}, D.~W. and {Eyer}, L. and {Guerra}, R. and {Hutton}, A. and {Jordi}, C. and {Klioner}, S.~A. and {Lammers}, U.~L. and {Lindegren}, L. and {Luri}, X. and {Mignard}, F. and {Panem}, C. and {Pourbaix}, D. and {Randich}, S. and {Sartoretti}, P. and {Soubiran}, C. and {Tanga}, P. and {Walton}, N.~A. and {Bailer-Jones}, C.~A.~L. and {Bastian}, U. and {Jansen}, F. and {Katz}, D. and {Lattanzi}, M.~G. and {van Leeuwen}, F. and {Bakker}, J. and {Cacciari}, C. and {Casta{\~n}eda}, J. and {De Angeli}, F. and {Fabricius}, C. and {Galluccio}, L. and {Guerrier}, A. and {Heiter}, U. and {Masana}, E. and {Messineo}, R. and {Mowlavi}, N. and {Nicolas}, C. and {Nienartowicz}, K. and {Pailler}, F. and {Panuzzo}, P. and {Riclet}, F. and {Roux}, W. and {Seabroke}, G.~M. and {Sordo}, R. and {Th{\'e}venin}, F. and {Gracia-Abril}, G. and {Portell}, J. and {Teyssier}, D. and {Altmann}, M. and {Audard}, M. and {Bellas-Velidis}, I. and {Benson}, K. and {Berthier}, J. and {Burgess}, P.~W. and {Busonero}, D. and {Busso}, G. and {C{\'a}novas}, H. and {Carry}, B. and {Cellino}, A. and {Cheek}, N. and {Damerdji}, Y. and {Davidson}, M. and {de Teodoro}, P. and {Nu{\~n}ez Campos}, M. and {Delchambre}, L. and {Dell'Oro}, A. and {Esquej}, P. and {Fern{\'a}ndez-Hern{\'a}ndez}, J. and {Fraile}, E. and {Garabato}, D. and {Garc{\'\i}a-Lario}, P. and {Gosset}, E. and {Haigron}, R. and {Halbwachs}, J. -L. and {Hambly}, N.~C. and {Harrison}, D.~L. and {Hern{\'a}ndez}, J. and {Hestroffer}, D. and {Hodgkin}, S.~T. and {Holl}, B. and {Jan{\ss}en}, K. and {Jevardat de Fombelle}, G. and {Jordan}, S. and {Krone-Martins}, A. and {Lanzafame}, A.~C. and {L{\"o}ffler}, W. and {Marchal}, O. and {Marrese}, P.~M. and {Moitinho}, A. and {Muinonen}, K. and {Osborne}, P. and {Pancino}, E. and {Pauwels}, T. and {Recio-Blanco}, A. and {Reyl{\'e}}, C. and {Riello}, M. and {Rimoldini}, L. and {Roegiers}, T. and {Rybizki}, J. and {Sarro}, L.~M. and {Siopis}, C. and {Smith}, M. and {Sozzetti}, A. and {Utrilla}, E. and {van Leeuwen}, M. and {Abbas}, U. and {{\'A}brah{\'a}m}, P. and {Abreu Aramburu}, A. and {Aerts}, C. and {Aguado}, J.~J. and {Ajaj}, M. and {Aldea-Montero}, F. and {Altavilla}, G. and {{\'A}lvarez}, M.~A. and {Alves}, J. and {Anders}, F. and {Anderson}, R.~I. and {Anglada Varela}, E. and {Antoja}, T. and {Baines}, D. and {Baker}, S.~G. and {Balaguer-N{\'u}{\~n}ez}, L. and {Balbinot}, E. and {Balog}, Z. and {Barache}, C. and {Barbato}, D. and {Barros}, M. and {Barstow}, M.~A. and {Bartolom{\'e}}, S. and {Bassilana}, J. -L. and {Bauchet}, N. and {Becciani}, U. and {Bellazzini}, M. and {Berihuete}, A. and {Bernet}, M. and {Bertone}, S. and {Bianchi}, L. and {Binnenfeld}, A. and {Blanco-Cuaresma}, S. and {Boch}, T. and {Bombrun}, A. and {Bossini}, D. and {Bouquillon}, S. and {Bragaglia}, A. and {Bramante}, L. and {Breedt}, E. and {Bressan}, A. and {Brouillet}, N. and {Brugaletta}, E. and {Bucciarelli}, B. and {Burlacu}, A. and {Butkevich}, A.~G. and {Buzzi}, R. and {Caffau}, E. and {Cancelliere}, R. and {Carballo}, R. and {Carlucci}, T. and {Carnerero}, M.~I. and {Carrasco}, J.~M. and {Casamiquela}, L. and {Castellani}, M. and {Chaoul}, L. and {Charlot}, P. and {Chiaramida}, V. and {Chiavassa}, A. and {Chornay}, N. and {Comoretto}, G. and {Contursi}, G. and {Cooper}, W.~J. and {Cornez}, T. and {Cowell}, S. and {Crifo}, F. and {Cropper}, M. and {Crosta}, M. and {Crowley}, C. and {Dafonte}, C. and {Dapergolas}, A. and {David}, P. and {de Laverny}, P. and {De Luise}, F. and {De March}, R. and {De Ridder}, J. and {de Souza}, R. and {de Torres}, A. and {del Peloso}, E.~F. and {del Pozo}, E. and {Delbo}, M. and {Delgado}, A. and {Delisle}, J. -B. and {Demouchy}, C. and {Dharmawardena}, T.~E. and {Di Matteo}, P. and {Diakite}, S. and {Diener}, C. and {Distefano}, E. and {Dolding}, C. and {Enke}, H. and {Fabre}, C. and {Fabrizio}, M. and {Faigler}, S. and {Fedorets}, G. and {Fernique}, P. and {Fournier}, Y. and {Fouron}, C. and {Fragkoudi}, F. and {Gai}, M. and {Garcia-Gutierrez}, A. and {Garcia-Reinaldos}, M. and {Garc{\'\i}a-Torres}, M. and {Garofalo}, A. and {Gavel}, A. and {Gavras}, P. and {Gerlach}, E. and {Geyer}, R. and {Giacobbe}, P. and {Gilmore}, G. and {Girona}, S. and {Giuffrida}, G. and {Gomel}, R. and {Gomez}, A. and {Gonz{\'a}lez-N{\'u}{\~n}ez}, J. and {Gonz{\'a}lez-Santamar{\'\i}a}, I. and {Gonz{\'a}lez-Vidal}, J.~J. and {Granvik}, M. and {Guillout}, P. and {Guiraud}, J. and {Guti{\'e}rrez-S{\'a}nchez}, R. and {Guy}, L.~P. and {Hatzidimitriou}, D. and {Hauser}, M. and {Haywood}, M. and {Helmer}, A. and {Helmi}, A. and {Sarmiento}, M.~H. and {Hidalgo}, S.~L. and {H{\l}adczuk}, N. and {Hobbs}, D. and {Holland}, G. and {Huckle}, H.~E. and {Jasniewicz}, G. and {Jean-Antoine Piccolo}, A. and {Jim{\'e}nez-Arranz}, {\'O}. and {Juaristi Campillo}, J. and {Julbe}, F. and {Karbevska}, L. and {Kervella}, P. and {Kordopatis}, G. and {Korn}, A.~J. and {K{\'o}sp{\'a}l}, {\'A}. and {Kostrzewa-Rutkowska}, Z. and {Kruszy{\'n}ska}, K. and {Kun}, M. and {Laizeau}, P. and {Lambert}, S. and {Lanza}, A.~F. and {Lasne}, Y. and {Le Campion}, J. -F. and {Lebreton}, Y. and {Lebzelter}, T. and {Leccia}, S. and {Leclerc}, N. and {Lecoeur-Taibi}, I. and {Liao}, S. and {Licata}, E.~L. and {Lindstr{\o}m}, H.~E.~P. and {Lister}, T.~A. and {Livanou}, E. and {Lorca}, A. and {Loup}, C. and {Madrero Pardo}, P. and {Magdaleno Romeo}, A. and {Managau}, S. and {Mann}, R.~G. and {Manteiga}, M. and {Marchant}, J.~M. and {Marconi}, M. and {Marcos}, J. and {Marcos Santos}, M.~M.~S. and {Mar{\'\i}n Pina}, D. and {Marinoni}, S. and {Marocco}, F. and {Martin Polo}, L. and {Mart{\'\i}n-Fleitas}, J.~M. and {Marton}, G. and {Mary}, N. and {Masip}, A. and {Massari}, D. and {Mastrobuono-Battisti}, A. and {Mazeh}, T. and {McMillan}, P.~J. and {Messina}, S. and {Michalik}, D. and {Millar}, N.~R. and {Mints}, A. and {Molina}, D. and {Molinaro}, R. and {Moln{\'a}r}, L. and {Monari}, G. and {Mongui{\'o}}, M. and {Montegriffo}, P. and {Montero}, A. and {Mor}, R. and {Mora}, A. and {Morbidelli}, R. and {Morel}, T. and {Morris}, D. and {Murphy}, C.~P. and {Musella}, I. and {Nagy}, Z. and {Noval}, L. and {Oca{\~n}a}, F. and {Ogden}, A. and {Ordenovic}, C. and {Osinde}, J.~O. and {Pagani}, C. and {Pagano}, I. and {Palaversa}, L. and {Palicio}, P.~A. and {Pallas-Quintela}, L. and {Panahi}, A. and {Payne-Wardenaar}, S. and {Pe{\~n}alosa Esteller}, X. and {Penttil{\"a}}, A. and {Pichon}, B. and {Piersimoni}, A.~M. and {Pineau}, F. -X. and {Plachy}, E. and {Plum}, G. and {Pr{\v{s}}a}, A. and {Pulone}, L. and {Racero}, E. and {Ragaini}, S. and {Rainer}, M. and {Raiteri}, C.~M. and {Ramos-Lerate}, M. and {Re Fiorentin}, P. and {Regibo}, S. and {Richards}, P.~J. and {Rios Diaz}, C. and {Riva}, A. and {Rix}, H. -W. and {Rixon}, G. and {Robichon}, N. and {Robin}, A.~C. and {Robin}, C. and {Roelens}, M. and {Rogues}, H.~R.~O. and {Rohrbasser}, L. and {Rowell}, N. and {Royer}, F. and {Ruz Mieres}, D. and {Rybicki}, K.~A. and {Sadowski}, G. and {S{\'a}ez N{\'u}{\~n}ez}, A. and {Sagrist{\`a} Sell{\'e}s}, A. and {Sahlmann}, J. and {Salguero}, E. and {Samaras}, N. and {Sanchez Gimenez}, V. and {Sanna}, N. and {Santove{\~n}a}, R. and {Sarasso}, M. and {Schultheis}, M.~S. and {Sciacca}, E. and {Segol}, M. and {Segovia}, J.~C. and {S{\'e}gransan}, D. and {Semeux}, D. and {Shahaf}, S. and {Siddiqui}, H.~I. and {Siebert}, A. and {Siltala}, L. and {Silvelo}, A. and {Slezak}, E. and {Slezak}, I. and {Smart}, R.~L. and {Snaith}, O.~N. and {Solano}, E. and {Solitro}, F. and {Souami}, D. and {Souchay}, J. and {Spagna}, A. and {Spina}, L. and {Spoto}, F. and {Steele}, I.~A. and {Steidelm{\"u}ller}, H. and {Stephenson}, C.~A. and {S{\"u}veges}, M. and {Surdej}, J. and {Szabados}, L. and {Szegedi-Elek}, E. and {Taris}, F. and {Taylor}, M.~B. and {Teixeira}, R. and {Tolomei}, L. and {Tonello}, N. and {Torra}, F. and {Torra}, J. and {Torralba Elipe}, G. and {Trabucchi}, M. and {Tsounis}, A.~T. and {Turon}, C. and {Ulla}, A. and {Unger}, N. and {Vaillant}, M.~V. and {van Dillen}, E. and {van Reeven}, W. and {Vanel}, O. and {Vecchiato}, A. and {Viala}, Y. and {Vicente}, D. and {Voutsinas}, S. and {Weiler}, M. and {Wevers}, T. and {Wyrzykowski}, {\L}. and {Yoldas}, A. and {Yvard}, P. and {Zhao}, H. and {Zorec}, J. and {Zucker}, S. and {Zwitter}, T.},
title = "{Gaia Data Release 3. Mapping the asymmetric disc of the Milky Way}",
journal = {\aap},
year = 2023,
month = jun,
volume = {674},
eid = {A37},
pages = {A37},
doi = {10.1051/0004-6361/202243797},
archivePrefix = {arXiv},
eprint = {2206.06207},
primaryClass = {astro-ph.GA},
adsurl = {https://ui.adsabs.harvard.edu/abs/2023A&A...674A..37G}
}

@ARTICLE{DuranCamacho2024,
       author = {{Dur{\'a}n-Camacho}, Eva and {Duarte-Cabral}, Ana and {Pettitt}, Alex R. and {Tre{\ss}}, Robin G. and {Clark}, Paul C. and {Klessen}, Ralf S. and {Bogue}, Kamran R.~J. and {Smith}, Rowan J. and {Sormani}, Mattia C.},
        title = "{Self-consistent modelling of the Milky Way structure using live potentials}",
      journal = {\mnras},
         year = 2024,
        month = jul,
       volume = {532},
       number = {1},
        pages = {126-148},
          doi = {10.1093/mnras/stae1469},
archivePrefix = {arXiv},
       eprint = {2405.09503},
       adsurl = {https://ui.adsabs.harvard.edu/abs/2024MNRAS.532..126D}
}

@ARTICLE{DuranCamacho2026,
       author = {{Dur{\'a}n-Camacho}, E. and {Duarte-Cabral}, A.},
        title = "{Insights into the structure and kinematics of a Milky Way-like galaxy}",
      journal = {\aap},
         year = 2026,
        month = apr,
       volume = {708},
          eid = {A370},
        pages = {A370},
          doi = {10.1051/0004-6361/202558565},
archivePrefix = {arXiv},
       eprint = {2506.13560},
 primaryClass = {astro-ph.GA},
       adsurl = {https://ui.adsabs.harvard.edu/abs/2026A&A...708A.370D}
}

@ARTICLE{Dwek1995,
       author = {{Dwek}, E. and {Arendt}, R.~G. and {Hauser}, M.~G. and {Kelsall}, T. and {Lisse}, C.~M. and {Moseley}, S.~H. and {Silverberg}, R.~F. and {Sodroski}, T.~J. and {Weiland}, J.~L.},
        title = "{Morphology, Near-Infrared Luminosity, and Mass of the Galactic Bulge from COBE DIRBE Observations}",
      journal = {\apj},
         year = 1995,
        month = jun,
       volume = {445},
        pages = {716},
          doi = {10.1086/175734},
       adsurl = {https://ui.adsabs.harvard.edu/abs/1995ApJ...445..716D}
}

@ARTICLE{Efstathiou1982,
       author = {{Efstathiou}, G. and {Lake}, G. and {Negroponte}, J.},
        title = "{The stability and masses of disc galaxies.}",
      journal = {\mnras},
         year = 1982,
        month = jun,
       volume = {199},
        pages = {1069-1088},
          doi = {10.1093/mnras/199.4.1069},
       adsurl = {https://ui.adsabs.harvard.edu/abs/1982MNRAS.199.1069E}
}

@ARTICLE{Elmegreen2007,
       author = {{Elmegreen}, Bruce G. and {Elmegreen}, Debra Meloy and {Knapen}, Johan H. and {Buta}, Ronald J. and {Block}, David L. and {Puerari}, Iv{\^a}nio},
        title = "{Variation of Galactic Bar Length with Amplitude and Density as Evidence for Bar Growth over a Hubble Time}",
      journal = {\apjl},
         year = 2007,
        month = dec,
       volume = {670},
       number = {2},
        pages = {L97-L100},
          doi = {10.1086/524359},
archivePrefix = {arXiv},
       eprint = {0711.3055},
 primaryClass = {astro-ph},
       adsurl = {https://ui.adsabs.harvard.edu/abs/2007ApJ...670L..97E}
}

@ARTICLE{Eskridge2000,
       author = {{Eskridge}, Paul B. and {Frogel}, Jay A. and {Pogge}, Richard W. and {Quillen}, Alice C. and {Davies}, Roger L. and {DePoy}, D.~L. and {Houdashelt}, Mark L. and {Kuchinski}, Leslie E. and {Ram{\'\i}rez}, Solange V. and {Sellgren}, K. and {Terndrup}, Donald M. and {Tiede}, Glenn P.},
        title = "{The Frequency of Barred Spiral Galaxies in the Near-Infrared}",
      journal = {\aj},
         year = 2000,
        month = feb,
       volume = {119},
       number = {2},
        pages = {536-544},
          doi = {10.1086/301203},
archivePrefix = {arXiv},
       eprint = {astro-ph/9910479},
 primaryClass = {astro-ph},
       adsurl = {https://ui.adsabs.harvard.edu/abs/2000AJ....119..536E}
}

@ARTICLE{Erwin2013,
       author = {{Erwin}, Peter and {Debattista}, Victor P.},
        title = "{Peanuts at an angle: detecting and measuring the three-dimensional structure of bars in moderately inclined galaxies}",
      journal = {\mnras},
         year = 2013,
        month = jun,
       volume = {431},
       number = {4},
        pages = {3060-3086},
          doi = {10.1093/mnras/stt385},
archivePrefix = {arXiv},
       eprint = {1301.0638},
 primaryClass = {astro-ph.CO},
       adsurl = {https://ui.adsabs.harvard.edu/abs/2013MNRAS.431.3060E}
}

@ARTICLE{Erwin2016,
       author = {{Erwin}, Peter and {Debattista}, Victor P.},
        title = "{Caught in the Act: Direct Detection of Galactic Bars in the Buckling Phase}",
      journal = {\apjl},
         year = 2016,
        month = jul,
       volume = {825},
       number = {2},
          eid = {L30},
        pages = {L30},
          doi = {10.3847/2041-8205/825/2/L30},
archivePrefix = {arXiv},
       eprint = {1607.01290},
 primaryClass = {astro-ph.GA},
       adsurl = {https://ui.adsabs.harvard.edu/abs/2016ApJ...825L..30E}
}

@ARTICLE{Erwin2017,
       author = {{Erwin}, Peter and {Debattista}, Victor P.},
        title = "{The frequency and stellar-mass dependence of boxy/peanut-shaped bulges in barred galaxies}",
      journal = {\mnras},
         year = 2017,
        month = jun,
       volume = {468},
       number = {2},
        pages = {2058-2080},
          doi = {10.1093/mnras/stx620},
archivePrefix = {arXiv},
       eprint = {1703.01602},
 primaryClass = {astro-ph.GA},
       adsurl = {https://ui.adsabs.harvard.edu/abs/2017MNRAS.468.2058E}
}

@ARTICLE{Erwin2018,
       author = {{Erwin}, Peter},
        title = "{The dependence of bar frequency on galaxy mass, colour, and gas content - and angular resolution - in the local universe}",
      journal = {\mnras},
         year = 2018,
        month = mar,
       volume = {474},
       number = {4},
        pages = {5372-5392},
          doi = {10.1093/mnras/stx3117},
archivePrefix = {arXiv},
       eprint = {1711.04867},
 primaryClass = {astro-ph.GA},
       adsurl = {https://ui.adsabs.harvard.edu/abs/2018MNRAS.474.5372E}
}

@ARTICLE{Erwin2019,
       author = {{Erwin}, Peter},
        title = "{What determines the sizes of bars in spiral galaxies?}",
      journal = {\mnras},
         year = 2019,
        month = nov,
       volume = {489},
       number = {3},
        pages = {3553-3564},
          doi = {10.1093/mnras/stz2363},
archivePrefix = {arXiv},
       eprint = {1908.08423},
 primaryClass = {astro-ph.GA},
       adsurl = {https://ui.adsabs.harvard.edu/abs/2019MNRAS.489.3553E}
}

@ARTICLE{li2023bucklingII,
       author = {{Li}, Xingchen and {Shlosman}, Isaac and {Heller}, Clayton and {Pfenniger}, Daniel},
        title = "{Stellar bars in spinning haloes: delayed buckling and absence of slowdown}",
      journal = {\mnras},
         year = 2023,
        month = dec,
       volume = {526},
       number = {2},
        pages = {1972-1986},
          doi = {10.1093/mnras/stad2799},
archivePrefix = {arXiv},
       eprint = {2211.04484},
 primaryClass = {astro-ph.GA},
       adsurl = {https://ui.adsabs.harvard.edu/abs/2023MNRAS.526.1972L}
}

@ARTICLE{Lokas2019,
       author = {{{\L}okas}, Ewa L.},
        title = "{Anatomy of a buckling galactic bar}",
      journal = {\aap},
         year = 2019,
        month = sep,
       volume = {629},
          eid = {A52},
        pages = {A52},
          doi = {10.1051/0004-6361/201936056},
archivePrefix = {arXiv},
       eprint = {1906.03916},
 primaryClass = {astro-ph.GA},
       adsurl = {https://ui.adsabs.harvard.edu/abs/2019A&A...629A..52L}
}

@ARTICLE{Lokas2020,
       author = {{{\L}okas}, Ewa L.},
        title = "{The effect of warm gas on the buckling instability in galactic bars}",
      journal = {\aap},
         year = 2020,
        month = feb,
       volume = {634},
          eid = {A122},
        pages = {A122},
          doi = {10.1051/0004-6361/201937165},
archivePrefix = {arXiv},
       eprint = {2001.07041},
 primaryClass = {astro-ph.GA},
       adsurl = {https://ui.adsabs.harvard.edu/abs/2020A&A...634A.122L}
}

@ARTICLE{Lokas2026,
       author = {{{\L}okas}, Ewa L.},
        title = "{Diverse lifestyles of bar-like galaxies and their coevolution with the brightest galaxy in the most massive cluster of TNG50}",
      journal = {\aap},
         year = 2026,
        month = apr,
       volume = {708},
          eid = {A341},
        pages = {A341},
          doi = {10.1051/0004-6361/202658996},
archivePrefix = {arXiv},
       eprint = {2604.01758},
 primaryClass = {astro-ph.GA},
       adsurl = {https://ui.adsabs.harvard.edu/abs/2026A&A...708A.341L}
}

@ARTICLE{PaulaLopez2025,
       author = {{L{\'o}pez}, Paula D. and {Fragkoudi}, Francesca and {Cora}, Sof{\'\i}a A. and {Scannapieco}, Cecilia and {Pakmor}, R{\"u}diger and {Grand}, Robert J.~J. and {G{\'o}mez}, Facundo and {Marinacci}, Federico},
        title = "{Formation and evolution of boxy/peanut bulges in the Auriga cosmological simulations}",
      journal = {\mnras},
         year = 2025,
        month = jul,
       volume = {540},
       number = {3},
        pages = {2031-2048},
          doi = {10.1093/mnras/staf818},
archivePrefix = {arXiv},
       eprint = {2505.13590},
 primaryClass = {astro-ph.GA},
       adsurl = {https://ui.adsabs.harvard.edu/abs/2025MNRAS.540.2031L}
}

@ARTICLE{Lutticke2000I,
       author = {{L{\"u}tticke}, R. and {Dettmar}, R.-J. and {Pohlen}, M.},
        title = "{Box- and peanut-shaped bulges. I. Statistics}",
      journal = {\aaps},
         year = 2000,
        month = sep,
       volume = {145},
        pages = {405-414},
          doi = {10.1051/aas:2000354},
archivePrefix = {arXiv},
       eprint = {astro-ph/0006359},
 primaryClass = {astro-ph},
       adsurl = {https://ui.adsabs.harvard.edu/abs/2000A&AS..145..405L}
}

@ARTICLE{Lutticke2000II,
       author = {{L{\"u}tticke}, R. and {Dettmar}, R.-J. and {Pohlen}, M.},
        title = "{Box- and peanut-shaped bulges. II. NIR observations}",
      journal = {\aap},
         year = 2000,
        month = oct,
       volume = {362},
        pages = {435-446},
          doi = {10.48550/arXiv.astro-ph/0008478},
archivePrefix = {arXiv},
       eprint = {astro-ph/0008478},
 primaryClass = {astro-ph},
       adsurl = {https://ui.adsabs.harvard.edu/abs/2000A&A...362..435L}
}

@ARTICLE{LyndenBell1972,
       author = {{Lynden-Bell}, D. and {Kalnajs}, A.~J.},
        title = "{On the generating mechanism of spiral structure}",
      journal = {\mnras},
         year = 1972,
        month = jan,
       volume = {157},
        pages = {1},
          doi = {10.1093/mnras/157.1.1},
       adsurl = {https://ui.adsabs.harvard.edu/abs/1972MNRAS.157....1L}
}

@ARTICLE{Masters2011,
       author = {{Masters}, Karen L. and {Nichol}, Robert C. and {Hoyle}, Ben and {Lintott}, Chris and {Bamford}, Steven P. and {Edmondson}, Edward M. and {Fortson}, Lucy and {Keel}, William C. and {Schawinski}, Kevin and {Smith}, Arfon M. and {Thomas}, Daniel},
        title = "{Galaxy Zoo: bars in disc galaxies}",
      journal = {\mnras},
         year = 2011,
        month = mar,
       volume = {411},
       number = {3},
        pages = {2026-2034},
          doi = {10.1111/j.1365-2966.2010.17834.x},
archivePrefix = {arXiv},
       eprint = {1003.0449},
 primaryClass = {astro-ph.CO},
       adsurl = {https://ui.adsabs.harvard.edu/abs/2011MNRAS.411.2026M}
}

@ARTICLE{MartinezValpuesta2006,
       author = {{Martinez-Valpuesta}, Inma and {Shlosman}, Isaac and {Heller}, Clayton},
        title = "{Evolution of Stellar Bars in Live Axisymmetric Halos: Recurrent Buckling and Secular Growth}",
      journal = {\apj},
         year = 2006,
        month = jan,
       volume = {637},
       number = {1},
        pages = {214-226},
          doi = {10.1086/498338},
archivePrefix = {arXiv},
       eprint = {astro-ph/0507219},
 primaryClass = {astro-ph},
       adsurl = {https://ui.adsabs.harvard.edu/abs/2006ApJ...637..214M}
}

@ARTICLE{mcmillan2016mass,
author = {{McMillan}, Paul J.},
title = "{The mass distribution and gravitational potential of the Milky Way}",
journal = {\mnras},
year = 2017,
month = feb,
volume = {465},
number = {1},
pages = {76-94},
doi = {10.1093/mnras/stw2759},
archivePrefix = {arXiv},
eprint = {1608.00971},
primaryClass = {astro-ph.GA},
adsurl = {https://ui.adsabs.harvard.edu/abs/2017MNRAS.465...76M}
}

@ARTICLE{McClure2025,
       author = {{McClure}, Rachel Lee and {Beane}, Angus and {D'Onghia}, Elena and {Filion}, Carrie and {Daniel}, Kathryne J.},
        title = "{The impact of classical bulges on stellar bars and boxy-peanut-X features in disc galaxies}",
      journal = {\mnras},
         year = 2025,
        month = feb,
       volume = {537},
       number = {2},
        pages = {1475-1488},
          doi = {10.1093/mnras/staf107},
archivePrefix = {arXiv},
       eprint = {2410.08277},
 primaryClass = {astro-ph.GA},
       adsurl = {https://ui.adsabs.harvard.edu/abs/2025MNRAS.537.1475M}
}

@ARTICLE{McWilliam2010,
       author = {{McWilliam}, Andrew and {Zoccali}, Manuela},
        title = "{Two Red Clumps and the X-shaped Milky Way Bulge}",
      journal = {\apj},
         year = 2010,
        month = dec,
       volume = {724},
       number = {2},
        pages = {1491-1502},
          doi = {10.1088/0004-637X/724/2/1491},
archivePrefix = {arXiv},
       eprint = {1008.0519},
 primaryClass = {astro-ph.GA},
       adsurl = {https://ui.adsabs.harvard.edu/abs/2010ApJ...724.1491M}
}

@ARTICLE{MendezAbreu2008II,
       author = {{M{\'e}ndez-Abreu}, J. and {Corsini}, E.~M. and {Debattista}, Victor P. and {De Rijcke}, S. and {Aguerri}, J.~A.~L. and {Pizzella}, A.},
        title = "{Confirmation of a Kinematic Diagnostic for Face-On Box/Peanut-shaped Bulges}",
      journal = {\apjl},
         year = 2008,
        month = jun,
       volume = {679},
       number = {2},
        pages = {L73},
          doi = {10.1086/589541},
archivePrefix = {arXiv},
       eprint = {0804.3589},
 primaryClass = {astro-ph},
       adsurl = {https://ui.adsabs.harvard.edu/abs/2008ApJ...679L..73M}
}

@ARTICLE{MendezAbreu2014,
       author = {{M{\'e}ndez-Abreu}, J. and {Debattista}, V.~P. and {Corsini}, E.~M. and {Aguerri}, J.~A.~L.},
        title = "{Secular- and merger-built bulges in barred galaxies}",
      journal = {\aap},
         year = 2014,
        month = dec,
       volume = {572},
          eid = {A25},
        pages = {A25},
          doi = {10.1051/0004-6361/201423955},
archivePrefix = {arXiv},
       eprint = {1409.2876},
 primaryClass = {astro-ph.GA},
       adsurl = {https://ui.adsabs.harvard.edu/abs/2014A&A...572A..25M}
}

@ARTICLE{MendezAbreu2019,
       author = {{M{\'e}ndez-Abreu}, J. and {de Lorenzo-C{\'a}ceres}, A. and {Gadotti}, D.~A. and {Fragkoudi}, F. and {van de Ven}, G. and {Falc{\'o}n-Barroso}, J. and {Leaman}, R. and {P{\'e}rez}, I. and {Querejeta}, M. and {S{\'a}nchez-Blazquez}, P. and {Seidel}, M.},
        title = "{Inner bars also buckle. The MUSE TIMER view of the double-barred galaxy NGC 1291}",
      journal = {\mnras},
         year = 2019,
        month = jan,
       volume = {482},
       number = {1},
        pages = {L118-L122},
          doi = {10.1093/mnrasl/sly196},
archivePrefix = {arXiv},
       eprint = {1811.03855},
 primaryClass = {astro-ph.GA},
       adsurl = {https://ui.adsabs.harvard.edu/abs/2019MNRAS.482L.118M}
}

@ARTICLE{MenendezDelmestre2007,
       author = {{Men{\'e}ndez-Delmestre}, Kar{\'\i}n and {Sheth}, Kartik and {Schinnerer}, Eva and {Jarrett}, Thomas H. and {Scoville}, Nick Z.},
        title = "{A Near-Infrared Study of 2MASS Bars in Local Galaxies: An Anchor for High-Redshift Studies}",
      journal = {\apj},
         year = 2007,
        month = mar,
       volume = {657},
       number = {2},
        pages = {790-804},
          doi = {10.1086/511025},
archivePrefix = {arXiv},
       eprint = {astro-ph/0611540},
 primaryClass = {astro-ph},
       adsurl = {https://ui.adsabs.harvard.edu/abs/2007ApJ...657..790M}
}

@ARTICLE{Merrifield1995,
       author = {{Merrifield}, Michael R. and {Kuijken}, Konrad},
        title = "{The pattern speed of the bar in NGC 936}",
      journal = {\mnras},
         year = 1995,
        month = jun,
       volume = {274},
       number = {3},
        pages = {933-938},
          doi = {10.1093/mnras/274.3.933},
archivePrefix = {arXiv},
       eprint = {astro-ph/9502069},
 primaryClass = {astro-ph},
       adsurl = {https://ui.adsabs.harvard.edu/abs/1995MNRAS.274..933M}
}

@ARTICLE{Merritt1994,
       author = {{Merritt}, David and {Sellwood}, J.~A.},
        title = "{Bending Instabilities in Stellar Systems}",
      journal = {\apj},
         year = 1994,
        month = apr,
       volume = {425},
        pages = {551},
          doi = {10.1086/174005},
       adsurl = {https://ui.adsabs.harvard.edu/abs/1994ApJ...425..551M}
}

@ARTICLE{Michel-Dansac2006,
       author = {{Michel-Dansac}, L. and {Wozniak}, H.},
        title = "{The length of stellar bars in SB galaxies and N-body simulations}",
      journal = {\aap},
         year = 2006,
        month = jun,
       volume = {452},
       number = {1},
        pages = {97-112},
          doi = {10.1051/0004-6361:20041038},
archivePrefix = {arXiv},
       eprint = {astro-ph/0503406},
 primaryClass = {astro-ph},
       adsurl = {https://ui.adsabs.harvard.edu/abs/2006A&A...452...97M}
}

@ARTICLE{Merrow2026,
       author = {{Merrow}, Alex and {Fragkoudi}, Francesca and {Grand}, Rob J.~J. and {Martig}, Marie},
        title = "{What drives bar rotation? The effect of internal properties and galaxy interactions on bar pattern speeds}",
      journal = {arXiv e-prints},
         year = 2026,
        month = jan,
          eid = {arXiv:2601.20941},
        pages = {arXiv:2601.20941},
          doi = {10.48550/arXiv.2601.20941},
archivePrefix = {arXiv},
       eprint = {2601.20941},
 primaryClass = {astro-ph.GA},
       adsurl = {https://ui.adsabs.harvard.edu/abs/2026arXiv260120941M}
}

@ARTICLE{Ness2012,
       author = {{Ness}, M. and {Freeman}, K. and {Athanassoula}, E. and {Wylie-De-Boer}, E. and {Bland-Hawthorn}, J. and {Lewis}, G.~F. and {Yong}, D. and {Asplund}, M. and {Lane}, R.~R. and {Kiss}, L.~L. and {Ibata}, R.},
        title = "{The Origin of the Split Red Clump in the Galactic Bulge of the Milky Way}",
      journal = {\apj},
         year = 2012,
        month = sep,
       volume = {756},
       number = {1},
          eid = {22},
        pages = {22},
          doi = {10.1088/0004-637X/756/1/22},
archivePrefix = {arXiv},
       eprint = {1207.0888},
 primaryClass = {astro-ph.GA},
       adsurl = {https://ui.adsabs.harvard.edu/abs/2012ApJ...756...22N}
}

@ARTICLE{Ostriker1973,
       author = {{Ostriker}, J.~P. and {Peebles}, P.~J.~E.},
        title = "{A Numerical Study of the Stability of Flattened Galaxies: or, can Cold Galaxies Survive?}",
      journal = {\apj},
         year = 1973,
        month = dec,
       volume = {186},
        pages = {467-480},
          doi = {10.1086/152513},
       adsurl = {https://ui.adsabs.harvard.edu/abs/1973ApJ...186..467O}
}

@ARTICLE{Fragkoudi2015,
       author = {{Fragkoudi}, F. and {Athanassoula}, E. and {Bosma}, A. and {Iannuzzi}, F.},
        title = "{The effects of Boxy/Peanut bulges on galaxy models}",
      journal = {\mnras},
         year = 2015,
        month = jun,
       volume = {450},
       number = {1},
        pages = {229-245},
          doi = {10.1093/mnras/stv537},
archivePrefix = {arXiv},
       eprint = {1503.03068},
 primaryClass = {astro-ph.GA},
       adsurl = {https://ui.adsabs.harvard.edu/abs/2015MNRAS.450..229F}
}

@ARTICLE{Fragkoudi2017,
       author = {{Fragkoudi}, F. and {Di Matteo}, P. and {Haywood}, M. and {G{\'o}mez}, A. and {Combes}, F. and {Katz}, D. and {Semelin}, B.},
        title = "{Bars and boxy/peanut bulges in thin and thick discs. I. Morphology and line-of-sight velocities of a fiducial model}",
      journal = {\aap},
         year = 2017,
        month = oct,
       volume = {606},
          eid = {A47},
        pages = {A47},
          doi = {10.1051/0004-6361/201630244},
archivePrefix = {arXiv},
       eprint = {1704.00734},
 primaryClass = {astro-ph.GA},
       adsurl = {https://ui.adsabs.harvard.edu/abs/2017A&A...606A..47F}
}

@ARTICLE{Fragkoudi2021,
       author = {{Fragkoudi}, F. and {Grand}, R.~J.~J. and {Pakmor}, R. and {Springel}, V. and {White}, S.~D.~M. and {Marinacci}, F. and {Gomez}, F.~A. and {Navarro}, J.~F.},
        title = "{Revisiting the tension between fast bars and the {\ensuremath{\Lambda}}CDM paradigm}",
      journal = {\aap},
         year = 2021,
        month = jun,
       volume = {650},
          eid = {L16},
        pages = {L16},
          doi = {10.1051/0004-6361/202140320},
archivePrefix = {arXiv},
       eprint = {2011.13942},
 primaryClass = {astro-ph.GA},
       adsurl = {https://ui.adsabs.harvard.edu/abs/2021A&A...650L..16F}
}

@ARTICLE{Fragkoudi2025,
       author = {{Fragkoudi}, Francesca and {Grand}, Robert J.~J. and {Pakmor}, R{\"u}diger and {G{\'o}mez}, Facundo and {Marinacci}, Federico and {Springel}, Volker},
        title = "{Bar formation and evolution in the cosmological context: inputs from the Auriga simulations}",
      journal = {\mnras},
         year = 2025,
        month = apr,
       volume = {538},
       number = {3},
        pages = {1587-1608},
          doi = {10.1093/mnras/staf389},
archivePrefix = {arXiv},
       eprint = {2406.09453},
 primaryClass = {astro-ph.GA},
       adsurl = {https://ui.adsabs.harvard.edu/abs/2025MNRAS.538.1587F}
}

@ARTICLE{Gadotti2015,
       author = {{Gadotti}, Dimitri A. and {Seidel}, Marja K. and {S{\'a}nchez-Bl{\'a}zquez}, Patricia and {Falc{\'o}n-Barroso}, Jesus and {Husemann}, Bernd and {Coelho}, Paula and {P{\'e}rez}, Isabel},
        title = "{MUSE tells the story of NGC 4371: The dawning of secular evolution}",
      journal = {\aap},
         year = 2015,
        month = dec,
       volume = {584},
          eid = {A90},
        pages = {A90},
          doi = {10.1051/0004-6361/201526677},
archivePrefix = {arXiv},
       eprint = {1509.00032},
 primaryClass = {astro-ph.GA},
       adsurl = {https://ui.adsabs.harvard.edu/abs/2015A&A...584A..90G}
}

@ARTICLE{Ghosh2021,
       author = {{Ghosh}, Soumavo and {Saha}, Kanak and {Di Matteo}, Paola and {Combes}, Francoise},
        title = "{Fate of stellar bars in minor merger of galaxies}",
      journal = {\mnras},
         year = 2021,
        month = apr,
       volume = {502},
       number = {2},
        pages = {3085-3100},
          doi = {10.1093/mnras/stab238},
archivePrefix = {arXiv},
       eprint = {2008.04942},
 primaryClass = {astro-ph.GA},
       adsurl = {https://ui.adsabs.harvard.edu/abs/2021MNRAS.502.3085G}
}

@ARTICLE{Ghosh2023,
       author = {{Ghosh}, Soumavo and {Fragkoudi}, Francesca and {Di Matteo}, Paola and {Saha}, Kanak},
        title = "{Bars and boxy/peanut bulges in thin and thick discs. II. Can bars form in hot thick discs?}",
      journal = {\aap},
         year = 2023,
        month = jun,
       volume = {674},
          eid = {A128},
        pages = {A128},
          doi = {10.1051/0004-6361/202245275},
archivePrefix = {arXiv},
       eprint = {2210.14244},
 primaryClass = {astro-ph.GA},
       adsurl = {https://ui.adsabs.harvard.edu/abs/2023A&A...674A.128G}
}

@ARTICLE{Ghosh2024,
       author = {{Ghosh}, Soumavo and {Fragkoudi}, Francesca and {Di Matteo}, Paola and {Saha}, Kanak},
        title = "{Bars and boxy/peanut bulges in thin and thick discs. III. Boxy/peanut bulge formation and evolution in the presence of thick discs}",
      journal = {\aap},
         year = 2024,
        month = mar,
       volume = {683},
          eid = {A196},
        pages = {A196},
          doi = {10.1051/0004-6361/202347831},
archivePrefix = {arXiv},
       eprint = {2308.16239},
 primaryClass = {astro-ph.GA},
       adsurl = {https://ui.adsabs.harvard.edu/abs/2024A&A...683A.196G}
}

@ARTICLE{Guo2019,
       author = {{Guo}, Rui and {Mao}, Shude and {Athanassoula}, E. and {Li}, Hongyu and {Ge}, Junqiang and {Long}, R.~J. and {Merrifield}, Michael and {Masters}, Karen},
        title = "{SDSS-IV MaNGA: pattern speeds of barred galaxies}",
      journal = {\mnras},
         year = 2019,
        month = jan,
       volume = {482},
       number = {2},
        pages = {1733-1756},
          doi = {10.1093/mnras/sty2715},
archivePrefix = {arXiv},
       eprint = {1810.03257},
 primaryClass = {astro-ph.GA},
       adsurl = {https://ui.adsabs.harvard.edu/abs/2019MNRAS.482.1733G}
}

@ARTICLE{Hohl1971,
       author = {{Hohl}, Frank},
        title = "{Numerical Experiments with a Disk of Stars}",
      journal = {\apj},
         year = 1971,
        month = sep,
       volume = {168},
        pages = {343},
          doi = {10.1086/151091},
       adsurl = {https://ui.adsabs.harvard.edu/abs/1971ApJ...168..343H}
}

@ARTICLE{iles2022differences,
author = {{Iles}, Elizabeth J. and {Pettitt}, Alex R. and {Okamoto}, Takashi},
title = "{Differences in star formation activity between tidally triggered and isolated bars: a case study of NGC 4303 and NGC 3627}",
journal = {\mnras},
year = 2022,
month = mar,
volume = {510},
number = {3},
pages = {3899-3916},
doi = {10.1093/mnras/stab3330},
archivePrefix = {arXiv},
eprint = {2111.09545},
primaryClass = {astro-ph.GA},
adsurl = {https://ui.adsabs.harvard.edu/abs/2022MNRAS.510.3899I}
}

@ARTICLE{kataria2018study,
author = {{Kataria}, Sandeep Kumar and {Das}, Mousumi},
title = "{A study of the effect of bulges on bar formation in disc galaxies}",
journal = {\mnras},
year = 2018,
month = apr,
volume = {475},
number = {2},
pages = {1653-1664},
doi = {10.1093/mnras/stx3279},
adsurl = {https://ui.adsabs.harvard.edu/abs/2018MNRAS.475.1653K}
}

@ARTICLE{Knapen2000,
       author = {{Knapen}, Johan H. and {Shlosman}, Isaac and {Peletier}, Reynier F.},
        title = "{A Subarcsecond Resolution Near-Infrared Study of Seyfert and ``Normal'' Galaxies. II. Morphology}",
      journal = {\apj},
         year = 2000,
        month = jan,
       volume = {529},
       number = {1},
        pages = {93-100},
          doi = {10.1086/308266},
archivePrefix = {arXiv},
       eprint = {astro-ph/9907379},
 primaryClass = {astro-ph},
       adsurl = {https://ui.adsabs.harvard.edu/abs/2000ApJ...529...93K}
}

@ARTICLE{kormendy2004,
       author = {{Kormendy}, John and {Kennicutt}, Jr., Robert C.},
        title = "{Secular Evolution and the Formation of Pseudobulges in Disk Galaxies}",
      journal = {\araa},
         year = 2004,
        month = sep,
       volume = {42},
       number = {1},
        pages = {603-683},
          doi = {10.1146/annurev.astro.42.053102.134024},
archivePrefix = {arXiv},
       eprint = {astro-ph/0407343},
 primaryClass = {astro-ph},
       adsurl = {https://ui.adsabs.harvard.edu/abs/2004ARA&A..42..603K}
}

@ARTICLE{Kruk2019,
       author = {{Kruk}, Sandor J. and {Erwin}, Peter and {Debattista}, Victor P. and {Lintott}, Chris},
        title = "{Revealing the cosmic evolution of boxy/peanut-shaped bulges from HST COSMOS and SDSS}",
      journal = {\mnras},
         year = 2019,
        month = dec,
       volume = {490},
       number = {4},
        pages = {4721-4739},
          doi = {10.1093/mnras/stz2877},
archivePrefix = {arXiv},
       eprint = {1910.04768},
 primaryClass = {astro-ph.GA},
       adsurl = {https://ui.adsabs.harvard.edu/abs/2019MNRAS.490.4721K}
}

@ARTICLE{pakmor2016improving,
author = {{Pakmor}, R{\"u}diger and {Springel}, Volker and {Bauer}, Andreas and {Mocz}, Philip and {Munoz}, Diego J. and {Ohlmann}, Sebastian T. and {Schaal}, Kevin and {Zhu}, Chenchong},
title = "{Improving the convergence properties of the moving-mesh code AREPO}",
journal = {\mnras},
year = 2016,
month = jan,
volume = {455},
number = {1},
pages = {1134-1143},
doi = {10.1093/mnras/stv2380},
archivePrefix = {arXiv},
eprint = {1503.00562},
primaryClass = {astro-ph.GA},
adsurl = {https://ui.adsabs.harvard.edu/abs/2016MNRAS.455.1134P}
}

@ARTICLE{Petersen2021,
       author = {{Petersen}, Michael S. and {Weinberg}, Martin D. and {Katz}, Neal},
        title = "{Using commensurabilities and orbit structure to understand barred galaxy evolution}",
      journal = {\mnras},
         year = 2021,
        month = jan,
       volume = {500},
       number = {1},
        pages = {838-858},
          doi = {10.1093/mnras/staa3202},
archivePrefix = {arXiv},
       eprint = {1902.05081},
 primaryClass = {astro-ph.GA},
       adsurl = {https://ui.adsabs.harvard.edu/abs/2021MNRAS.500..838P}
}

@ARTICLE{pettitt2015morphology,
author = {{Pettitt}, Alex R. and {Dobbs}, Clare L. and {Acreman}, David M. and {Bate}, Matthew R.},
title = "{The morphology of the Milky Way - II. Reconstructing CO maps from disc galaxies with live stellar distributions}",
journal = {\mnras},
year = 2015,
month = jun,
volume = {449},
number = {4},
pages = {3911-3926},
doi = {10.1093/mnras/stv600},
archivePrefix = {arXiv},
eprint = {1503.05563},
primaryClass = {astro-ph.GA},
adsurl = {https://ui.adsabs.harvard.edu/abs/2015MNRAS.449.3911P}
}

@ARTICLE{pfenniger1991structure,
author = {{Pfenniger}, D. and {Friedli}, D.},
title = "{Structure and dynamics of 3D N-body barred galaxies.}",
journal = {\aap},
year = 1991,
month = dec,
volume = {252},
pages = {75-93},
adsurl = {https://ui.adsabs.harvard.edu/abs/1991A&A...252...75P}
}

@ARTICLE{pfenniger1993,
       author = {{Pfenniger}, D. and {Friedli}, D.},
        title = "{Computational issues connected with 3D N-body simulations}",
      journal = {\aap},
         year = 1993,
        month = mar,
       volume = {270},
       number = {1-2},
        pages = {561-572},
       adsurl = {https://ui.adsabs.harvard.edu/abs/1993A&A...270..561P}
}

@ARTICLE{queiroz2021milky,
author = {{Queiroz}, A.~B.~A. and {Chiappini}, C. and {Perez-Villegas}, A. and {Khalatyan}, A. and {Anders}, F. and {Barbuy}, B. and {Santiago}, B.~X. and {Steinmetz}, M. and {Cunha}, K. and {Schultheis}, M. and {Majewski}, S.~R. and {Minchev}, I. and {Minniti}, D. and {Beaton}, R.~L. and {Cohen}, R.~E. and {da Costa}, L.~N. and {Fern{\'a}ndez-Trincado}, J.~G. and {Garcia-Hern{\'a}ndez}, D.~A. and {Geisler}, D. and {Hasselquist}, S. and {Lane}, R.~R. and {Nitschelm}, C. and {Rojas-Arriagada}, A. and {Roman-Lopes}, A. and {Smith}, V. and {Zasowski}, G.},
title = "{The Milky Way bar and bulge revealed by APOGEE and Gaia EDR3}",
journal = {\aap},
year = 2021,
month = dec,
volume = {656},
eid = {A156},
pages = {A156},
doi = {10.1051/0004-6361/202039030},
archivePrefix = {arXiv},
eprint = {2007.12915},
primaryClass = {astro-ph.GA},
adsurl = {https://ui.adsabs.harvard.edu/abs/2021A&A...656A.156Q}
}

@ARTICLE{Quillen2002,
       author = {{Quillen}, A.~C.},
        title = "{Growth of a Peanut-shaped Bulge via Resonant Trapping of Stellar Orbits in the Vertical Inner Lindblad Resonances}",
      journal = {\aj},
         year = 2002,
        month = aug,
       volume = {124},
       number = {2},
        pages = {722-732},
          doi = {10.1086/341753},
archivePrefix = {arXiv},
       eprint = {astro-ph/0203170},
 primaryClass = {astro-ph},
       adsurl = {https://ui.adsabs.harvard.edu/abs/2002AJ....124..722Q}
}

@ARTICLE{Raha1991,
       author = {{Raha}, N. and {Sellwood}, J.~A. and {James}, R.~A. and {Kahn}, F.~D.},
        title = "{A dynamical instability of bars in disk galaxies}",
      journal = {\nat},
         year = 1991,
        month = aug,
       volume = {352},
       number = {6334},
        pages = {411-412},
          doi = {10.1038/352411a0},
       adsurl = {https://ui.adsabs.harvard.edu/abs/1991Natur.352..411R}
}

@ARTICLE{rosas-guevara2022,
       author = {{Rosas-Guevara}, Yetli and {Bonoli}, Silvia and {Dotti}, Massimo and {Izquierdo-Villalba}, David and {Lupi}, Alessandro and {Zana}, Tommaso and {Bonetti}, Matteo and {Nelson}, Dylan and {Springel}, Volker and {Hernquist}, Lars and {Vogelsberger}, Mark},
        title = "{The evolution of the barred galaxy population in the TNG50 simulation}",
      journal = {\mnras},
         year = 2022,
        month = jun,
       volume = {512},
       number = {4},
        pages = {5339-5357},
          doi = {10.1093/mnras/stac816},
archivePrefix = {arXiv},
       eprint = {2110.04537},
 primaryClass = {astro-ph.GA},
       adsurl = {https://ui.adsabs.harvard.edu/abs/2022MNRAS.512.5339R}
}

@ARTICLE{Sanders2019,
       author = {{Sanders}, Jason L. and {Smith}, Leigh and {Evans}, N. Wyn},
        title = "{The pattern speed of the Milky Way bar from transverse velocities}",
      journal = {\mnras},
         year = 2019,
        month = oct,
       volume = {488},
       number = {4},
        pages = {4552-4564},
          doi = {10.1093/mnras/stz1827},
archivePrefix = {arXiv},
       eprint = {1903.02009},
 primaryClass = {astro-ph.GA},
       adsurl = {https://ui.adsabs.harvard.edu/abs/2019MNRAS.488.4552S}
}

@ARTICLE{Sellwood1984,
       author = {{Sellwood}, J.~A. and {Carlberg}, R.~G.},
        title = "{Spiral instabilities provoked by accretion and star formation}",
      journal = {\apj},
         year = 1984,
        month = jul,
       volume = {282},
        pages = {61-74},
          doi = {10.1086/162176},
       adsurl = {https://ui.adsabs.harvard.edu/abs/1984ApJ...282...61S}
}

@ARTICLE{Sellwood1993,
       author = {{Sellwood}, J.~A. and {Wilkinson}, A.},
        title = "{Dynamics of barred galaxies}",
      journal = {Reports on Progress in Physics},
         year = 1993,
        month = feb,
       volume = {56},
       number = {2},
        pages = {173-256},
          doi = {10.1088/0034-4885/56/2/001},
archivePrefix = {arXiv},
       eprint = {astro-ph/0608665},
 primaryClass = {astro-ph},
       adsurl = {https://ui.adsabs.harvard.edu/abs/1993RPPh...56..173S}
}

@ARTICLE{Sellwood2014barslowdown,
       author = {{Sellwood}, J.~A. and {Debattista}, Victor P.},
        title = "{Re-interpretation of ``Bar slowdown and the distribution of dark matter in barred galaxies'' by Athanassoula}",
      journal = {arXiv e-prints},
         year = 2014,
        month = oct,
          eid = {arXiv:1410.0834},
        pages = {arXiv:1410.0834},
          doi = {10.48550/arXiv.1410.0834},
archivePrefix = {arXiv},
       eprint = {1410.0834},
 primaryClass = {astro-ph.GA},
       adsurl = {https://ui.adsabs.harvard.edu/abs/2014arXiv1410.0834S}
}

@ARTICLE{Sellwood2014,
       author = {{Sellwood}, J.~A. and {Carlberg}, R.~G.},
        title = "{Transient Spirals as Superposed Instabilities}",
      journal = {\apj},
         year = 2014,
        month = apr,
       volume = {785},
       number = {2},
          eid = {137},
        pages = {137},
          doi = {10.1088/0004-637X/785/2/137},
archivePrefix = {arXiv},
       eprint = {1403.1135},
 primaryClass = {astro-ph.GA},
       adsurl = {https://ui.adsabs.harvard.edu/abs/2014ApJ...785..137S}
}

@ARTICLE{Sellwood.Gerhard.2020,
       author = {{Sellwood}, J.~A. and {Gerhard}, Ortwin},
        title = "{Three mechanisms for bar thickening}",
      journal = {\mnras},
         year = 2020,
        month = jul,
       volume = {495},
       number = {3},
        pages = {3175-3191},
          doi = {10.1093/mnras/staa1336},
archivePrefix = {arXiv},
       eprint = {2005.05184},
 primaryClass = {astro-ph.GA},
       adsurl = {https://ui.adsabs.harvard.edu/abs/2020MNRAS.495.3175S}
}

@ARTICLE{Shen2010,
       author = {{Shen}, Juntai and {Rich}, R. Michael and {Kormendy}, John and {Howard}, Christian D. and {De Propris}, Roberto and {Kunder}, Andrea},
        title = "{Our Milky Way as a Pure-disk Galaxy{\textemdash}A Challenge for Galaxy Formation}",
      journal = {\apjl},
         year = 2010,
        month = sep,
       volume = {720},
       number = {1},
        pages = {L72-L76},
          doi = {10.1088/2041-8205/720/1/L72},
archivePrefix = {arXiv},
       eprint = {1005.0385},
 primaryClass = {astro-ph.CO},
       adsurl = {https://ui.adsabs.harvard.edu/abs/2010ApJ...720L..72S}
}

@ARTICLE{Smirnov2018,
       author = {{Smirnov}, Anton A. and {Sotnikova}, Natalia Ya},
        title = "{What determines the flatness of X-shaped structures in edge-on galaxies?}",
      journal = {\mnras},
         year = 2018,
        month = dec,
       volume = {481},
       number = {3},
        pages = {4058-4076},
          doi = {10.1093/mnras/sty2423},
archivePrefix = {arXiv},
       eprint = {1809.06167},
 primaryClass = {astro-ph.GA},
       adsurl = {https://ui.adsabs.harvard.edu/abs/2018MNRAS.481.4058S}
}

@ARTICLE{Smirnov2021,
       author = {{Smirnov}, Anton A. and {Tikhonenko}, Iliya S. and {Sotnikova}, Natalia Ya},
        title = "{Face-on structure of barlenses and boxy bars: an insight from spectral dynamics}",
      journal = {\mnras},
         year = 2021,
        month = apr,
       volume = {502},
       number = {4},
        pages = {4689-4707},
          doi = {10.1093/mnras/stab327},
archivePrefix = {arXiv},
       eprint = {2007.09090},
 primaryClass = {astro-ph.GA},
       adsurl = {https://ui.adsabs.harvard.edu/abs/2021MNRAS.502.4689S}
}

@ARTICLE{springel2005modelling,
author = {{Springel}, Volker and {Di Matteo}, Tiziana and {Hernquist}, Lars},
title = "{Modelling feedback from stars and black holes in galaxy mergers}",
journal = {\mnras},
year = 2005,
month = aug,
volume = {361},
number = {3},
pages = {776-794},
doi = {10.1111/j.1365-2966.2005.09238.x},
archivePrefix = {arXiv},
eprint = {astro-ph/0411108},
primaryClass = {astro-ph},
adsurl = {https://ui.adsabs.harvard.edu/abs/2005MNRAS.361..776S}
}

@ARTICLE{springel2010pur,
author = {{Springel}, Volker},
title = "{E pur si muove: Galilean-invariant cosmological hydrodynamical simulations on a moving mesh}",
journal = {\mnras},
year = 2010,
month = jan,
volume = {401},
number = {2},
pages = {791-851},
doi = {10.1111/j.1365-2966.2009.15715.x},
archivePrefix = {arXiv},
eprint = {0901.4107},
primaryClass = {astro-ph.CO},
adsurl = {https://ui.adsabs.harvard.edu/abs/2010MNRAS.401..791S}
}

@ARTICLE{stanek1994,
author = {{Stanek}, K.~Z. and {Mateo}, M. and {Udalski}, A. and {Szymanski}, M. and {Kaluzny}, J. and {Kubiak}, M. and {Krzeminski}, W.},
title = "{Evidence for the Galactic Bar from the Two Color Photometry of the Bulge Red Clump Stars}",
journal = {arXiv e-prints},
year = 1994,
month = oct,
eid = {astro-ph/9410044},
pages = {astro-ph/9410044},
doi = {10.48550/arXiv.astro-ph/9410044},
archivePrefix = {arXiv},
eprint = {astro-ph/9410044},
primaryClass = {astro-ph},
adsurl = {https://ui.adsabs.harvard.edu/abs/1994astro.ph.10044S}
}

@ARTICLE{Tremaine1984,
       author = {{Tremaine}, S. and {Weinberg}, M.~D.},
        title = "{A kinematic method for measuring the pattern speed of barred galaxies.}",
      journal = {\apjl},
         year = 1984,
        month = jul,
       volume = {282},
        pages = {L5-L7},
          doi = {10.1086/184292},
       adsurl = {https://ui.adsabs.harvard.edu/abs/1984ApJ...282L...5T}
}

@ARTICLE{wegg2015structure,
author = {{Wegg}, Christopher and {Gerhard}, Ortwin and {Portail}, Matthieu},
title = "{The structure of the Milky Way's bar outside the bulge}",
journal = {\mnras},
year = 2015,
month = jul,
volume = {450},
number = {4},
pages = {4050-4069},
doi = {10.1093/mnras/stv745},
archivePrefix = {arXiv},
eprint = {1504.01401},
primaryClass = {astro-ph.GA},
adsurl = {https://ui.adsabs.harvard.edu/abs/2015MNRAS.450.4050W}
}

@ARTICLE{Wheeler2023,
       author = {{Wheeler}, Vance and {Valluri}, Monica and {Beraldo e Silva}, Leandro and {Dattathri}, Shashank and {Debattista}, Victor P.},
        title = "{Early-growing Supermassive Black Holes Strengthen Bars and Boxy/Peanut Bulges}",
      journal = {\apj},
         year = 2023,
        month = dec,
       volume = {958},
       number = {2},
          eid = {119},
        pages = {119},
          doi = {10.3847/1538-4357/ace962},
archivePrefix = {arXiv},
       eprint = {2306.07424},
 primaryClass = {astro-ph.GA},
       adsurl = {https://ui.adsabs.harvard.edu/abs/2023ApJ...958..119W}
}



\appendix

\section{Rotation curves}
\label{app:rotation_curves}

\begin{figure*}[htp]
    \centering
    \includegraphics[width=\textwidth]{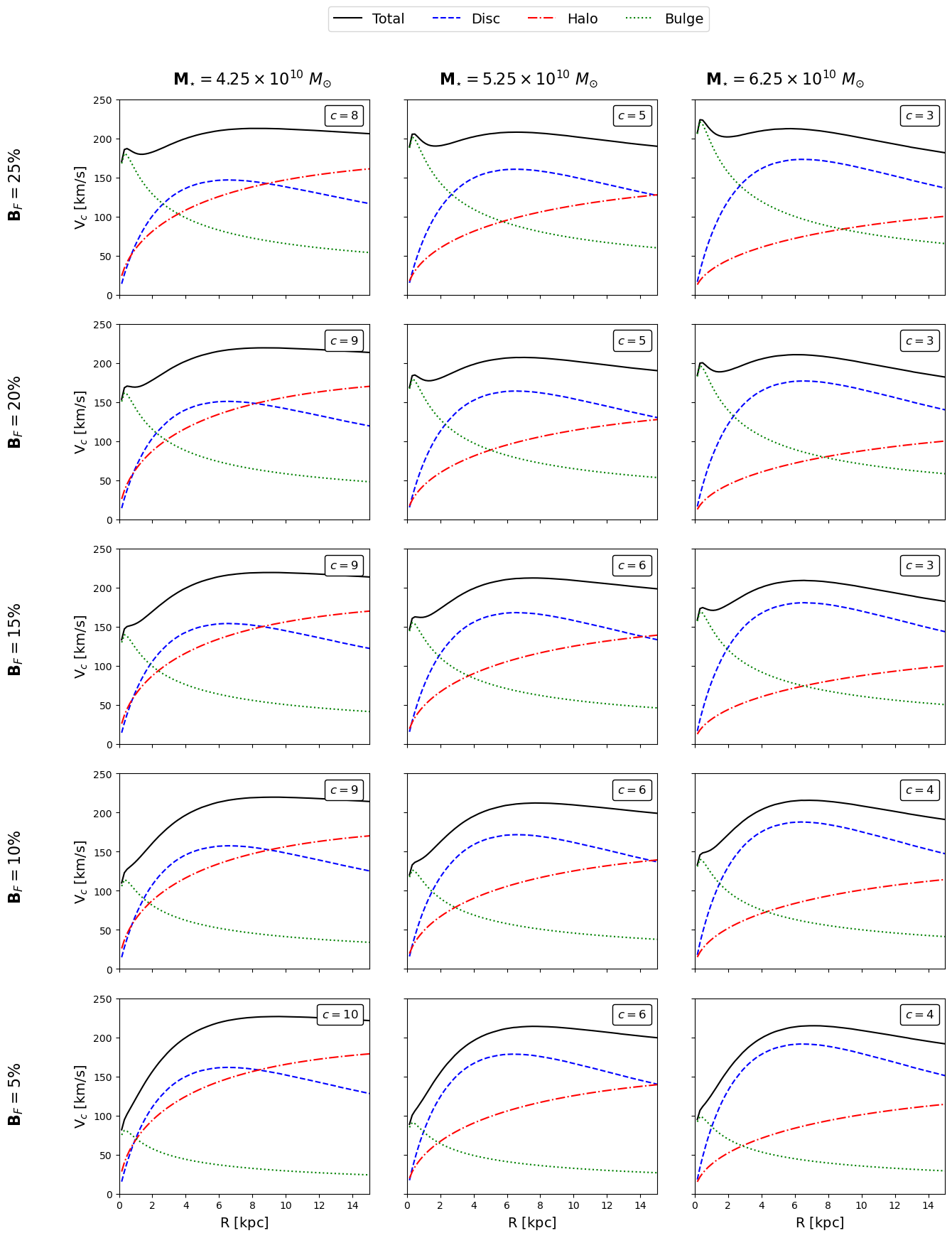} 
    \caption{
        Rotation curves for the fifteen MW-like models. Columns correspond to different initial stellar masses ($M_{\star}$), while the different rows represent different bulge mass fractions ($B_{F}$). In each plot, the black solid line represents the total rotation curve, the blue dashed line shows the disc contribution, the red dashed line represents the halo, and the green dotted line corresponds to the bulge. The halo concentration parameter ($c$) for each model is indicated in the top-right corner of each subplot.
    }
    \label{fig:rotation_curve}
\end{figure*}

Figure~\ref{fig:rotation_curve} presents the initial rotation curves of all fifteen models, highlighting the diversity of mass distributions explored in this work. The contributions from the stellar disc, bulge, and dark matter halo are shown separately. Increasing the bulge fraction primarily affects the inner rotation curve, producing a steeper central rise, while increasing the stellar mass reduces the halo contribution at large radii. 

\section{Initial parameters}
\label{appendix initial conditions}

Table~\ref{table B} summarises the parameters used to construct the fifteen MW-like galaxy models analysed in this work. The suite spans three values of the total stellar mass and five values of the stellar bulge mass fraction, while the remaining structural parameters are held fixed. Variations in stellar mass are compensated by corresponding changes in the dark matter halo mass, and the halo concentration is adjusted to reproduce realistic MW-like rotation curves.

In order to characterise the initial dynamical state, Table~\ref{table B} also reports for each model the minimum stellar Toomre parameter $Q_{\rm min}$ and the radius at which it occurs, $R_{Q_{\rm min}}$, its value at $R=2.2h$, the \cite{Efstathiou1982} parameter $\varepsilon=V_{\rm max}/(GM_{\rm d}/h)^{1/2}$, and the \cite{Ostriker1973} ratio $T_{\rm rot}/|W|$, all evaluated at the start of the simulations. $Q$ is computed for the stellar disc alone, since the bulge is a dispersion-supported component and the gas does not self-gravitate in these runs. $\Omega$ and $\kappa$ are obtained as described in Appendix~\ref{appendix resonances}.

All fifteen models lie on the unstable side of all three criteria. The minimum Toomre parameter is $Q_{\rm min} = 0.60-0.89$, occurring at $R_{Q_{\rm min}} = 4.9-8.9$ kpc, and $Q(\rm{2.2}h) = 0.60-0.93$, so the stellar discs are locally unstable to axisymmetric perturbations at the beginning of the simulations. In addition, $\varepsilon= 0.69-0.95$, below the $\varepsilon \simeq 1.1$ threshold for the global bar mode, and $T_{\rm rot}/|W| = 0.18-0.20$, above the Ostriker–Peebles value of $\simeq 0.14$. The models therefore begin in approximate equilibrium but far from stability. $Q_{\rm min}$ decreases systematically with increasing stellar disc mass, so the more massive discs begin colder relative to their own self-gravity. We note that the Ostriker–Peebles ratio should be treated with caution for a disc embedded in a massive live halo, as it was originally formulated for isolated discs.

All three criteria order the models in the same sense as the measured bar formation times, where larger $Q_{\rm min}$ and $\varepsilon$, and smaller $T_{\rm rot}/|W|$ correspond to later bar formation. In particular, the two models with the most stable initial conditions, M1 and M2, whose diagnostics agree to within 0.003 in $\varepsilon$ and 0.01 in $Q_{\rm min}$, are also the two latest to form a bar.

\begin{table*}
    \centering
    \caption{
        Initial conditions of the fifteen galaxy models generated with the {\sc makenewdisk} code, together with the stability diagnostics measured from the first snapshot of each run.
        $M_{\star}$ is the total stellar mass, $BF$ the stellar bulge mass fraction, $M_{d}^{\star}=M_{\star}(1-BF)$ the stellar disc mass, $M_{b}^{\star}=M_{\star}BF$ the stellar bulge mass, $c$ the dark matter halo concentration and $M_{h}^{dm}$ the dark matter halo mass.
        $V_{\rm max}$ is the peak circular velocity, $Q_{\rm min}$ the minimum stellar Toomre parameter and $R_{Q_{\rm min}}$ the radius at which it occurs, $Q(2.2h)$ its value at $R=2.2h$, $\varepsilon$ is the parameter of \cite{Efstathiou1982} and $T_{\rm rot}/|W|$ that of \cite{Ostriker1973}.
        All models share the same disc scale length ($3.0$ kpc), disc scale height ($0.29$ kpc) and bulge scale length ($0.35$ kpc).
    }
    \renewcommand{\arraystretch}{1.2}
    \setlength{\tabcolsep}{4pt}
    \begin{tabular}{c c c c c c c c c c c c c}
        \toprule
        & \multicolumn{6}{c}{Initial structure} & \multicolumn{6}{c}{Stability diagnostics at $t=0$} \\
        \cmidrule(r){2-7} \cmidrule(l){8-13}
        Model & $M_{\star}$ & $BF$ & $M_{d}^{\star}$ & $M_{b}^{\star}$ & $c$ & $M_{h}^{dm}$ & $V_{\rm max}$ & $Q_{\rm min}$ & $R_{Q_{\rm min}}$ & $Q(2.2h)$ & $\varepsilon$ & $T_{\rm rot}/|W|$ \\
        & [$10^{10}$ M$_{\odot}$] & [$\%$] & [$10^{10}$ M$_{\odot}$] & [$10^{10}$ M$_{\odot}$] & & [$10^{10}$ M$_{\odot}$] & [km\,s$^{-1}$] & & [kpc] & & & \\
        \midrule
        M1  & 4.25 & 25 & 3.19 & 1.06 & 8  & 89.75 & 202.9 & 0.89 & 5.38 & 0.93 & 0.94 & 0.18 \\
        M2  & 4.25 & 20 & 3.40 & 0.85 & 9  & 89.75 & 210.2 & 0.89 & 5.38 & 0.93 & 0.95 & 0.18 \\
        M3  & 4.25 & 15 & 3.61 & 0.64 & 9  & 89.75 & 210.5 & 0.86 & 4.88 & 0.89 & 0.92 & 0.18 \\
        M4  & 4.25 & 10 & 3.83 & 0.43 & 9  & 89.75 & 210.8 & 0.83 & 4.88 & 0.86 & 0.91 & 0.18 \\
        M5  & 4.25 & 5  & 4.04 & 0.21 & 10 & 89.75 & 218.0 & 0.83 & 6.12 & 0.84 & 0.90 & 0.18 \\
        \midrule
        M6  & 5.25 & 25 & 3.94 & 1.31 & 5  & 88.75 & 197.6 & 0.74 & 5.38 & 0.76 & 0.83 & 0.19 \\
        M7  & 5.25 & 20 & 4.20 & 1.05 & 5  & 88.75 & 195.7 & 0.72 & 5.38 & 0.73 & 0.79 & 0.19 \\
        M8  & 5.25 & 15 & 4.46 & 0.79 & 6  & 88.75 & 202.2 & 0.72 & 5.38 & 0.73 & 0.79 & 0.19 \\
        M9  & 5.25 & 10 & 4.73 & 0.53 & 6  & 88.75 & 202.1 & 0.70 & 5.88 & 0.70 & 0.77 & 0.19 \\
        M10 & 5.25 & 5  & 4.99 & 0.26 & 6  & 88.75 & 203.6 & 0.67 & 5.62 & 0.67 & 0.76 & 0.19 \\
        \midrule
        M11 & 6.25 & 25 & 4.69 & 1.56 & 3  & 87.75 & 215.0 & 0.65 & 8.88 & 0.67 & 0.82 & 0.20 \\
        M12 & 6.25 & 20 & 5.00 & 1.25 & 3  & 87.75 & 199.1 & 0.63 & 8.88 & 0.65 & 0.74 & 0.20 \\
        M13 & 6.25 & 15 & 5.31 & 0.94 & 3  & 87.75 & 198.0 & 0.61 & 8.88 & 0.62 & 0.71 & 0.20 \\
        M14 & 6.25 & 10 & 5.63 & 0.63 & 4  & 87.75 & 204.4 & 0.62 & 7.88 & 0.62 & 0.71 & 0.20 \\
        M15 & 6.25 & 5  & 5.94 & 0.31 & 4  & 87.75 & 203.8 & 0.60 & 7.62 & 0.60 & 0.69 & 0.20 \\
        \bottomrule
    \end{tabular}
    \label{table B}
\end{table*}

\section{Bar detection}
\label{appendix bar detection}

We identify the strength of stellar bars using the $m=2$ Fourier mode of the face-on stellar surface density distribution. For each snapshot, stellar particles with $|z|<2$ kpc are divided into radial annuli of width 0.2 kpc, we compute the normalised Fourier amplitude as

\begin{equation}
    A_2(R) = \frac{\sqrt{a_2^2(R)+b_2^2(R)}}{A_0(R)},
\end{equation}

where

\begin{equation}
    a_2(r)=\sum_i m_i \cos(2\phi_i),
    \qquad
    b_2(r)=\sum_i m_i \sin(2\phi_i),
\end{equation}

and $A_0(r)=\sum_i m_i$. Here, $m_i$ and $\phi_i$ are the mass and azimuthal angle of stellar particle $i$ within a given radial annulus. The bar strength is then defined as the maximum value of $A_2$ within the inner 8 kpc. The phase angle at each radius is computed as

\begin{equation}
    \phi_2(R)=\frac{1}{2}\tan^{-1}\left(\frac{b_2(R)}{a_2(R)}\right).
\end{equation}

We then compare the phase angle at all radii within this region to the phase angle measured at the radius of maximum bar strength.

A snapshot is classified as containing a clear bar when the phase angle remains within $3^\circ$ of the phase at the bar strength maximum over at least ten radial bins. Snapshots satisfying the amplitude criterion but exhibiting phase coherence over only five to nine radial bins are classified as possible bars, while all remaining snapshots are classified as unbarred. The resulting bar classification flag is

\begin{equation}
    {\rm flag} =
        \begin{cases}
            1, & A_{2,\rm max} > 0.25 \ \mathrm{and}\ N_{\phi} \geq 10, \\
            0.5, & A_{2,\rm max} > 0.25 \ \mathrm{and}\ 5 \leq N_{\phi} < 10, \\
            0, & A_{2,\rm max} \leq 0.25 \ \mathrm{or}\ N_{\phi} < 5,
        \end{cases}
\end{equation}

where $N_\phi$ is the number of radial bins for which the phase angle remains within $3^\circ$ of the phase measured at $A_{2,\rm max}$.

Finally, to avoid identifying transient spiral features or short-lived asymmetries as bars, we require a clear bar classification (${ \rm flag}=1$) to persist for at least $3$ consecutive snapshots before assigning a bar formation time.

\section{Single snapshot pattern speed}
\label{appendix pattern speed}

The bar pattern speed was measured using the single-snapshot Fourier method introduced by \citet{Dehnen2023}. Unlike traditional approaches that determine the pattern speed by tracking the bar orientation between consecutive snapshots, this method estimates the instantaneous angular frequency directly from the stellar positions and velocities within a single snapshot. As a result, independent measurements of $\Omega_{\rm bar}$ and their associated uncertainties can be obtained at every output time.

For each snapshot, we selected stellar particles within $R<10$ kpc and $|z|\leq0.25$ kpc. The method first performs a Fourier decomposition of the face-on stellar density distribution in adaptive radial bins. A candidate bar region is identified from the coherent phase structure of the $m=2$ mode together with the strength of the even Fourier components, following the procedure described in Appendix C of \citet{Dehnen2023}.

Once the bar region is identified, the pattern speed is determined from the time derivative of the phase of the $m=2$ Fourier mode,

\begin{equation}
    \Omega_{\rm bar} = \frac{{\rm d}\phi_2}{{\rm d}t},
\end{equation}

where $\phi_2$ denotes the phase angle of the bisymmetric density perturbation. The method evaluates this quantity using both the stellar positions and velocities, thereby avoiding the need to compare multiple snapshots.

In the public implementation used here (\texttt{patternSpeed.py} and \texttt{variance.py}), the bar region is defined by requiring a sufficiently strong even Fourier signal and a nearly constant $m=2$ phase over a finite radial interval. The pattern speed is then measured using a smooth radial weighting function, which avoids biases associated with sharp, time-dependent particle selections at the bar boundary. Statistical uncertainties are estimated internally from the covariance matrix of the Fourier coefficients and propagated following the formalism presented by \citet{Dehnen2023}.

\section{Resonance calculations}
\label{appendix resonances}

The resonance analysis requires estimates of the axisymmetric gravitational potential, circular frequency, and epicyclic frequency at each simulation snapshot. These quantities are derived directly from the potential outputs of the simulations. For each snapshot, we combine stellar, gaseous, and dark matter particles and select tracers within a cylindrical volume of radius $R<15$ kpc and height $|z|<0.25$ kpc, with radial bins of $0.25$ kpc width.

The axisymmetric potential profile, ($\Phi_0(R)$), is obtained by computing the mass-weighted average potential within each annulus,

\begin{equation}
    \Phi_0(R_i)= \frac{\sum_j m_j \Phi_j} {\sum_j m_j}
\end{equation}

where the sum extends over all particles located within the radial bin centred on ($R_i$). Bins containing fewer than 50 tracers are discarded. Missing bins are filled using linear interpolation and the resulting profile is smoothed with a Gaussian kernel of width ($\sigma=2$) radial bins to suppress numerical noise.

The circular velocity profile is then calculated from the radial derivative of the smoothed potential,

\begin{equation}
    V_{\rm c}(R)=   \sqrt{R\left|\frac{\mathrm d\Phi_0}{\mathrm dR}\right|}
\end{equation}

from which the circular frequency follows as

\begin{equation}
    \Omega(R)=\frac{V_{\rm c}(R)}{R}
\end{equation}

The epicyclic frequency is computed using

\begin{equation}
    \kappa^2(R)=4\Omega^2(R)+2R,\Omega(R)\frac{\mathrm d\Omega}{\mathrm dR},
\end{equation}

The corotation (CR), inner Lindblad (ILR), and outer Lindblad resonance (OLR) radii are obtained numerically from the intersections

\begin{equation}
    \Omega(R)=\Omega_{\rm bar},\qquad\Omega(R)-\frac{\kappa(R)}{2}=\Omega_{\rm bar},\qquad\Omega(R)+\frac{\kappa(R)}{2}=\Omega_{\rm bar},
\end{equation}

respectively. When multiple intersections are present, all solutions are retained, allowing for the possibility of multiple ILRs. The resonance radii derived in this manner are subsequently used to study the secular evolution of the barred galaxies in our simulation suite.

\section{Quantifying the Boxy/Peanut Structure}
\label{app:BP_method}

To quantify the growth of boxy/peanut (B/P) structures we follow the method introduced by \citet{Ghosh2024}, which measures the relative increase in the vertical thickness of the stellar distribution with respect to the initial disc configuration. Similar approaches have recently been applied to cosmological simulations by \citet{PaulaLopez2025}.

For each simulation snapshot, stellar particles are first rotated such that the bar major axis lies along the $x$-direction. Following \citet{Ghosh2024}, we then select particles within a narrow slice around the bar major axis. In our simulations, we identify stellar particles within $|y| < 1.2~{\rm kpc}$, $|z| < 2~{\rm kpc}$ and a radius of 6 kpc, which encompasses the full radial extent of the B/P structure in all models. 

The B/P strength profile is then defined as

\begin{equation}
    S(R,t)=\frac{\langle |z|(R,t)\rangle}{\langle |z|(R,t_0)\rangle},
    \label{eq:BP_strength}
\end{equation}

where $\langle |z|(R,t)\rangle$ denotes the mean absolute vertical height of stars within a radial bin at time $t$, and $t_0$ corresponds to the initial snapshot of the simulation. Values of $S=1$ therefore indicate no net vertical thickening relative to the initial disc, while $S>1$ corresponds to the development of a vertically extended structure.

The radial location of the maximum of $S(R,t)$ traces the region where the peanut shape is strongest, while the peak value of the profile provides a measure of the overall B/P strength.

\begin{figure}
    \centering
    \includegraphics[width=\columnwidth]{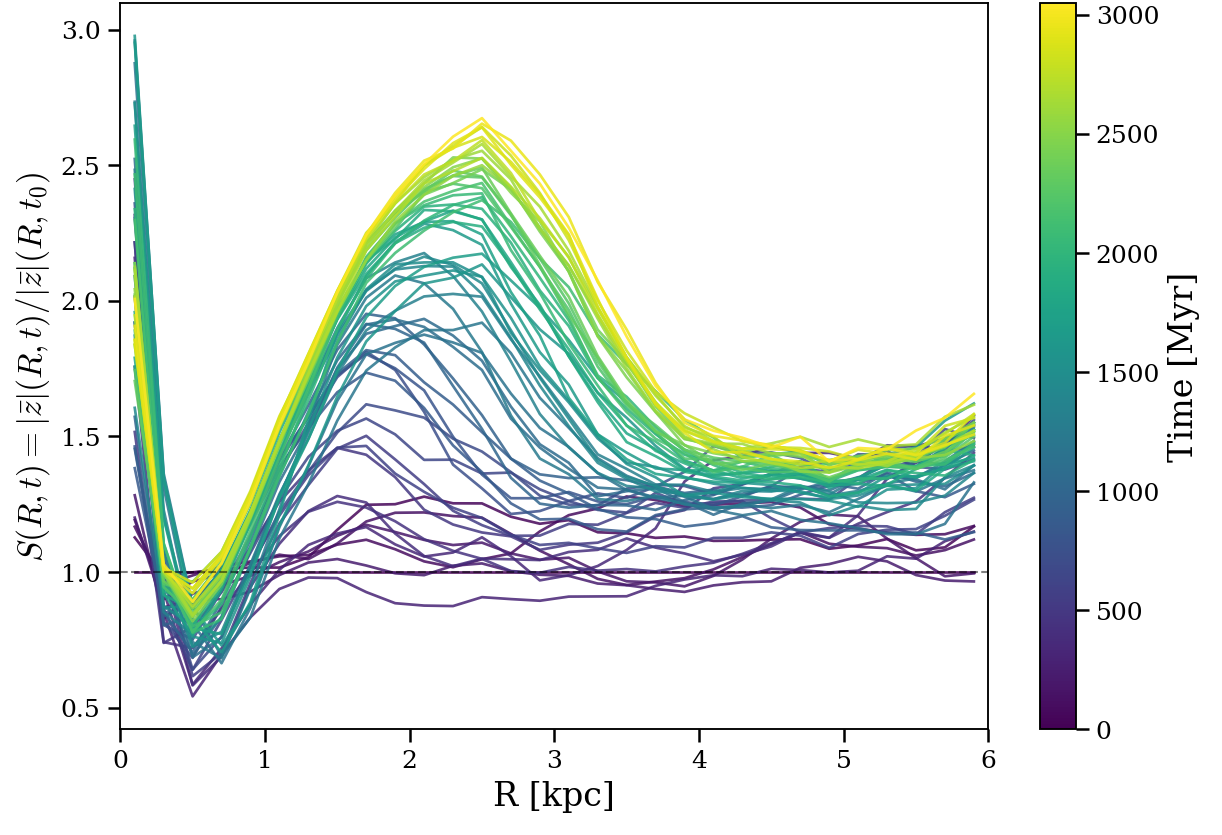}
    \caption{
        Example evolution of the boxy/peanut (B/P) strength profile for model M12. Colours indicate time, increasing from dark purple to yellow. The quantity $S(R,t)$ measures the mean absolute vertical height of stars at radius $R$, normalised by the corresponding value in the initial disc (Equation~\ref{eq:BP_strength}). The horizontal dashed line marks $S=1$, corresponding to the initial vertical thickness of the disc.
    }
    \label{fig:BP_example}
\end{figure}

An illustrative example of the evolution of the B/P strength profile is shown in Figure~\ref{fig:BP_example} for model M12. At early times the profile remains close to $S=1$, indicating a vertically thin disc. As the bar grows and undergoes secular evolution, a prominent peak develops at $R\sim2$--3 kpc, demonstrating the emergence of a vertically thickened boxy/peanut structure. The increase of the peak amplitude with time reflects the strengthening of the B/P bulge, while its radial position identifies the region most affected by vertical heating and orbit trapping.

\section{Determination of B/P formation times}
\label{appendix:BPformation}

The vertical thickening diagnostic described in Appendix~\ref{app:BP_method} provides a radial profile

\begin{equation}
    S(R,t)=\frac{|\bar{z}|(R,t)}{|\bar{z}|(R,t_0)},
\end{equation}

for every simulation snapshot. While the peak value of this profile measures the strength of the B/P structure, the shape of the profile itself contains information on when the peanut morphology first emerges. To quantify this transition, we first smooth each radial profile using a third-order Butterworth filter with a characteristic scale of $1.2$ kpc. The first and second radial derivatives are then computed numerically,

As the B/P structure develops, the profile acquires a characteristic concave shape, producing a pronounced negative minimum in ${\rm d}^2S/{\rm d}R^2$. For each snapshot, we therefore identify the strongest local minimum of ${\rm d}^2S/{\rm d}R^2$ within $0.4<R<5$ kpc.

The B/P formation time is defined as the first time for which

\begin{equation}
    \min \left(\frac{{\rm d}^2S}{{\rm d}R^2}\right) < -1.19
\end{equation}

for at least five consecutive snapshots. The threshold value was calibrated through visual inspection of the simulations and was found to robustly identify the emergence of visually recognisable B/P structures across the model suite. 

\begin{figure}
    \centering
    \includegraphics[width=\columnwidth]{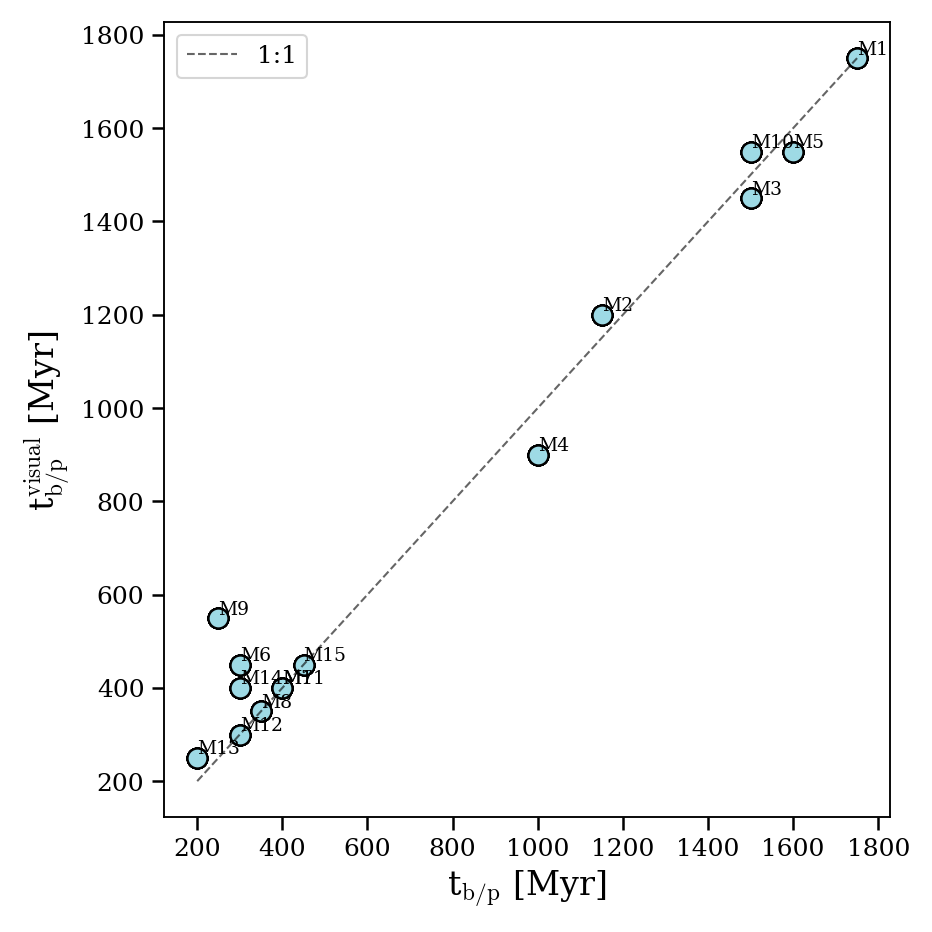}
    \caption{
    Comparison between the B/P formation times determined using the automated curvature criterion described in this appendix ($t_{\rm b/p}$) and those obtained from visual inspection of the edge-on stellar density maps ($t_{\rm b/p}^{\rm visual}$). The dashed line shows the one-to-one relation. Most models lie close to this line, demonstrating that the threshold $\min({\rm d}^2S/{\rm d}R^2)<-1.19$ for five consecutive snapshots provides a reliable and objective estimate of the onset of the B/P structure. The remaining scatter reflects both the finite temporal resolution of the simulations and the subjective nature of visual identification.
    }
    \label{fig:BP visual vs code}
\end{figure}

We independently estimated the B/P formation time through visual inspection of the edge-on stellar density maps to assess the reliability of this automated criterion. The visual estimate, $t_{\rm b/p}^{\rm visual}$, was defined as the earliest snapshot at which a clear boxy or peanut-shaped morphology could be identified. Figure~\ref{fig:BP visual vs code} compares these visually determined formation times with those obtained from the curvature criterion described above.

The agreement is generally very good. Most models lie close to the one-to-one relation, indicating that the automated method successfully recovers the onset of the B/P structure within approximately one simulation output interval. Small deviations are expected because the visual identification of a B/P morphology is inherently subjective and because the curvature criterion is sensitive to the earliest stages of vertical thickening, which may precede the appearance of a fully developed peanut shape. Nevertheless, the close correspondence between the two estimates demonstrates that the adopted threshold provides a robust and objective measure of the B/P formation time throughout the model suite.

\end{document}